\documentclass[journal,comsoc]{IEEEtran}

\usepackage[T1]{fontenc}

\usepackage[switch,columnwise]{lineno}
\usepackage{cite}
\usepackage{url}
\usepackage{subfig}
\usepackage{caption}
\usepackage{amsmath,amsthm,amssymb}
\usepackage{algorithmic}
\usepackage{graphicx}
\usepackage{float}
\usepackage{textcomp}
\usepackage{color}
\usepackage{mathtools}
\usepackage{mathrsfs}
\usepackage{bm}
\usepackage{balance}
\usepackage{acronym}
\usepackage{epstopdf}
\usepackage{threeparttable}
\usepackage[table, dvipsnames]{xcolor}
\usepackage{colortbl}
\usepackage{multicol,lipsum}
\usepackage[ruled,linesnumbered]{algorithm2e}
\usepackage{cite}
\usepackage{tabularx}
\usepackage{booktabs}
\usepackage{tikz}
\usepackage{bbding}
\usepackage{makecell}
\usepackage{multirow} 
\usepackage{rotating}
\usepackage{longtable}
\usepackage{adjustbox}
\usepackage{pdflscape}

\newcommand{\beq}{\begin{equation}}
	\newcommand{\eeq}{\end{equation}}
\newcommand{\beqa}{\begin{eqnarray}}
	\newcommand{\eeqa}{\end{eqnarray}}

\begin{document}


\title{Reliability and Availability in Virtualized Networks: A Survey on Standards, Modeling Approaches, and Research Challenges\footnotemark}



\author{
	Mario~Di~Mauro, Walter~Cerroni, Fabio~Postiglione, Massimo~Tornatore, Kishor~S.~Trivedi
 
		\IEEEcompsocitemizethanks{\IEEEcompsocthanksitem M. Di Mauro and F. Postiglione are with University of Salerno, Italy and with National Inter-University Consortium for Telecommunications (CNIT), Italy (\{mdimauro,fpostiglione\}@unisa.it).
        \\
        \indent W. Cerroni is with the Department of Electrical, Electronic, and Information Engineering ``G. Marconi,'' University of Bologna, Italy (walter.cerroni@unibo.it).
        \\
        \indent M. Tornatore is with Politecnico di Milano, Italy (massimo.tornatore@polimi.it).
        \\
        \indent K.S. Trivedi is  is with the Department of Electrical and Computer Engineering, Duke University, USA (ktrivedi@duke.edu)
		\protect\\
		
	}
}

\maketitle
\setcounter{footnote}{1}
\footnotetext{This manuscript is a preprint and has not been peer-reviewed. A shorter version has been submitted to IEEE Communications Surveys \& Tutorials (COMST). Copyright may be transferred upon acceptance.}

\begin{abstract}


The rise of Network Function Virtualization (NFV) has transformed network infrastructures by replacing fixed hardware with software-based Virtualized Network Functions (VNFs), enabling greater agility, scalability, and cost efficiency. While inspired by cloud computing innovations, this shift introduces new challenges in ensuring reliability and availability. 
Virtualization increases the distribution of system components and introduces stronger interdependencies. As a result, failures become harder to predict, monitor, and manage compared to traditional monolithic networks.
Reliability, i.e. the ability of a system to perform regularly under specified conditions, and availability, i.e. the probability of a system of being ready to use, are critical requirements that must be guaranteed to maintain seamless network operations. Accurate modeling of these aspects is crucial for designing robust, fault-tolerant virtualized systems that can withstand service disruptions, ensuring continuous user access.
Accordingly, this survey focuses on reliability and availability attributes of virtualized networks from a modeling perspective. After introducing the NFV standard architecture and basic definitions, we discuss the standardization efforts of the European Telecommunications Standards Institute (ETSI), which provides guidelines, recommendations, and architectural solutions through a series of standard documents focusing on reliability and availability (denoted by the acronym ETSI-REL).  
Next, we explore several formalisms proposed in the literature for characterizing reliability and availability, with a focus on their application to modeling the failure and repair behavior of virtualized networks through practical examples. As a relevant contribution, we overview numerous references demonstrating how different authors adopt specific methods to characterize reliability and/or availability of architectural components in a virtualized network system. Moreover, we present a selection of the most valuable software tools that support modeling of reliable virtualized networks. Finally, we discuss a set of challenges and open problems with the aim to encourage readers to explore further advances in the field, and contribute to the ongoing development of reliable virtualized networks.
\end{abstract}


\IEEEpeerreviewmaketitle

\section*{List of Acronyms}
\vspace{3 mm}
\begin{acronym}[11111111111]
	\setlength{\parskip}{0.5ex}
	\setlength{\itemsep}{0.2ex}
	\acro{AMF}{Access and Mobility Management Function}
	\acro{CNF}{Containerized Network Function}
	\acro{CNR}{Containerized Network Replica}
	\acro{CNT}{Container}
	\acro{CSCF}{Call Session Control Function}
	\acro{CSPN}{Coloured Stochastic Petri Networks}
	\acro{CTMC}{Continuous-Time Markov Chain}
	\acro{DCK}{Docker}
	\acro{DTMC}{Discrete-Time Markov Chain}
	\acro{EMS}{Element Management System}
	\acro{ETA}{Event Tree Analysis}
	\acro{ETSI}{European Telecommunications Standards Institute}
	\acro{FMEA}{Failure Modes and Effects Analysis}
	\acro{FMECA}{Failure Modes, Effects, and Criticality Analysis}	
	\acro{FT}{Fault Tree}
	\acro{FTA}{Fault Tree Analysis}
	\acro{GSPN}{Generalized Stochastic Petri Networks}
	\acro{HA}{High Availability}
 	\acro{HSS}{Home Subscriber Server}
	\acro{HYP}{Hypervisor}
	\acro{HW}{Hardware}
	\acro{IaaS}{Infrastructure as a Service}
	\acro{IMS}{IP Multimedia Subsystem}
	\acro{MDD}{Multi-valued Decision Diagram}
	\acro{MEC}{Mobile Edge Computing}
	\acro{MRGP}{Markov Regenerative Process}
	\acro{MRM}{Markov Reward Model}
	\acro{MSS}{Multi-State System}
	\acro{MTTF}{Mean Time to Failure}
	\acro{MTTR}{Mean Time to Repair}
 	\acro{MUGF}{Multidimensional Universal Generating Function}
	\acro{NFV}{Network Function Virtualization}
 	\acro{NFV-REL}{NFV Reliability (ETSI standard)}
	\acro{NFVI}{Network Function Virtualization Infrastructure}
	\acro{NFV-MANO}{NFV Management and Network Orchestration}
	\acro{NFVO}{Network Function Virtualization Orchestration}
	\acro{NMP}{Non-Markovian Process}
	\acro{NS}{Network Slicing}
	\acro{OSS/BSS}{Operations/Business Support Systems}
    \acro{PTE}{Phase-Type Expansion}
      \acro{QoS}{Quality of Service}
	\acro{RAMS}{Reliability, Availability, Maintainability, and Safety}
	\acro{RBD}{Reliability Block Diagram}
	\acro{SAN}{Stochastic Activity Networks}
	\acro{SDN}{Software Defined Networking}
 	\acro{SF}{Service Function}
	\acro{SFC}{Service Function Chain}
	\acro{SLA}{Service Level Agreement}	
	\acro{SMF}{Session Management Function}
	\acro{SMP}{Semi-Markov Process}
	\acro{SPN}{Stochastic Petri Networks}
	\acro{SRN}{Stochastic Reward Networks}
 	\acro{UE}{User Equipment}
	\acro{UGF}{Universal Generating Function}
	\acro{UPF}{User Plane Function}
	\acro{VIM}{Virtualized Infrastructure Manager}
	\acro{VM}{Virtual Machine}
	\acro{VNF}{Virtual Network Function}
 	\acro{VNF-FG}{VNF Forwarding Graph}
	\acro{VNFM}{Virtual Network Function Manager}
\end{acronym}

\section{Introduction and Organization}
\label{sec:intro}

Network Function Virtualization (NFV) refers to the transformation of traditional hardware-based network functions into software-based implementations of service components, also known as Virtualized Network Functions (VNFs), that can run on commodity computing hardware~\cite{nfv_1,nfv_4}. Unlike traditional devices such as routers, firewalls, and switches, VNFs can be dynamically instantiated, scaled, and migrated across different hardware platforms, bringing to network management and service provisioning the agility and efficiency levels typical of cloud computing paradigms~\cite{nfv_2,nfv_3}. 

In virtualized network scenarios, \textit{reliability} and \textit{availability} are very critical aspects, as virtualized systems are inherently characterized by increased complexity and component heterogeneity, making them particularly vulnerable to unplanned downtime due to not only hardware faults, but also software bugs, malfunctioning and misconfigurations~\cite{intel}. Reliability refers to the system's ability to perform its required functions under stated conditions without failure during a given interval of time, while availability is the probability that a system is operational when required, often expressed as a percentage of uptime over a given period.

Accurate reliability/availability modeling of virtualized infrastructures is crucial for designing fault-tolerant systems resilient to service disruption, enabling users to access resources as needed. This is evidenced by standardization efforts undertaken by the European Telecommunications Standards Institute (ETSI), which has proposed a set of dedicated standard documents focusing on reliability and availability problems specifically related to NFV-based infrastructures.

However, translating the pragmatic needs related to reliability/availability design of virtualized networks into abstract formal models is challenging because it requires a combination of both practical and methodological skills. This process involves not only understanding theoretical frameworks and mathematical models but also applying hands-on experience gained from working on real-world problems.

Inspired by these challenges, in this survey we delve into multiple facets of reliability and availability in virtualized networks, exploring standards, formalisms, and relevant modeling software packages. Our analysis aims to strike a balance, providing insights into formalisms without overwhelming readers with excessive mathematical detail. Instead, we focus on practical implications and their relevance to real-world applications, ensuring accessibility and applicability for a wide audience.

\begin{figure}[t]
	\centering
	\captionsetup{justification=centering}
	\includegraphics[scale=0.48]{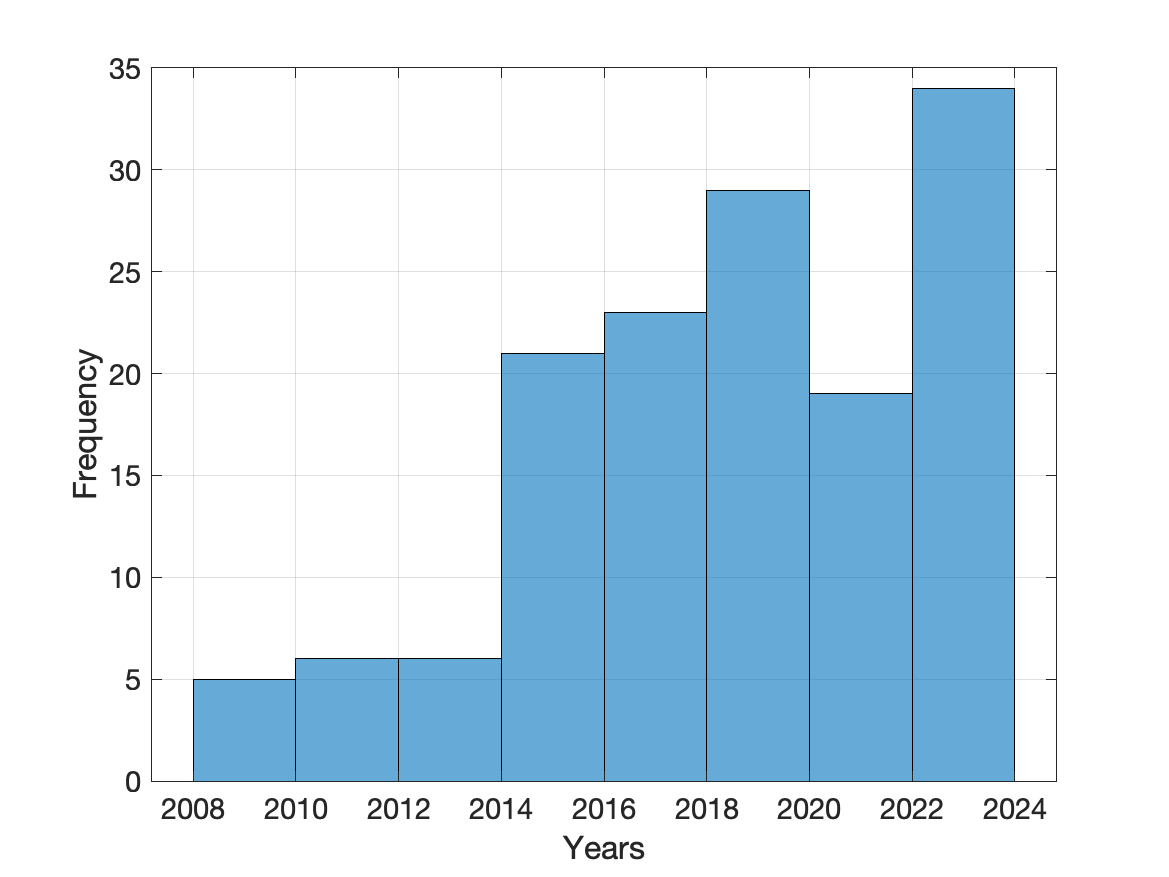}
	\caption{Distribution of years in the selected references.}
	\label{fig:year_distr}
\end{figure} 

\begin{figure*}[t]
	\centering
	\captionsetup{justification=centering}
	\includegraphics[scale=0.32]{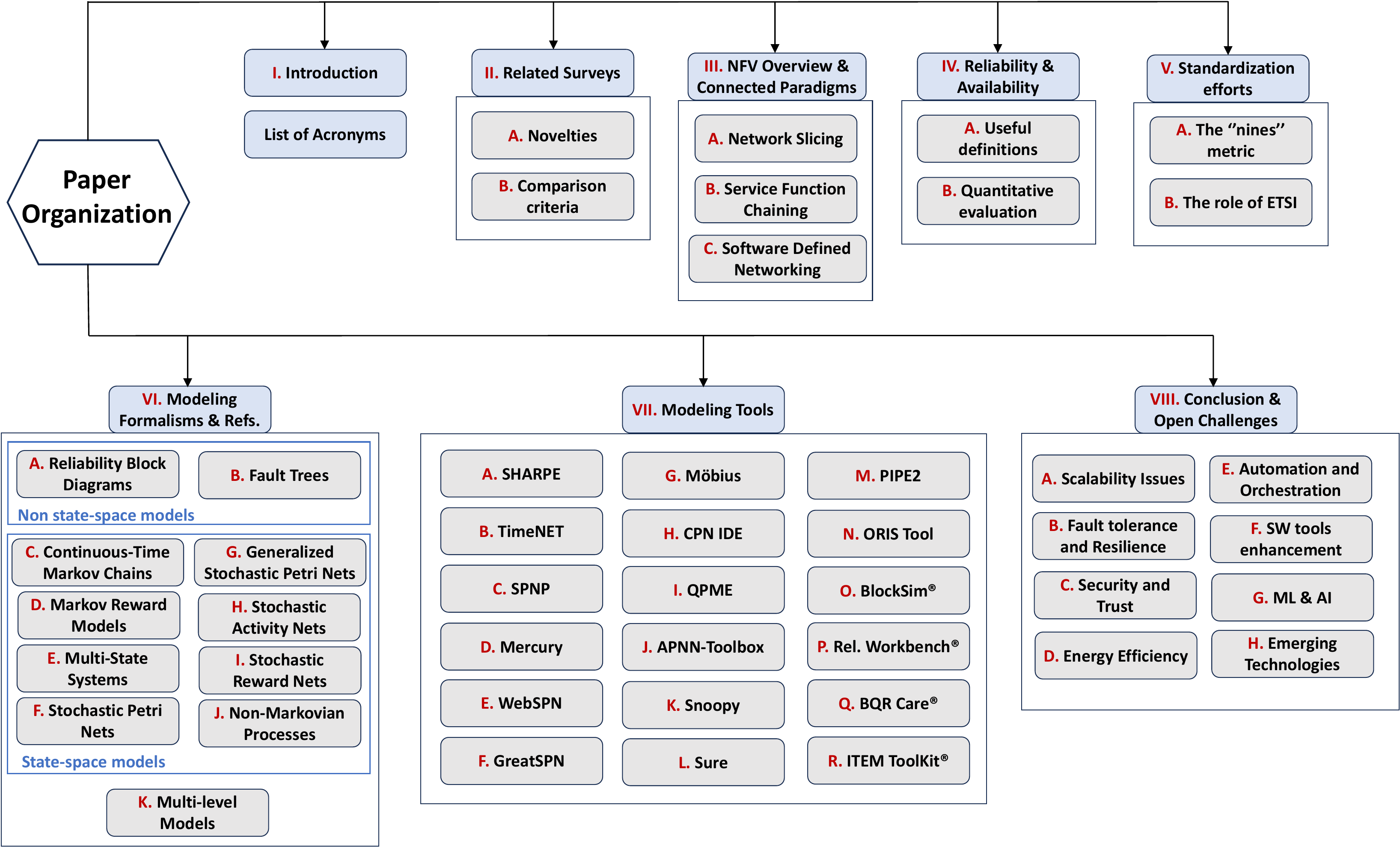}
	\caption{Paper Organization.}
	\label{fig:graph_org}
\end{figure*} 

The content of the paper is organized as follows. 

Section~\ref{sec:relsurv} discusses existing surveys that focus on various aspects of reliability and availability in virtualized and cloud-based network infrastructures. Each survey offers a unique perspective on the topic, such as resiliency techniques in cloud networks, dependability problems for the NFV orchestrator, or high-availability aspects in software-defined networks. This section also emphasizes the original perspective adopted in our survey, distinguishing it from related literature.

Section~\ref{sec:overview} provides a brief overview of the NFV ecosystem, along with related paradigms such as network slicing, service function chaining, and software-defined networks, highlighting the main reliability and availability challenges.

In Sect.~\ref{sec:rel_ava_concepts} we introduce formal definitions related to reliability and availability, which will be extensively referenced throughout the survey.

Section~\ref{sec:standard} addresses the reliability aspects of virtualized networks from a standardization point of view, discussing the guidelines released by ETSI (and denoted by the acronym ETSI NFV-REL).

Section~\ref{sec:formalisms} introduces a taxonomy of the most well-known modeling formalisms, categorized as non-state space, state-space, and multi-level, according to credited literature \cite{trivedi-bobbio}. This section outlines the pros and cons of each formalism and includes an extensive \textit{mapping} of $142$ selected references ($76$ conferences, $62$ journals, $4$ book chapters), demonstrating how to apply the existing models to characterize virtualized networks from a reliability and availability viewpoint. The distribution of the references in terms of year of appearance is shown in Fig.~\ref{fig:year_distr}. 

In Sect.~\ref{sec:tools}, we present a selection of the most well-known software tools, both commercial and open-source, that support the formal modeling of reliable virtualized networks.

Section~\ref{sec:conclusions} concludes the survey by also identifying a series of challenges and open issues designed to stimulate readers to pursue further advances in the field of reliable virtualized networks.

For the reader's convenience, the whole paper structure is outlined in Fig.~\ref{fig:graph_org}.

\section{Related Surveys and Novelties of the Current Work}
\label{sec:relsurv}

The primary objective of this survey is to offer a systematic compendium for researchers and practitioners interested in modeling reliability and availability  of virtualized network systems, and in becoming familiar with the formalisms, methodologies, and software tools used for modeling.
Although there are no existing surveys offering a similar angle, it is useful to analyze other related surveys that offer alternative perspectives. 
For the sake of conciseness, in this section, we consider only a selection of surveys that are closely related to reliability and availability topics applied to virtualized network infrastructures. 

We find it convenient to group these surveys based on specific application domains of network virtualization: Cloud computing environments, NFV components/modules, Software-defined systems, Domain-specific environments.

It is also worth noting that, although the authors of the following surveyed papers may define some concepts slightly differently, they generally follow Avizienis' terminology~\cite{avizienis}, where: \textit{failure} is an event that occurs when the delivered service deviates from the correct service; \textit{fault} is the hypothesized cause of an error that leads to a service failure; \textit{fault tolerance} refers to the ability to prevent service failures; \textit{resiliency} is often considered synonymous with fault tolerance; and \textit{robustness} describes the system’s ability to continue functioning despite external faults.

\newpage
 \textit{Cloud Computing environments}


A survey on resiliency techniques in cloud computing is proposed in~\cite{survey_meix}. 
The authors offer a qualitative overview of various types of failures that can occur in cloud infrastructures including physical and virtual layers. Then, a number of techniques for cloud computing resiliency are surveyed. Such techniques are categorized into different groups: $i)$ resiliency in facilities (e.g., power supplies, business continuity); $ii)$ resiliency in servers (e.g., physical and virtual server, RAID and backup strategies); $iii)$ resiliency in networks (e.g., resilient design and traffic engineering, topology redundancy); $iv)$ resiliency in cloud integrated infrastructures (e.g., reliable placement of  virtual resources, failure  detection and mitigation); $v)$ resiliency in cloud middleware infrastructures (e.g., checkpoint management, resilient load balancing); $vi)$ measurement of resiliency (e.g. empirical measurements, passive measurements). The last part of this survey focuses on the resiliency in application design and development, where core objects (logic of the application) are separately analyzed from data objects (content used and/or produced by the logic).

A survey of fault tolerance methods for cloud computing environments is proposed in~\cite{survey_mukwevho}, where the authors present a general three-category taxonomy: Reactive methods (RAMs), Proactive methods (PRMs), and Resilient methods (RSMs). Reactive methods are employed to mitigate the effect of failures once they have occurred, and include techniques such as restarting, replication, and exception handling, task resubmission. Typically, RAMs are not well-suited for short running tasks (e.g., real-time) since the recovery from a failure adds non-negligible delays to the overall response times. 
Proactive methods are based on the prediction of failures and include software rejuvenation, self-healing, preemptive migration, and active monitoring. Differently from RAMs, PRMs are suitable for real-time applications.
Resilient methods enable a system to continue providing critical services even in presence of failures. 
The main purpose of RSMs is to predict faults and deploy strategies to prevent or mitigate their impact.

RSMs can benefit from the application of reinforcement learning algorithms, aimed at continuously learning the environment and adapting to fault tolerance situations. 

In~\cite{survey_amiri}, the authors present a review of resilient and dependable approaches in cloud-based distributed environments, examining aspects such as fault tolerance and redundancy. The review primarily categorizes papers based on criteria such as the publisher (e.g., IEEE, ACM, Elsevier), frameworks and software tools used in simulations/analytics/testbeds (e.g., Matlab, Raspberry Pi, JavaScript), attributes considered (e.g., resiliency, scalability, dependability), and architectures studied (e.g., Cloud computing, Fog computing, IoT).

Finally, a survey on availability mechanisms and strategies adopted in cloud computing systems is presented in~\cite{survey_nabi}. To identify the most relevant papers for their analysis, the authors of the survey developed a taxonomy that considers metrics related to failures, failure detection, and replication.
\newline
\newline
\textit{NFV components/modules}

A survey on fault management problems affecting NFV-based environments is presented in~\cite{survey_cherrared}, where the authors pinpoint several issues pertaining to: $i)$ \textit{scalability}, meant as the ability of a network to maintain its performance as traffic increases, guaranteed by the distribution of VNFs which implies an increasing robustness to service failures; $ii)$ \textit{topology}, dealing with the possibility for an NFV infrastructure to flexibly adapt its topology as network failures occur; $iii)$ \textit{granularity}, referring to the possibility of precisely identifying a fault into an NFV layer due to the hardware/software decoupling logic; $iv)$ \textit{fault tolerance}, referring to the possibility of avoiding the single point of failure problem through a smart redundancy of virtualized nodes; $v)$ \textit{performance}, meant as the ability of guaranteeing a given Quality of Service (QoS) level.  

The survey proposed in~\cite{survey_gonzalez} focuses on a specific component of virtualized networks represented by the NFV Orchestrator (NFVO), whose dependability is analyzed according to the concept of dependability defined in~\cite{avizienis}. In particular, the authors of this survey stress the fact that the NFVO should be characterized by a strong robustness. To guarantee such a condition, fault detection, redundancy strategies, diagnosis and detection policies must be adequately planned. 
Accordingly, some crucial principles to design a dependable NFVO are discussed, including: $i)$ avoiding single point of failures by adopting modularity and redundancy strategies; $ii)$ implementing redundant control policies such as distributed-load or master/slave; $iii)$ considering failure-independent domains aimed at avoiding the propagation of failures across the various NFV modules. 

Reliability and availability issues of the NFV ecosystem are also highlighted in~\cite{survey_mijumbi}, where the authors clarify that such features are not only related to the customers expectation, but are also part of regulatory requirements. This is because virtualized infrastructures are today part of critical national networks, thus, legal obligations for service assurance and business continuity are in place.  

A taxonomy of various aspects of network services dependability is proposed in~\cite{survey_azadiabad1}. Specifically, the authors focus on service availability, reliability, and continuity as key attributes of dependability. They review 102 papers, categorized according to different criteria, including protection mechanisms, life-cycle phases, and solution approaches to address network service dependability in the context of NFV.
\newline
\newline
\textit{Software-defined systems}

Authors in~\cite{survey_fonseca} adopt a perspective mainly focused on Software Defined Networking (SDN), which provides the main separation logic between control and data planes into the realm of virtualized networks. In particular, fault tolerance issues are categorized per \textit{layers}. Fault tolerance problems at the \textit{infrastructure layer} are mostly related to issues also present in traditional networks and include nodes and links failures. The difference is that an SDN infrastructure makes it easier to manage: $i)$ network failure detection and location, since an SDN controller has a logically centralized view of all SDN switches affected by specific failures, and $ii)$ network failure recovery, since an SDN controller can manage traffic engineering, links backup, routes reorganization. Fault tolerance problems at the \textit{control layer} involve both infrastructural aspects, since multiple/distributed controllers may be deployed to avoid the single point of failure, and protocol aspects, since the OpenFlow protocol natively supports particular messages to assign different roles to the controller (Master/Slave/Equal). Fault tolerance problems at the \textit{application layer} typically involve the application design that should take into account fault tolerance requirements, and the application correctness which implies a strong test phase. 

Considering again the challenges involving the SDN fault management, authors in~\cite{survey_yu} focus on four tasks: $i)$ \textit{system monitoring}, related to the activity of monitoring the SDN system behavior and collecting statistical data related to monitoring metrics; $ii)$ \textit{fault diagnosis}, related to the possibility of detecting faults along with the location of the pertinent root cause; $iii)$ \textit{fault recovery and repair}, related to the techniques aimed at reconfiguring a system after a fault occurred; $iv)$ \textit{fault tolerance}, related to proactively preventing the faults with the goal of guaranteeing service continuity.
\newline
\newline
\textit{Domain-specific environments}

Aspects related to reliability and high availability in virtualized networks for domain-specific environments, such as mobile core networks and service function chain (SFC) deployments, are also addressed in surveys~\cite{survey_nguyen} and~\cite{survey_kaur}, respectively. As already highlighted in many other works, the presence of a single controller in an SDN network is risky since it is a potential single point of failure. Thus, according to~\cite{survey_nguyen}, a hot-standby design for the SDN controller with proper redundancy is highly recommended to increase availability and reliability in mobile core networks. Similar issues arise in an NFV environment, where the problem of maintaining the state of virtualized elements can be faced through a proper redundancy strategy that takes into account the reliability/performance trade-off.

The survey in~\cite{survey_kaur} focuses on the SFC provisioning in NFV/SDN architectures, where the whole section $5.3$ is devoted to the availability-aware deployment of SFCs. Authors survey a number of works dealing with recovery schemes, backup strategies, failure detection and localization, algorithms to provide resilient SFC services with minimum resource redundancy, reliability-aware VNF placement to optimize the SFC resilience, etc.

\subsection{Main contributions of this survey}
In our work, we provide a set of unique contributions that distinctly differentiate it from the surveys analyzed before.
First, we give readers a clear understanding of the standardization efforts related to reliability and availability in virtualized networks, by examining and commenting on a series of ETSI documents and guidelines. In this overview, we discuss how these standards influence the design and management of reliable and available virtualized networks.
Only few of the existing surveys propose a systematic classification of ETSI documents, which are crucial for the reader to capture the standard guidelines in this field.

Second, we present various formalisms and methods used to model reliability and availability in virtualized networks. 
This part is written to be accessible to readers who may not be highly skilled in mathematics.  Key elements include: $i)$ taxonomy of formalisms and methods, where we introduce a range of modeling approaches, each explained in a clear and accessible language, $ii)$ pros and cons, outlining the advantages and limitations of each formalism and method, helping readers to understand the contexts in which each approach is most effective, $iii)$ 
a structured overview of selected references, designed to help readers easily determine which formalism is best suited for modeling specific aspects of virtualized networks.
Existing surveys only mention the techniques used to model the failure/repair behavior of virtualized networks, but they do not delve into the specifics of the formalisms or explain why the authors choose them.

Finally, for readers interested in applying these formalisms in the field, we outline the main features of the most well-known software tools, both open source and commercial. 
Most existing surveys do not propose a systematic categorization of software tools useful for modeling the failure/repair behavior of virtualized networks. Instead, those surveys mention only highly customized software tools created ad hoc to solve very specific problems in the field.

By integrating the aforementioned contributions, our survey aims to offer a balanced and comprehensive resource that elucidates the theoretical, standardization, and practical aspects of reliability and availability in virtualized networks.

\subsection{Comparison criteria}
A concise comparison of our work with the surveys analyzed above is provided in Table~\ref{tab:comparison}. While each study has its unique characteristics, we have identified a set of general criteria to evaluate our work against the discussed technical literature. These criteria are as follows:
\begin{itemize}
    \item \textit{Techniques \& metrics}: The work examines techniques, formalisms, models, and metrics related to the reliability and availability of virtualized networks;  
    \item \textit{Standard guidelines}: The work examines the standardization processes governing the reliability and availability of virtualized networks;
    \item \textit{Tools \& SW support}: The work assesses software tools that can effectively model the reliability and availability aspects of virtualized networks, highlighting their key features;
    \item \textit{Practical examples \& case studies}: The work provides an extensive collection of practical examples and case studies, helping readers identify the best methodologies for analyzing their specific scenarios;
    \item \textit{Tutorial-like characteristics}: The work includes actionable insights on the use of formalisms in reliability and availability assessments of virtualized networks, while avoiding excessive technical or mathematical complexity.
\end{itemize}
To help the reader to visualize the positioning of our work with respect to existing surveys, in Table~\ref{tab:comparison} we assigned scores to each criterion as follows: sufficient coverage (\checkmark), good coverage (\checkmark \checkmark), and high coverage (\checkmark \checkmark \checkmark).

\begin{table*}[t]  
\centering  
\caption{Summary and comparison of related survey papers} 
\scalebox{0.5}{ 
\Large
\begin{tabular}{c c c c c c c}    
\toprule    
\textbf{Survey} & \textbf{\makecell{Techniques \\ \& metrics}} & \textbf{\makecell{Standard \\ guidelines}} & \textbf{\makecell{Tools \& \\ SW support}} & \textbf{\makecell{Practical examples  \\ \& case studies}} & \textbf{\makecell{Tutorial-like \\ characteristics}} & \textbf{\makecell{Brief description}} \\    
\midrule    
Colman-Meixner et al.~\cite{survey_meix} & \checkmark \checkmark \checkmark  & \checkmark & \checkmark \checkmark & \checkmark \checkmark \checkmark & \checkmark \checkmark &  \makecell{Qualitative overview  of failures in virtualized networks, \\ and analysis of cloud computing resiliency techniques}   \\   
\midrule   
Mukwevho and Celik~\cite{survey_mukwevho} & \checkmark \checkmark \checkmark & \checkmark & \checkmark \checkmark & \checkmark & \checkmark & \makecell{Three-category taxonomy of fault-tolerance methods \\ in cloud computing environments}  \\  
\midrule  
Amiri et al.~\cite{survey_amiri} & \checkmark \checkmark & \checkmark \checkmark & \checkmark \checkmark \checkmark & \checkmark & \checkmark & \makecell{Resilient and dependable approaches \\ in cloud networks} \\ 
\midrule   
Nabi et al.~\cite{survey_nabi} & \checkmark \checkmark \checkmark & \checkmark \checkmark & \checkmark & \checkmark \checkmark & \checkmark \checkmark & \makecell{Availability strategies and fault tolerance mechanisms \\ in cloud computing networks} \\ 
\midrule 
Cherrared et al.~\cite{survey_cherrared} & \checkmark & \checkmark \checkmark & \checkmark \checkmark & \checkmark & \checkmark & \makecell{Fault management issues and approaches \\ in NFV-based architectures} \\  
\midrule   
 Gonzalez et al.~\cite{survey_gonzalez} & \checkmark \checkmark \checkmark &  \checkmark & \checkmark & \checkmark \checkmark & \checkmark & \makecell{Focus on the NFV Orchestrator dependability, \\ with principles for its robust planning}  \\    
\midrule  
Mijumbi et al.~\cite{survey_mijumbi} & \checkmark & \checkmark \checkmark \checkmark & \checkmark \checkmark \checkmark & \checkmark \checkmark & \checkmark & \makecell{Research challenges in NFV-based systems \\ including reliability and availability aspects}  \\  
\midrule  
Azadiabad et al.~\cite{survey_azadiabad1} & \checkmark \checkmark \checkmark & \checkmark & \checkmark & \checkmark & \checkmark \checkmark & \makecell{Availability, reliability, and continuity challenges \\ in NFV architectures} \\  
\midrule     
Fonseca and Mota~\cite{survey_fonseca} & \checkmark \checkmark \checkmark &  \checkmark \checkmark & \checkmark \checkmark & \checkmark \checkmark \checkmark & \checkmark & \makecell{Fault tolerance issues in SDN-based architectures \\ and network management approaches}  \\ 
\midrule   
Yu et al.~\cite{survey_yu} & \checkmark \checkmark & \checkmark \checkmark & \checkmark \checkmark \checkmark & \checkmark \checkmark & \checkmark & \makecell{Systematic classification of SDN faults \\ and network reliability issues} \\  
\midrule   
Nguyen et al.~\cite{survey_nguyen} & \checkmark & \checkmark \checkmark & \checkmark \checkmark & \checkmark \checkmark \checkmark & \checkmark & \makecell{Deployment and reliability issues into \\ SDN/NFV mobile core networks}  \\  
\midrule   
Kaur et al.~\cite{survey_kaur} & \checkmark & \checkmark \checkmark \checkmark & \checkmark \checkmark \checkmark & \checkmark \checkmark & \checkmark & \makecell{Availability issues in SFCs \\ and research challenges for SFCs deployment}  \\  
\midrule   
This work & \checkmark \checkmark \checkmark  & \checkmark \checkmark \checkmark  & \checkmark \checkmark \checkmark & \checkmark \checkmark \checkmark & \checkmark \checkmark \checkmark & \makecell{Reliability/availability modeling and taxonomy in virtualized networks \\ balanced with tutorial-like explanation of techniques, \\ discussion on ETSI standards and available software tools} \\ 
\bottomrule    
\end{tabular}
}
\label{tab:comparison}
\end{table*}%





\section{Brief Overview on Network Function Virtualization and Connected Paradigms}
\label{sec:overview}

\begin{figure*}[t]
	\centering
	\captionsetup{justification=centering}
	\includegraphics[scale=0.44]{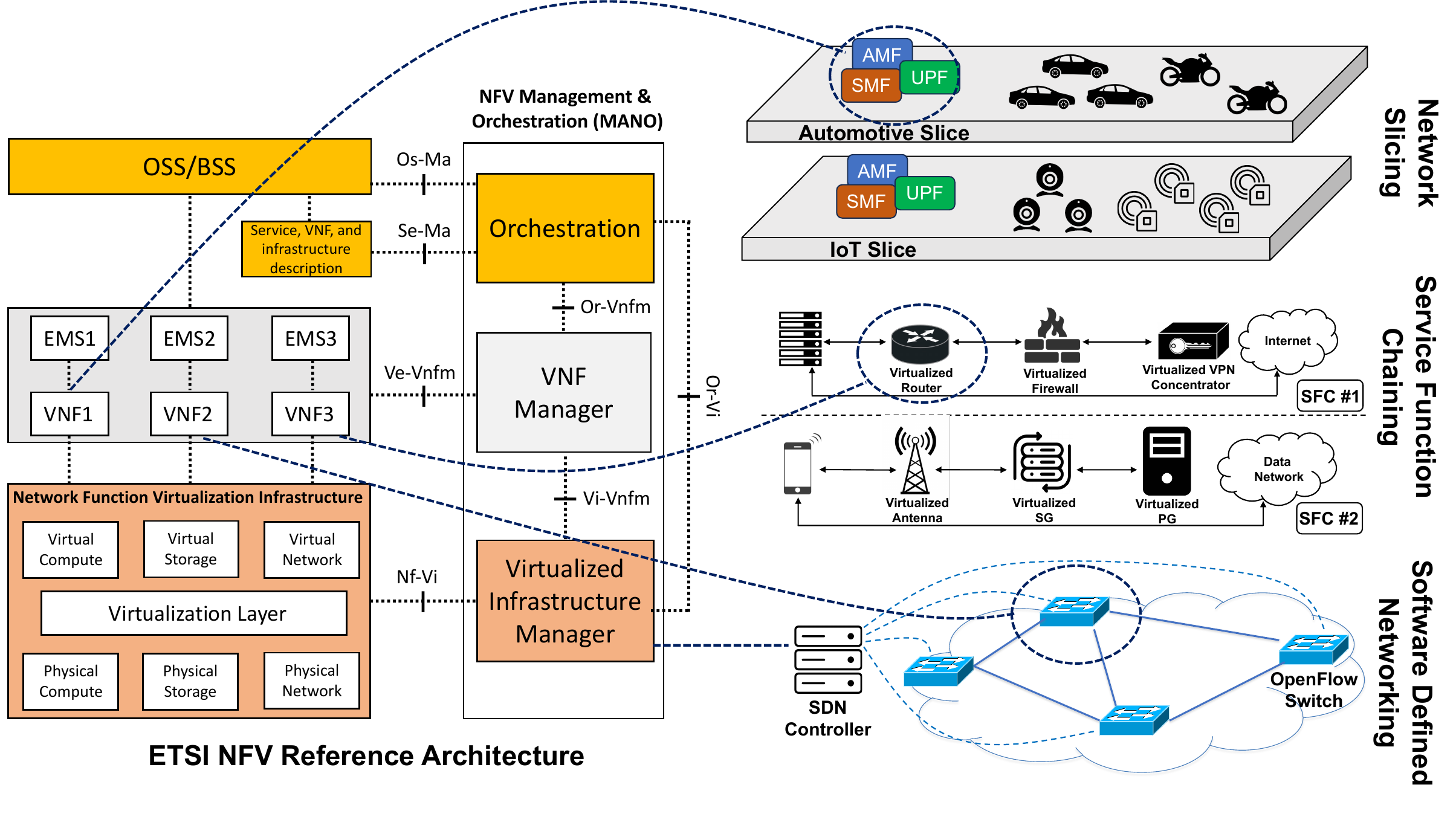}
	\caption{The standard ETSI-NFV architecture (on the left) with main connected paradigms, including: network slicing (top-right part), service function chaining (middle-right part), software defined networking (bottom-right part).}
	\label{fig:superfig}
\end{figure*} 

To improve accessibility to the content of the following Section, we provide here an overview on the standard ETSI NFV architecture along with the main connected paradigms. A unified vision is offered in Fig.~\ref{fig:superfig}, where the standard ETSI NFV architecture is depicted on the left part, and whose main blocks are summarized as follows (more accurate details are available in~\cite{etsi_base}).
The Network Function Virtualization Infrastructure (NFVI) block contains both physical resources (e.g. computing units, storage, network) and virtual resources (virtual machines) connected via the virtualization layer (the hypervisor). 
On top of this block are located the VNFs which represent the individual network functionalities to be virtualized (e.g., routers, firewalls, load  balancers, intrusion detection systems, etc.). The VNF instances are supervised by the Element Management Systems (EMS) which account for operations including fault configuration, accounting, performance, security management.
On top of EMS/VNF block there is the Operations/Business Support System (OSS/BSS) playing the role of an operator back-end in charge of handling services, customers, etc.
The NFV Management and Orchestration (MANO) is a crucial part of the architecture and includes: $i)$ an Orchestrator in charge of managing the lifecycle of network services (including instantiation, scale-out/in, performance measurements, etc.), $ii)$ a VNF manager (VNFM) in charge of managing the lifecycle of VNFs, and $iii)$ a Virtualized Infrastructure Manager (VIM) aimed at controlling the NFVI compute, storage, and network resources. 
The ETSI NFV reference architecture can serve as the basis to support other services or paradigms such as Network Slicing, Service Function Chaining, and Software Defined Networking (right part of Fig.~\ref{fig:superfig}). We briefly describe these three paradigms.

\subsection{Network Slicing (NS)}

Network slicing offers a novel approach to networking, shifting from the traditional one-size-fits-all model to a highly flexible, agile, and customized network infrastructure~\cite{netw_slic}. At its core, network slicing enables the creation of multiple virtual networks, or ``slices'' within a single (and shared) physical network infrastructure although they are mutually isolated and with independent control and management~\cite{netw_slic2}.
Remarkably, network slicing holds promise beyond the classic realm of applications, extending its applicability to various industries such as healthcare~\cite{ns_health1,ns_health2}, transportation~\cite{ns_transp1,ns_transp2}, smart cities~\cite{ns_smart1,ns_smart2}, critical services~\cite{ns_crit1,ns_crit2}, optical networks~\cite{ns_opt1,ns_opt2}. 

In the top-right part of Fig.~\ref{fig:superfig}, two separate slices (automotive slice and IoT slice) are governed by two separate groups of nodes: $i)$ the Access and Mobility Management Function (AMF) node, which handles operations such as registration/de-registration procedures of UEs, connection management, mobility management to ensure that UEs are always reachable and available; $ii)$ the Session Management Function (SMF) node, responsible for operations such as creating and updating Protocol Data Unit (PDU) sessions to provide end-to-end user connectivity, allocating IP addresses, enforcing QoS policies during session establishment;  and $iii)$ the User Plane Function (UPF) node, serving as the anchor point for data packets and responsible for routing and forwarding operations. Interestingly, such nodes can be realized as VNFs and benefit from the flexibility offered by the NFV paradigm. 

Remarkably, addressing availability and reliability in network slicing is vital to enable the coexistence of virtual networks with diverse requirements on a shared physical infrastructure. For instance, designing fault-tolerance strategies for the physical layer is crucial to prevent common faults in the slices that reside on top~\cite{etsi_nfvrel10}. Additionally, ensuring high availability and reliability within each slice is essential to meet distinct service level agreements (SLAs) and QoS demands~\cite{ns1,ns2}.
 
\subsection{Service Function Chaining (SFC)}
This paradigm enables the creation of dynamic chains of network services to fulfill specific application requirements\cite{sfc_ietf}. In SFCs, service functions (SFs) such as firewalls, load balancers, or intrusion detection systems can be orchestrated and sequenced to form customizable service chains tailored to the unique needs of each application flow. 
Obviously, the NFV standard architecture and the SFC concept are strongly interconnected. Leveraging the functionalities offered by NFV-MANO, it is possible to arrange the VNFs to build specific SFCs and dynamically composing different services. 
In the exemplary scenario shown in the mid-right part of Fig.~\ref{fig:superfig}, two SFCs are drawn. The first one (SFC \#1) represents a classic chain that can be used to support access to a data network, including a series of virtualized nodes, as a virtualized router in charge of forwarding data towards specific destinations, a virtualized firewall aimed at controlling/blocking some undesired traffic, and a virtualized VPN concentrator to handle Virtual Private Networks. The second one (SFC \#2) is a chain in charge to manage cellular traffic and is composed by: a virtualized antenna (where many functions are realized completely in software according to the Software Defined Radio paradigm), a virtualized Serving Gateway (SG) to forward and route packets to and from the antenna, and a virtualized Packet Gateway (PG) acting as interface between the cellular network and other packet data networks. All the virtualized nodes of both SFCs can be managed through the standard NFV architecture, thus highlighting the interconnections between the NFV and SFC paradigms. 

In Sect.~\ref{sec:formalisms}, we will see that many works deal with reliability/availability aspects of SFCs, since a failure or a disruption of a single ``block'' of the chain may lead to a malfunctioning of the whole SFC.

\subsection{Software Defined Networking (SDN)}

Software Defined Networking (SDN) refers to a networking paradigm aimed at separating control plane (\textit{routing} process) from the data plane (\textit{forwarding} process) in network devices, thus centralizing network control and enabling programmable network management. Specifically, the control plane has a direct control on the state of network elements involved in the data plane (i.e., switches, routers etc.)~\cite{sdn_base} through some Application Programming Interfaces (APIs).
Two basic elements are recognized into the SDN architecture~\cite{sdn_openflow}: the \textit{controller}, which handles the control logic, and the \textit{switches}, which execute the control logic. 
The idea is that a controller can instruct the switches with customized rules embodying 
new routing protocols, new addressing schemes or innovative security models.
In the bottom-right part of Fig.~\ref{fig:superfig}, an SDN controller controls $4$ switches through the OpenFlow protocol. Although SDN was conceived before NFV and was referring to controlling physical switches, two interesting facts should be taken into account in terms of integration with virtualized networks. First, the SDN controller is considered to be a module of the VIM of the ETSI NFV architecture. This possibility is explicitly envisaged in~\cite{etsi_nfv_sdn}, where SDN controller functionalities can be merged with the VIM functionalities such that the two functions are not distinguishable. Second, OpenFlow switches can be implemented as virtualized components and treated as VNFs, thus they can be managed by the NFV-MANO in a scalable way. 

In Sect.~\ref{sec:formalisms}, we will see that reliability and availability aspects in SDN environments are particularly focused on the SDN controller. Any failure or downtime in the SDN controller, in fact, can lead to network disruptions, compromised performance, and potential security vulnerabilities.

\section{Reliability, Availability, and connected attributes}
\label{sec:rel_ava_concepts}

\begin{figure}[t]
	\centering
	\captionsetup{justification=centering}
	\includegraphics[scale=0.42]{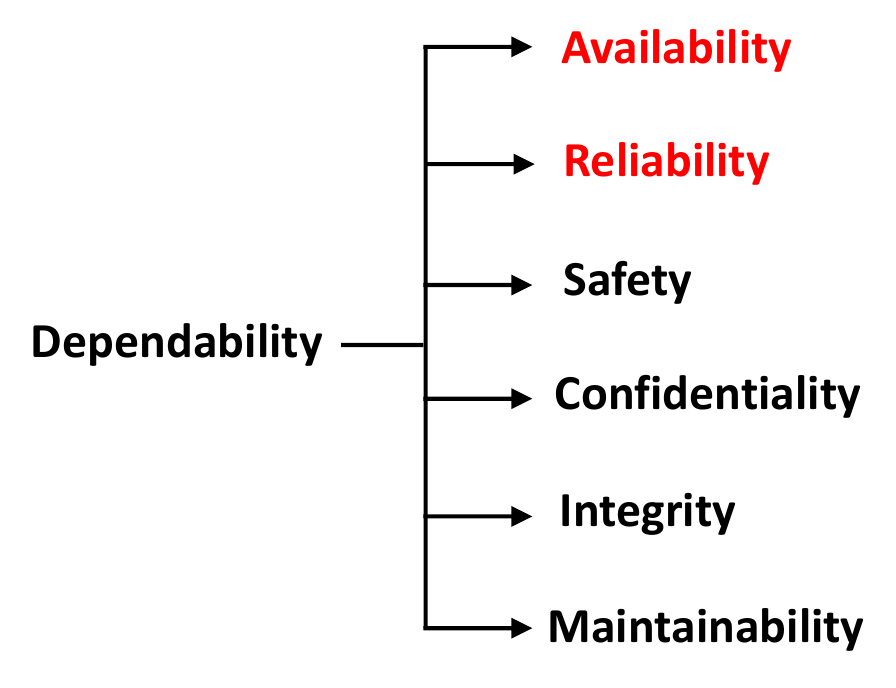}
	\caption{Availability and Reliability as attributes of Dependability according to the taxonomy in~\cite{avizienis}.}
	\label{fig:aviz}
\end{figure}

One of the earliest attempts to provide a taxonomy of concepts such as availability and reliability applied to computer science is credited to Avizienis et al.~\cite{avizienis}. 
In this work, dependability serves as a comprehensive umbrella term, encapsulating essential attributes as shown in Fig.~\ref{fig:aviz}. 
An explanatory definition of dependability provided in~\cite{avizienis} is the following: \textit{the dependability of a system is the ability to avoid service failures that are more frequent and more severe than is acceptable}. 
In this context, a system can be: a single component (e.g., a VNF), an ensemble of components (e.g., an SFC), a complex infrastructure (e.g., the NFV-MANO governing a whole SFC).

In our work we focus on availability and reliability, which are considered the most relevant attributes when evaluating system behavior in virtualized environments, especially where uptime, failover mechanisms, and redundancy are critical.
Moreover, being availability and reliability officially recognized as \textit{standardized metrics} (see next Section), for these two attributes we provide both definitions from~\cite{avizienis} and from ETSI standard documents. According to ETSI~\cite{etsi_nfvrel1}, \textit{availability} is defined as the capability of a component to be in a state to perform a required function or service at a given time. This definition is coherent with the definition of availability in~\cite{avizienis} defined as the \textit{readiness for correct service}. Similarly, ETSI~\cite{etsi_nfvrel1} defines \textit{reliability} as the probability that a component can perform a required function or service for a given time interval, whereas the definition in~\cite{avizienis} is the \textit{continuity of correct service}. 
As for the other attributes, in~\cite{avizienis} \textit{safety} is defined as ``absence of catastrophic consequences on the user(s) and the environment'', \textit{confidentiality} is defined as the ``absence of unauthorized disclosure of information'',  \textit{integrity} is defined as the ``absence of improper system alterations'', and \textit{maintainability} is defined as the ``ability to undergo modifications and repairs''.

As observed in~\cite{trivedi-bobbio}, dependability can be defined as a qualitative property, namely the aptitude of a system to conform to user expectations and to design specifications. 
Nonetheless, it is important to precisely quantify availability and reliability attributes through specific mathematical techniques.

\begin{figure}[t]
	\centering
	\captionsetup{justification=centering}
	\includegraphics[scale=0.28]{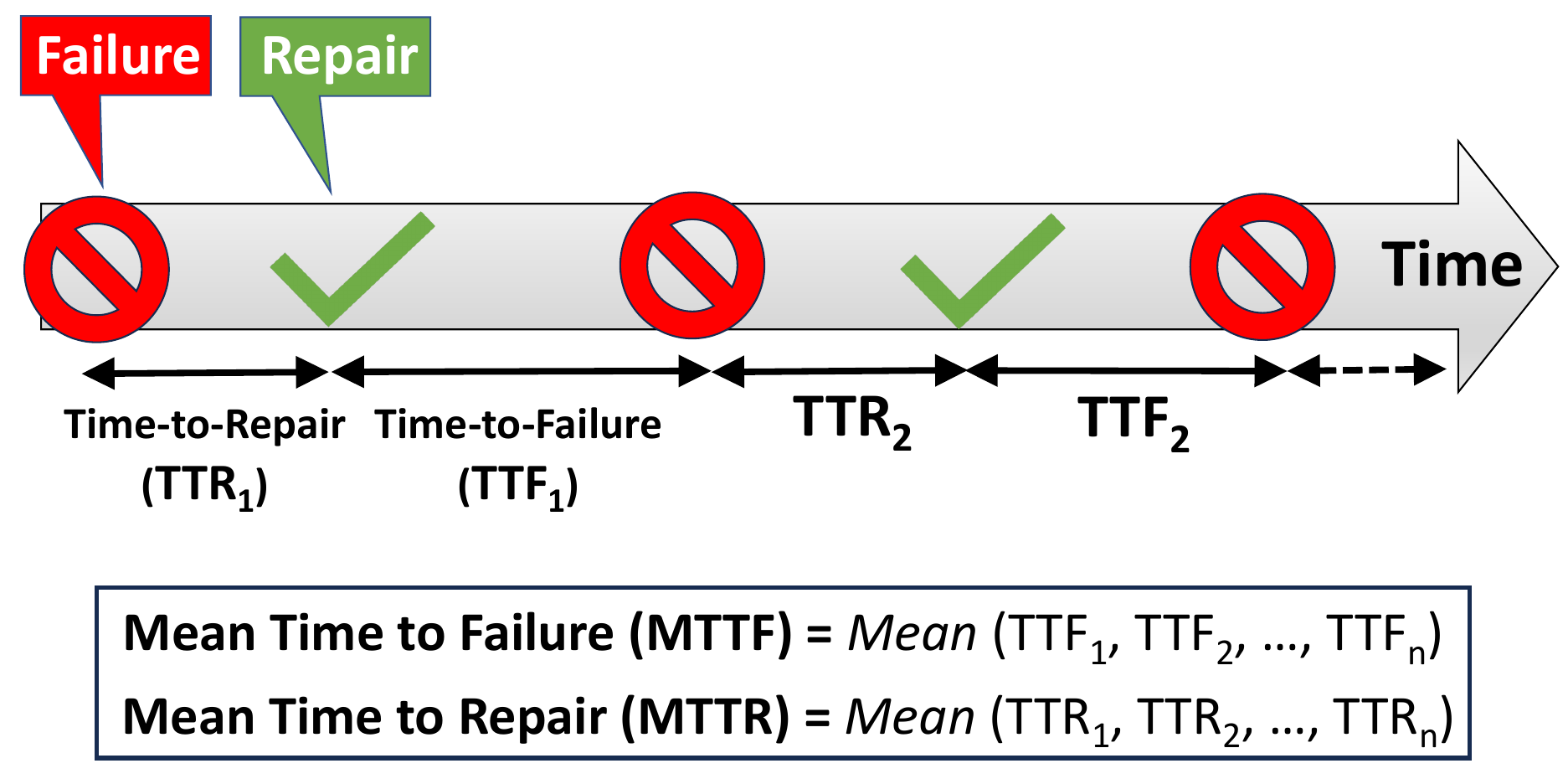}
	\caption{Mean Time to Failure (MTTF) and \\ Mean Time to Repair (MTTR).}
	\label{fig:mttf_mttr}
\end{figure} 

\subsection{Useful definitions}
\begin{figure}[t]
	\centering
	\captionsetup{justification=centering}
	\includegraphics[scale=0.41]{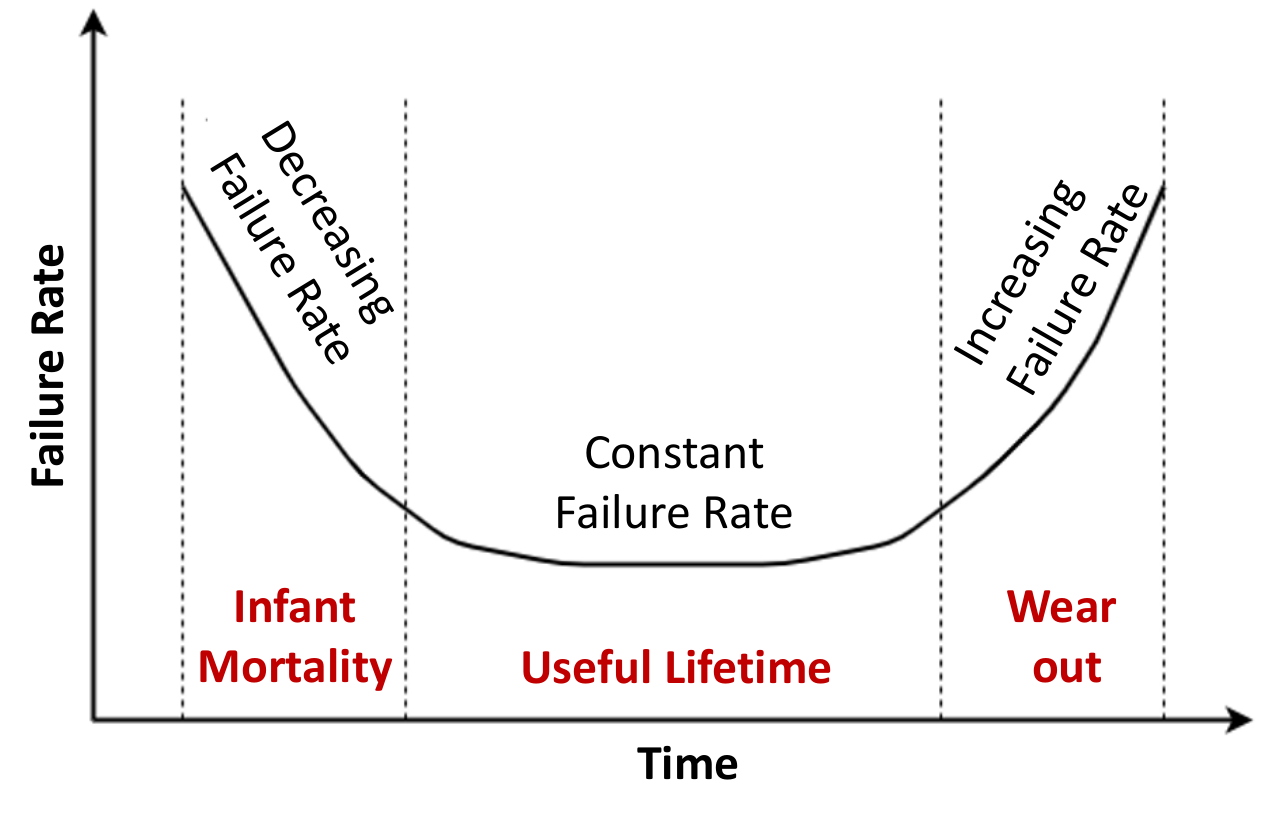}
	\caption{The \textit{bathtub} curve with three zones. Typically, the central zone models the behavior of a component during its useful lifetime with a constant failure rate $\lambda$.}
	\label{fig:bathtub}
\end{figure}

A fundamental difference between availability and reliability is that availability takes into account both failure and repair processes, whereas reliability characterizes the failure process of a component or of a system.  
Moreover, the availability concept is often considered when a system operates during its regime condition. In such a case  we refer to the \textit{steady-state availability}.
Typically, a (virtualized) system is analyzed by breaking it down into its constituent \textit{repairable} components, i.e. components characterized by alternating cycles of uptime periods (time-to-failures) and  downtime periods (time-to-repair) as illustrated in Fig.~\ref{fig:mttf_mttr}. Thus, the average duration of time to failure is called \textit{Mean Time to Failure} (MTTF), whereas the average duration of time to repair is called \textit{Mean Time to Repair} (MTTR). 
The steady-state availability is the probability that the system is working in regime condition computed as
\beq
A= \frac{\textnormal{Uptime}}{\textnormal{Uptime~+~Downtime}} = \frac{\textnormal{MTTF}}{\textnormal{MTTF}+\textnormal{MTTR}}.
\label{eq:ava}
\eeq
Conversely, the \textit{reliability} of a system is defined as the probability that the system has not failed after a time period $t$. Under the assumption that failure times are exponentially distributed with a parameter known as failure rate, $\lambda=1/MTTF$, the reliability is:
\beq
R(t)=e^{-\lambda t} = e^{-\frac{t}{MTTF}}.
\label{eq:rel}
\eeq
The exponential assumption is very common in many practical reliability problems due to the \textit{memoryless} property, meaning that the probability of a system failing within a certain time period is independent of how long it has already operated before that time period. This property aligns well with many real-world scenarios where failures occur randomly and are not influenced by the past history of the system. Moreover, the exponential distribution has a simple mathematical form, characterized by a single parameter $\lambda$ which is supposed to be constant. This assumption stems from the fact that, in many cases, we are interested to evaluate the reliability during the \textit{useful life} of a component where the failure rate $\lambda$ can be assumed to be constant. Figure~\ref{fig:bathtub} shows the so-called \textit{bathtub curve}, which represents the typical life cycle of a component (including the virtualized ones) where three zones can be recognized: the \textit{infant mortality} zone, where failures usually occur due to design issues, deployment issues, and/or initial configuration issues; the \textit{useful lifetime} zone, representing the steady-state behavior of a component, where failures are typically due to hidden issues (e.g. software bugs) and occur at normal and low rate; the \textit{wear out} zone, where the increasing failure rate is due to the natural aging of components (e.g., software not adequately updated).

\subsection{Quantitative evaluation}
We have seen that to characterize availability and reliability attributes we need to evaluate failure and repair quantities. In virtualized networks, examples of failures include: VNF malfunctioning, crash of the hypervisor, unreachability of an SFC element, and many others. In contrast, examples of repairs include: VNF reboots, resource reallocation, repairman intervention, and many others.  
In real-world deployments, the evaluation of failures and repairs may be not trivial due to a number of constraints (real data not available, systems not physically accessible, etc.)
Accordingly, two approaches to obtain failure and repair information are described in~\cite{trivedi-bobbio}. 
The first approach is referred to as \textit{measurement-based evaluation}, and consists in gathering failure/repair information from the observation of a system both during its normal functioning and under controlled test conditions. When it is not possible to test a system directly on the field, we can operate within an emulated environment or through accelerated tests. The latter refers to tests in which the stress levels are chosen to exceed standard operational conditions so as to observe the system failures in a shorter time frame. This approach has a series of benefits including: $i)$ \textit{efficiency and speed}, as emulated environments and accelerated tests allow for the rapid simulation of various failure scenarios and repair procedures; $ii)$ 
\textit{cost-effectiveness}, as setting up and running emulated environments or accelerated tests is often more cost-effective compared to conducting experiments in a live production network; $iii)$ \textit{controlled environment}, as emulated environments provide a controlled setting where specific failure scenarios can be induced systematically (in this case we refer to fault injection techniques); $iv)$ \textit{reproducibility}, as emulated environments and accelerated tests offer the advantage of reproducibility, enabling researchers and engineers to repeat experiments consistently to validate results or conduct comparative analyses.
This approach is typically discarded when the cost associated to set up the emulated environment is high, or when the considered system is in a very early stage of its lifetime.

The second approach is referred to as \textit{model-based evaluation}, and involves a mathematical abstraction of a system with the aim to provide a simplification of its behavior. Two phases can be recognized in this approach. First, a \textit{modeling phase} that consists in building an abstraction of the system preserving its main characteristics. During this phase, the designer should choose an appropriate formal language to describe the model. Then, a \textit{solution phase} that consists in solving the system model through specific analytical or simulative methods aimed at evaluating the quantities of interest. In general, the model-based evaluation approach is faster and less expensive than the measurement-based evaluation approach, but produces less accurate results.
 
\section{Standardization Efforts}
\label{sec:standard}

The standardization process plays a crucial role in defining metrics for measuring availability and reliability in network virtualization. By establishing consistent and universally accepted guidelines, standards ensure that different systems and solutions can be compared and evaluated on a common basis. In network virtualization, where various vendors and technologies coexist, standardized metrics allow for objective assessment and benchmarking of performance, enabling stakeholders to make informed decisions. 
\subsection{The ``nines'' metric}

We recall that a highly available system is a system with a low probability of breaking down (high reliability) and a high probability of being quickly repaired (high maintainability). This also justifies the choice of the availability as an evaluation metric into the industrial field, where the \textit{number of nines} is chosen as a standardized benchmark~\cite{itu} as also highlighted by ETSI~\cite{etsi_nfvrel1}. For example, the \textit{five nines} availability ($99.999\%$ or high availability) benchmark means that a system can tolerate a maximum yearly downtime of about five minutes and 15 seconds (see Table~\ref{tab:nines}). 

The necessity of reaching five nines availability for virtualized network infrastructures stems from several compelling factors. First, as businesses increasingly rely on digital services to drive revenue and productivity, any downtime or service disruption can have significant financial implications. For industries such as finance, healthcare, and e-commerce, where real-time transactions and data exchange are critical, even a few minutes of downtime can lead to substantial revenue losses and damage to brand reputation. 
Secondly, the interconnected nature of applications and services amplifies the impact of network outages or performance degradation. A failure in one part of the network can cascade across multiple services, affecting users, applications, and business processes. In virtualized environments, where network functions are dynamically instantiated and interconnected, the risk of such cascading failures becomes more pronounced, underscoring the importance of robust availability measures.

\subsection{The role of ETSI}
In the last decade, many agencies in charge of defining globally applicable standards for ICT systems have devoted efforts to define standard guidelines for high availability of virtualized networks. In particular, ETSI has released a set of technical documents (identified by the common acronym ETSI NFV-REL) which define specifications, guidelines, techniques, use cases related to the reliability and availability of virtualized networks. In this section, we will provide details on the most relevant ETSI documents.

\begin{table}[t]  \centering  \caption{Common steady-state availability values expressed in terms of \textit{number of nines}.}    
	\resizebox{\columnwidth}{!}{%
		\begin{tabular}{ c c c c c }    \toprule    \rowcolor[rgb]{ .91,  .91,  .91} \textbf{Availability} & \textbf{Downtime/Year} & \textbf{Downtime/Month} & \textbf{Downtime/Week} & \textbf{Downtime/Day} \\    \midrule    \textbf{99.999\%} & 5.26 minutes & 0.438 minutes & 0.101 minutes & 0.014 minutes \\    \midrule    \textbf{99.995\%} & 26.28 minutes & 2.19 minutes & 0.505 minutes & 0.072 minutes \\    \midrule    \textbf{99.990\%} & 52.56 minutes & 4.38 minutes & 1.011 minutes & 0.144 minutes \\    \midrule    \textbf{99.950\%} & 4.38 hours & 21.9 minutes & 5.054 minutes & 0.72 minutes \\    \midrule    \textbf{99.900\%} & 8.76 hours & 43.8 minutes & 10.108 minutes & 1.44 minutes \\    \midrule    \textbf{99.500\%} & 43.8 hours & 3.65 hours & 50.538 minutes & 7.2 minutes \\    \midrule    \textbf{99.250\%} & 65.7 hours & 5.475 hours & 75.808 minutes & 10.8 minutes \\    \midrule    \textbf{99.000\%} & 87.6 hours & 7.3 hours & 101.077 minutes & 14.4 minutes \\    \bottomrule    
		\end{tabular}
		\label{tab:nines}
	}
\end{table}%

\begin{figure}[t]
	\centering
	\captionsetup{justification=centering}
	\includegraphics[scale=0.42]{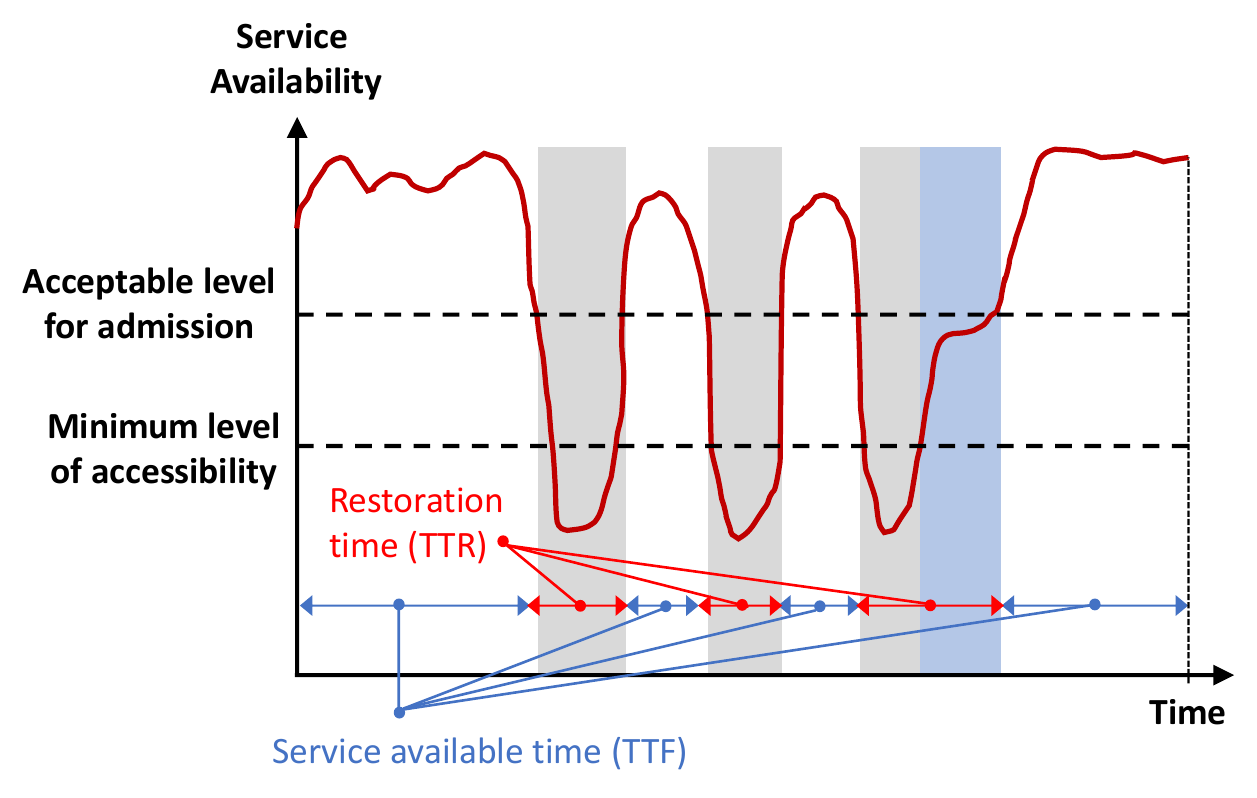}
	\caption{NFV Service availability as interpreted by ETSI GS NFV-REL 001~\cite{etsi_nfvrel1}.}
	\label{fig:serv_ava_etsi}
\end{figure}

According to ETSI GS NFV-REL 001 guidelines~\cite{etsi_nfvrel1}, NFV service availability is provided by the result of \textit{accessibility} and \textit{admission control}.  
Accessibility is defined as the ability of a service to access the (physical) resources needed to provide that service. As shown in Fig.~\ref{fig:serv_ava_etsi}, if the target service satisfies a minimum level of accessibility, it is possible to provide this service to end users. Admission control refers to the administrative decision (managed by the operators) to actually provide that service. To offer a more stable and reliable service, admission control may require better performance and/or additional resources than the minimum required to provide the service. Figure~\ref{fig:serv_ava_etsi} depicts the TTF intervals when the service is not available due to either lack of accessibility or admission control.
In the ETSI GS NFV-REL 001 document, three main factors that may affect service availability are highlighted:
\begin{itemize}
	\item \textit{Service Type:} different service types can have different accessibility requirements, such as: $i)$ network control traffic (e.g., control plane traffic), $ii)$ real-time services (e.g., VoIP, VoLTE, interactive services), $iii)$ data services (e.g., transactional services and messaging), $iv)$ ISP services (e.g., Web browsing, e-mail).
	\item  \textit{Customer Agreements}: specific Government regulatory demands can impact service availability requirements. Examples include: $i)$ emergency telecommunication services (allowing Governments to activate critical procedures in response to severe disaster situations such as calamities, terrorist attacks, etc.), $ii)$ human life and safety services (e.g., air traffic control, railroad signalling, etc.), $iii)$ regional $112/911/119$ etc. (local emergency service allowing to contact authorities requesting assistance).
	\item \textit{Availability Factors and Relevant Standards}: even if operators design their networks to satisfy all the availability requirements, some unpredictable conditions may happen, including congestion conditions (e.g., large bursts of traffic may flood network resources), or failure conditions implying to design accurate redundancy strategies and restoration priority policies.   
\end{itemize}

Moreover, borrowing some fundamental concepts from classic literature (see, e.g.,~\cite{avizienis}) and adapting them to the NFV, in the ETSI GS NFV-REL 001 document, the so-called \textit{Fault $\rightarrow$ Error $\rightarrow$ Failure} chain is introduced. 
Then, a set of issues that can activate faults are pinpointed: $i)$ malicious attacks from cyber-criminals against the service or the system, including attacks against VNFs, VNFM, VIM, Orchestrator; $ii)$ large-scale disasters causing hardware disruption including NFVI components; $iii)$ unusual (but legitimate) traffic which exceeds the current system capabilities; $iv)$ environmental challenges including wireless channel interference, increased temperature; $v)$ failure of a dependent system or service such as a VIM malfunctioning that can negatively affect the Orchestrator; $vi)$ accidents and/or mistakes including systems misconfigurations or bad policies specifications.

\begin{figure}[t]
	\centering
	\captionsetup{justification=centering}
	\includegraphics[scale=0.3]{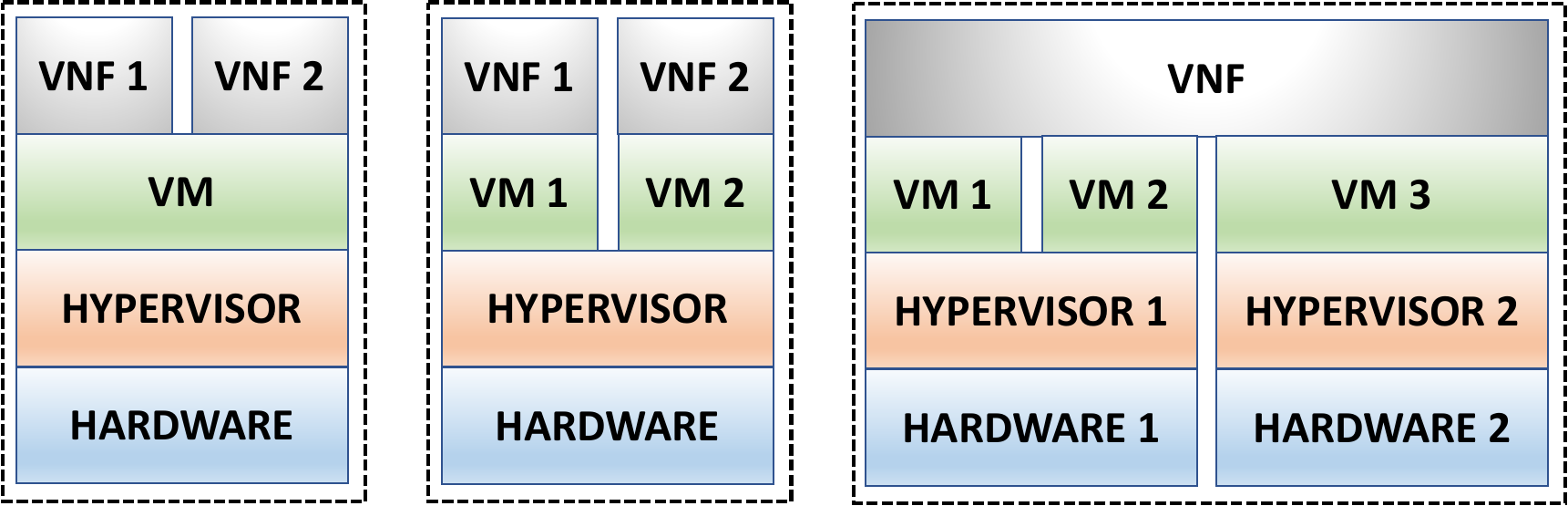}
	\caption{Three different deployment options for VNFs: all-shared resources (left), shared hypervisor and hardware (middle), VNF spanning across multiple physical/virtual resources (right).}
	\label{fig:vnf_models}
\end{figure}

Due to the \textit{nested} structure of NFV-based elements, some considerations onto VNF failure modes are drawn into the ETSI GS NFV-REL 001 document. Depending on the type of VNF deployment, in fact, the impact of failure may vary, and different restoration methods may be implemented. In Fig.~\ref{fig:vnf_models}, some possible VNF deployment options are shown. In case all the \textit{layers} (virtual machine, hypervisor, hardware) are shared between the two VNFs on top (Fig.~\ref{fig:vnf_models}, left panel), a failure on any of the shared resource impacts on both VNFs. In case hypervisor and hardware are shared (but not the two different VMs - see Fig.~\ref{fig:vnf_models}, middle panel), a common failure on hypervisor and/or hardware will affect the VNFs functioning. In contrast, a failure only onto one of the two VMs will just affect the corresponding VNF. Finally, new failure modes are introduced in case a single VNF is spanned across physical and virtual resources (see Fig.~\ref{fig:vnf_models}, right panel). In this case, it is enough that a single layer (or sub-layer) fails to cause a service failure of the VNF on top.

Moreover, in the ETSI GS NFV-REL 001 document, some crucial requirements are identified:
\begin{itemize}
	\item A VNF needs to ensure the availability of its part of the end-to-end service.
	\item The VNF designer must define the requirements of the NFVI, such as geo-redundancy requirements.
	\item The NFV-MANO function shall provide the necessary mechanisms to recreate VNF automatically after a failure.
	\item The NFV-MANO function shall support failure notification mechanisms at run time. 
	\item Failures in the NFVI shall be handled (i.e., detection) in the NFVI layer or in the NFV-MANO (e.g., hardware failure, loss of connectivity, etc.). 
	\item The NFVI shall provide the necessary functionality to enable high availability at the VNF level, such as failure notification and remediation. 
\end{itemize} 
ETSI GS NFV-REL 002 document~\cite{etsi_nfvrel2} provides a
discussion on how to exploit scalability in NFV environments to achieve high availability (HA). The usage of scalable architectures involves: $i)$ distribution of functionalities across multiple locations; $ii)$ duplication of functionalities across different locations such that failure in one location does not impact the processing of services; $iii)$ load balancing such that a given network location does not experience heavier loads than others; $iv)$ managing dynamically the VNFs through horizontal scaling, i.e. \textit{scale-in} (reducing the connected resources to optimize efficiency or reduce costs) and \textit{scale-out} (increasing the connected resources to handle increased workload or demand).  

 \begin{figure*}[t]
	\centering
	\captionsetup{justification=centering}
	\includegraphics[scale=0.44]{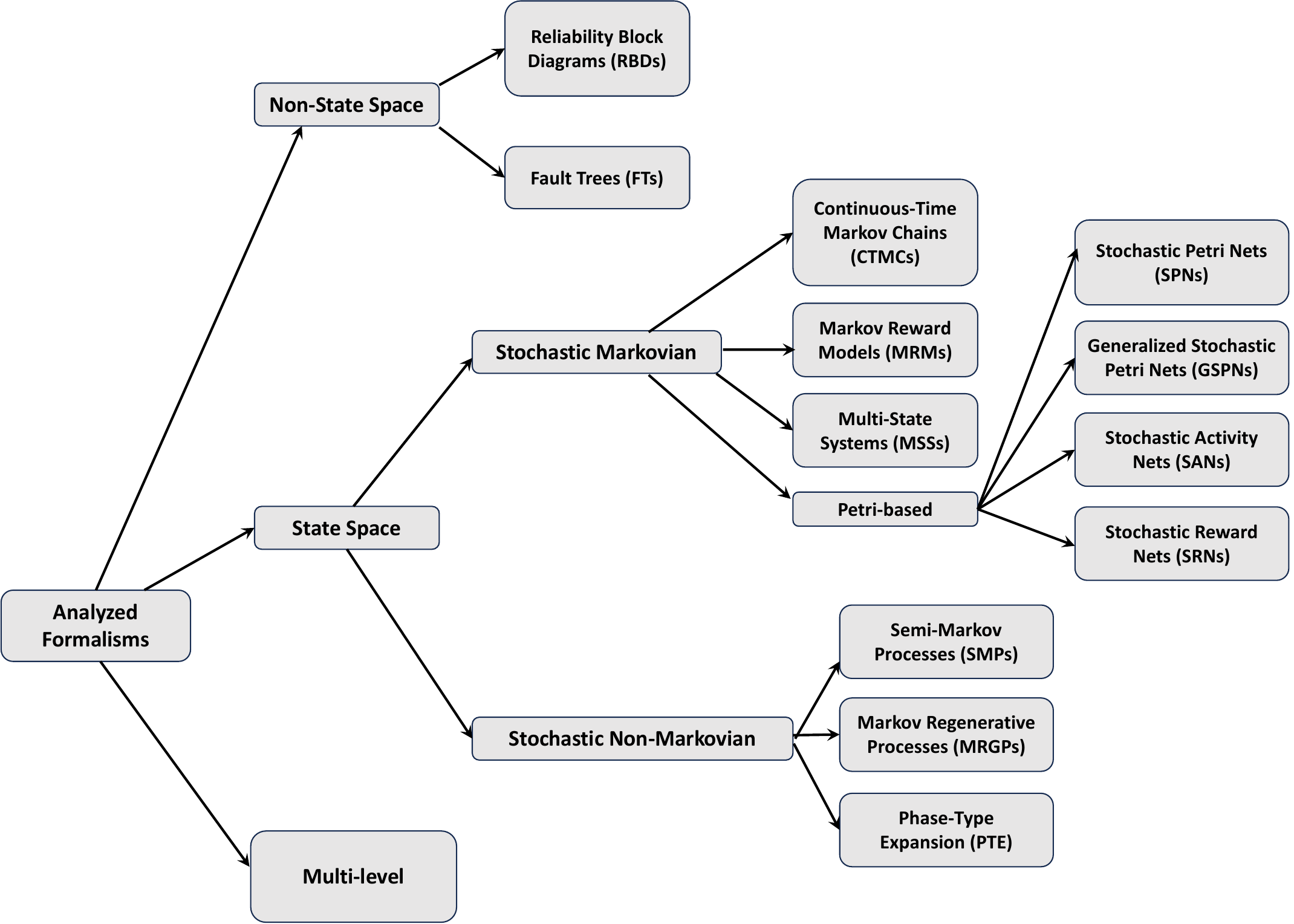}
	\caption{Taxonomy of formalisms used to model reliability and availability aspects in virtualized networks~\cite{trivedi-bobbio}.}
	\label{fig:models}
\end{figure*} 

ETSI GS NFV-REL 003 document~\cite{etsi_nfvrel3} provides formal definitions, models, and techniques to address availability issues of virtualized infrastructures. This report introduces the formal definitions of steady-state availability and reliability exactly as provided in~\eqref{eq:ava} and~\eqref{eq:rel}, respectively. Then, formulas and techniques to evaluate the reliability and the availability of systems made of many sub-components (as in the case of virtualized structures) are provided. For instance, service chains can be seen as softwarized structures with several software modules connected in series, where starting from the availability of each single module, it is possible to evaluate the availability of the whole chain. More details about these aspects will be provided in Sect.~\ref{sec:rbd} with the introduction of the Reliability Block Diagram (RBD) formalism. The ETSI GS NFV-REL 003 document also focuses on some practical redundancy/protection solutions for VNFs including \textit{active-standby} (the primary VNF works and the secondary serves as backup), \textit{active-active} (both primary and secondary VNFs work simultaneously), \textit{load balancing} (the load is distributed between the two VNFs). ETSI GS NFV-REL 003 also addresses the problem of VNFs software upgrading, as this procedure is similar to a failure which makes the system unavailable for a certain time period. 

ETSI GS NFV-REL 004 report~\cite{etsi_nfvrel4} focuses on possible procedures to perform active monitoring of an NFV infrastructure. In particular, three monitoring entities are mentioned: the \textit{virtual test agent} (provided by OSS/BSS via NFVO), i.e., a software entity in charge of providing failure and logging messages, the \textit{test controller}, in charge of managing the test agents and providing them test instructions, the \textit{test results analysis module} in charge of analyzing the collected logs. The two latter modules should be managed by the OSS/BSS of the ETSI NFV architecture. Moreover, the document mentions the \textit{fault injection} as a simulation technique allowing to evaluate the resilience of NFV-based infrastructures by deliberately introducing faults. 

ETSI GS NFV-REL 006~\cite{etsi_nfvrel6} focuses onto maintaining NFV service availability and continuity upon software modification. In particular, some interesting use cases are considered such as stateless VNF upgrade, stateful VNF upgrade, VNF software upgrade, network service update with simultaneous software upgrade of multiple VNF instances.  

ETSI GR NFV-REL 007~\cite{etsi_nfvrel7} mainly describes the role of the NFV-MANO in ensuring VNFs resiliency from their instantiation through their operation and recovery from failures. Based on the type of failure, the MANO should provide appropriate recovery or remediation such as VNF  re-instantiation/re-creation, VNF scaling (i.e., adding a VNF instance to provide more capacity so that the service is not degraded), VNF migration (i.e., moving the VNF to another server during maintenance or hardware failures), or failure containment (i.e., keep it from propagating further and negatively impact other VNFs), while ensuring service continuity. 

ETSI GR NFV-REL 010~\cite{etsi_nfvrel10} covers aspects related to resilience issues of network slices. A set of recommendations is provided for the deployment of a network service where multiple slices are involved and share the same underlying infrastructure. For instance, the overload of one network slice instance should not impact the performance of the other slice instances, and the failure of one network slice instance should not cause operation anomalies onto other slices. 

ETSI GR NFV-REL 014~\cite{etsi_nfvrel14} focuses on the study of reliability aspects for supporting the management of cloud-native VNFs, namely VNFs deployed in a container-based environment where new elements are present. One of these is Kubernetes, an open-source container orchestration system that holds significant promise as platform to guarantee system reliability, as it is responsible
for automating deployment, scaling, and management of containerized applications. 

\begin{table}[h]
    \centering
    \footnotesize
    \renewcommand{\arraystretch}{1.2}
    \caption{ETSI NFV-REL documents and their main focus.}
    \label{tab:ETSI-NFV-REL}
    \begin{tabular}{|c|p{4.5cm}|}
        \hline
        \rowcolor[rgb]{ .91,  .91,  .91} \textbf{ETSI Document} & \textbf{Brief Description} \\
        \hline
        ETSI GS NFV-REL 001~\cite{etsi_nfvrel1} & Definitions, failure causes and modes, factors affecting service availability \\
        \hline
        ETSI GS NFV-REL 002~\cite{etsi_nfvrel2} & Scalability and High Availability \\
        \hline
        ETSI GS NFV-REL 003~\cite{etsi_nfvrel3} & Models and techniques for end-to-end reliability and availability, redundancy, protection, software upgrade \\
        \hline
        ETSI GS NFV-REL 004~\cite{etsi_nfvrel4} & Active monitoring and failure detection, fault injection \\
        \hline
        ETSI GS NFV-REL 006~\cite{etsi_nfvrel6} & Maintaining service availability and continuity upon software modification \\
        \hline
        ETSI GR NFV-REL 007~\cite{etsi_nfvrel7} & Role and capabilities of NFV-MANO for ensuring VNF resiliency, availability, recovery, migration \\
        \hline
        ETSI GR NFV-REL 010~\cite{etsi_nfvrel10} & Resiliency for network slicing and slice isolation \\
        \hline
        ETSI GR NFV-REL 014~\cite{etsi_nfvrel14} & Reliability for cloud-native, containerized VNFs \\
        \hline 
    \end{tabular}
\end{table}

While the ETSI NFV-REL standards, summarized in Table~\ref{tab:ETSI-NFV-REL}, offer foundational guidelines for addressing availability and reliability in virtualized network environments, they fall short in providing comprehensive, formal specifications for these critical aspects. The standards touch upon certain formalisms, such as RBDs, but do not delve into detailed methodologies or endorse specific modeling tools to effectively analyze and ensure reliability and availability. This gap necessitates a deeper exploration of methods and tools that can be employed to model and evaluate the resilience of NFV architectures. Hence, in the next section we provide a description of methods and formalisms (without excessive mathematical details) that can be profitably exploited to model and analyze availability and reliability in virtualized networks.


%
%
%
%


\section{Modeling Formalisms and Selected References}
\label{sec:formalisms}

This section aims to investigate the taxonomy (borrowed by a milestone work on reliability and availability engineering~\cite{trivedi-bobbio}) of the most common formalisms used to model availability/reliability in the field of virtualized network infrastructures (see Fig.~\ref{fig:models}). 
The choice of the most suitable formalism to employ highly depends on the particular virtualized system or component to model, and can be guided by various factors including: $i)$ complexity of the system to model, $ii)$ fault/repair data available, $iii)$ familiarity of the modeler with the particular formalism. 

The modeling formalisms can be classified as:
\begin{itemize}
    \item \textit{Non-state-space models}: these models can be quickly solved with the assumptions that components are statistically independent. While they offer high analytical tractability by not requiring the generation of an underlying state space, their modeling capability is limited due to the exclusion of dependencies. 
   \item \textit{State-space models}: these models have the capability to incorporate statistical dependencies, as well as dependencies on state and time, thus exhibiting significant modeling capacity. However, they tend to have lower analytical tractability due to the curse of dimensionality. State-space approaches are generally categorized into Markovian (homogeneous and non-homogeneous) and non-Markovian methods. 
   \item \textit{Multi-level models}: these models combine both state-space and non-state-space models benefiting from the modeling power of the former and from the efficiency of the latter.
\end{itemize} 

Along with each formalism we provide a set of pertinent references in the field of virtualization technologies.

By organizing the literature in this manner, readers can easily map which formalisms are best suited to address particular challenges and characteristics inherent to virtualized architectures, thereby streamlining the process of finding relevant and effective modeling techniques.

We start with the description of Non-state-space models in subsection~\ref{sec:rbd} and~\ref{sec:ft}. Then, we describe State-space models in subsections from~\ref{sec:ctmc} to~\ref{sec:nmp}, and Multi-level models in subsection~\ref{sec:multilev}.

 \subsection{Reliability Block Diagrams (RBDs)}
\label{sec:rbd}

\begin{figure}[t]
	\centering
	\captionsetup{justification=centering}
	\includegraphics[scale=0.34]{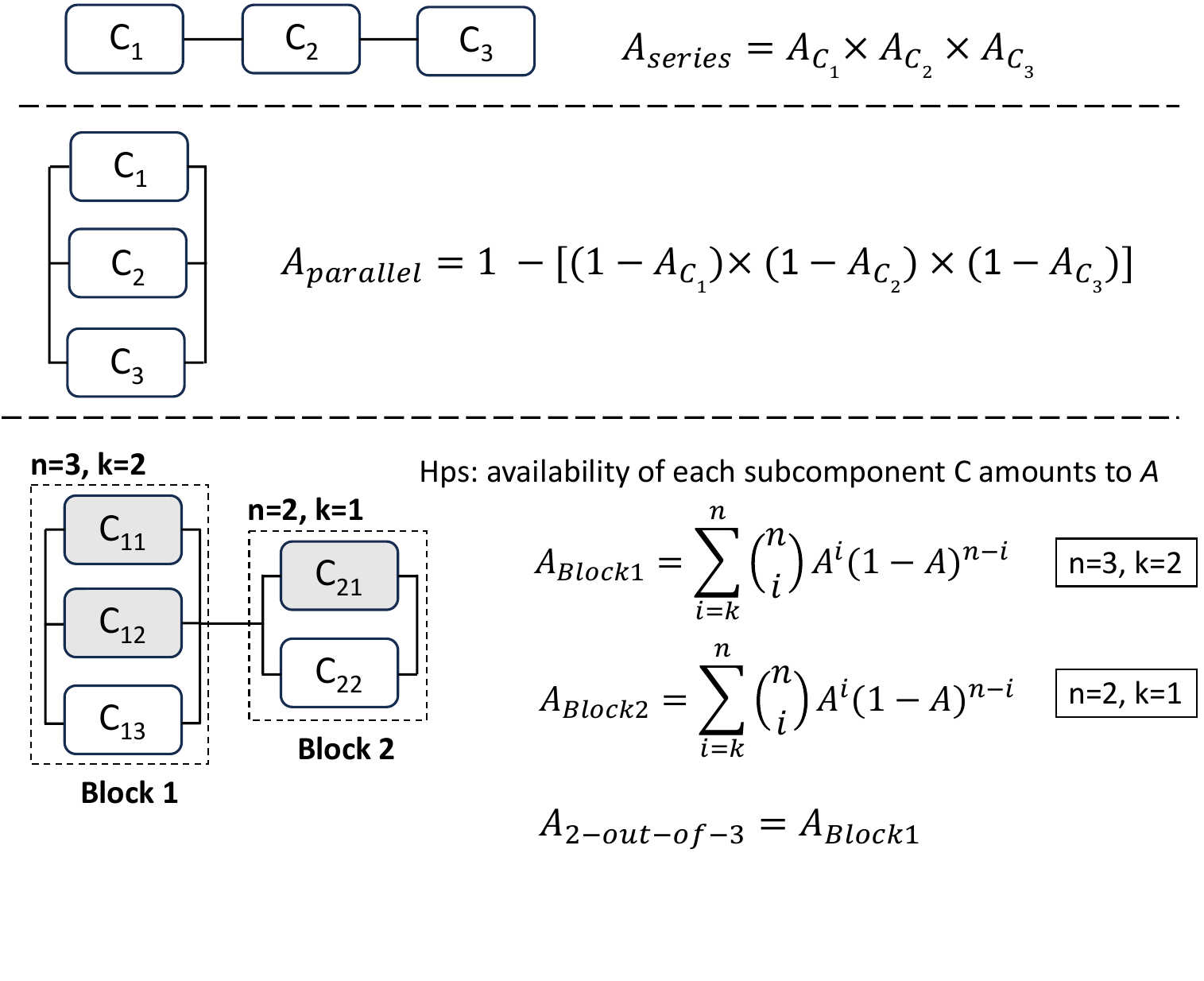}
	\caption{Reliability Block Diagrams (RBDs) for various configurations: series (top panel), parallel (middle panel), $k$-$out$-$of$-$n$ (bottom panel).}
	\label{fig:rbd}
\end{figure} 

Reliability block diagrams are formally described in the international standard IEC 1078, where RBD is defined as a representation useful to capture the logical connections of components necessary for the system to function successfully. At any given moment, a component is considered to be in one of two states: operational or faulty. Each block in the diagram represents a component or a subsystem, and the connections between blocks depict the logical flow.
When dealing with the RBD methodology, three typical configurations are possible (see Fig.~\ref{fig:rbd}):
\begin{itemize}
	\item \textit{Series configuration}: all components must function for the system to succeed. The system fails if any single component fails, making it highly sensitive to individual component reliability. Blocks are connected end-to-end in a single path. The overall system availability is the product of the availabilities of all individual components (see Fig.~\ref{fig:rbd} - (top panel)).
	\item \textit{Parallel configuration}: the system can succeed as long as at least one component functions. This setup provides redundancy, improving overall system reliability. Blocks are connected in parallel paths that converge at the end. The system availability is the probability that at least one component does not fail (see Fig.~\ref{fig:rbd} - (middle panel)).
	\item \textit{k-out-of-n configuration}: the system succeeds if at least $k$ out of $n$ components are operational. This configuration provides a balance between series and parallel systems, offering partial redundancy. Blocks can be visualized as forming groups, with the system requiring a specific number of operational blocks for success. The probability that exactly $k$ components out of $n$ are operational follows a binomial distribution (see Fig.~\ref{fig:rbd} - (bottom panel)).
\end{itemize}
Due to its high level of abstraction, this formalism is widely used to model virtualized systems in terms of logical blocks.

\begin{figure}[t]
 	\centering
 	\captionsetup{justification=centering}
 	\includegraphics[scale=0.45]{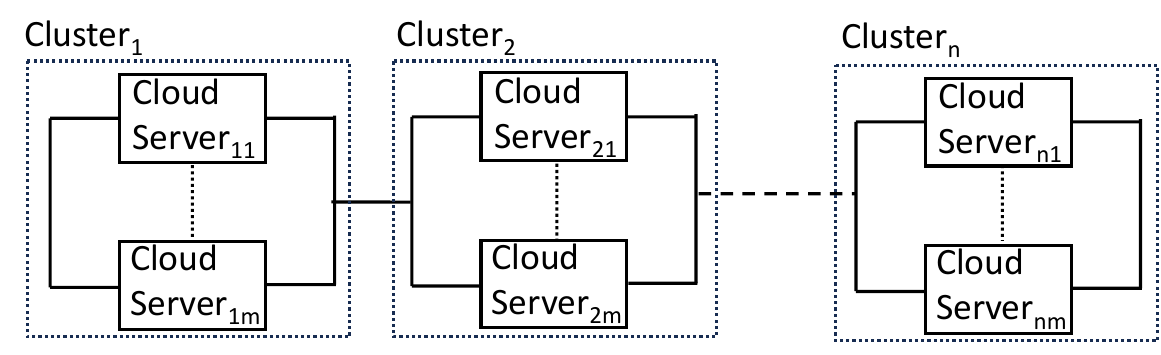}
 	\caption{Reliability Block Diagram (RBD) representation of a virtual data center (VDC)~\cite{rbd1}.}
 	\label{fig:rbd2_super}
 \end{figure}

Authors in~\cite{rbd1} propose an RBD formalization to perform a reliability assessment of a virtual data center (VDC) into a cloud computing architecture. Since the VDC can be seen as a series of network components, its overall reliability is analyzed through a nested series-parallel RBD configuration with the outer RBD modeling the connection of clusters (set of cloud servers), and the inner RBD modeling more specific virtualized resources. A pictorial representation of the VDC cloud RBD model is shown in Fig.~\ref{fig:rbd2_super}, where $n$ clusters are connected in series, whereas $m$ cloud servers in parallel form each cluster. 
Through this representation, the availability of the overall VDC can be easily calculated by using series and parallel RBD formulas. Denoting by $A_{Clu_i}$ the availability of a generic ($i$-th) cluster, and by $A_{CS_j}$ the availability of a generic ($j$-th) cloud server, we can use the RBD parallel formula to derive the availability of a generic cluster as:
\beq
A_{Clu_i}= 1- \prod_{j=1}^{m} {(1-A_{CS_j})},
\eeq
and the RBD series formula to derive the availability of the overall VDC as:
\beq
A_{VDC}=\prod_{i=1}^{n} {A_{Clu_i}}.
\eeq

Other works have employed the RBD methodology to characterize the availability of virtualized infrastructures. In~\cite{rbd2}, authors utilize RBD to evaluate the steady-state availability of an SFC implemented as a chain of virtualized nodes connected in series. Similarly, redundancy schemes to increase the overall availability are implemented by exploiting the straightforward rules of RBD parallel connections. 
The series/parallel RBD rules are again employed in~\cite{rbd2bis} to model both virtualized and non-virtualized nodes, and to evaluate the impact on reliability when introducing the hypervisor. Series/parallel schemes are also used in~\cite{rbd2ter} to build an availability model of VNFs forming an SFCs, in~\cite{rbd2quater} to deal with reliable broadcasting in 5G NFV-based networks, in~\cite{rbd2quinquies} where a novel redundancy scheme to improve the SFC reliability is proposed, and in~	\cite{rbd2sexies,rbd2septies,rbd2octies} to characterize the SFCs in a MEC-based environment.
Authors in~\cite{rbd3} rely on an RBD representation to model the NFV data center components. In particular, the authors start from a baseline data center architecture (built through RBD), and evaluate different variants categorized per availability importance.  

Due to their ability to represent only high-level connections, RBDs are rarely used alone but rather in conjunction with other formalisms, allowing to capture more details, as we will see in Sect.~\ref{sec:multilev}. 

We conclude this section by highlighting some pros and cons associated with the RBD representation.
 
 \begin{itemize}
 	\item [\textbf{Pros}]
 	\item RBDs provide a clear and straightforward way to visualize the configuration and reliability of a system, making it easier to understand the interdependencies of components.
 	\item RBDs allow for the analysis of individual components and their contributions to the overall system reliability being characterized by a high degree of modularity.
 	\item RBDs can effectively model different redundancy configurations (e.g., series, parallel, k-out-of-n systems) to evaluate the impact on overall system reliability.
    \newline
 	\item [\textbf{Cons}]
 	\item RBDs typically assume that component failures are statistically independent, which may not be accurate in systems with significant dependencies or common cause failures.
 	\item While RBDs can scale to complex systems, the diagrams can become difficult to manage for very large systems with numerous components and interactions.
 	\item RBDs often represent components in a binary state (either working or failed), which may oversimplify the actual behavior of components that can degrade gradually or have multiple failure modes.
 \end{itemize}

 \subsection{Fault Trees (FTs)}
 \label{sec:ft}

\begin{figure}[t]
	\centering
	\captionsetup{justification=centering}
	\includegraphics[scale=0.5]{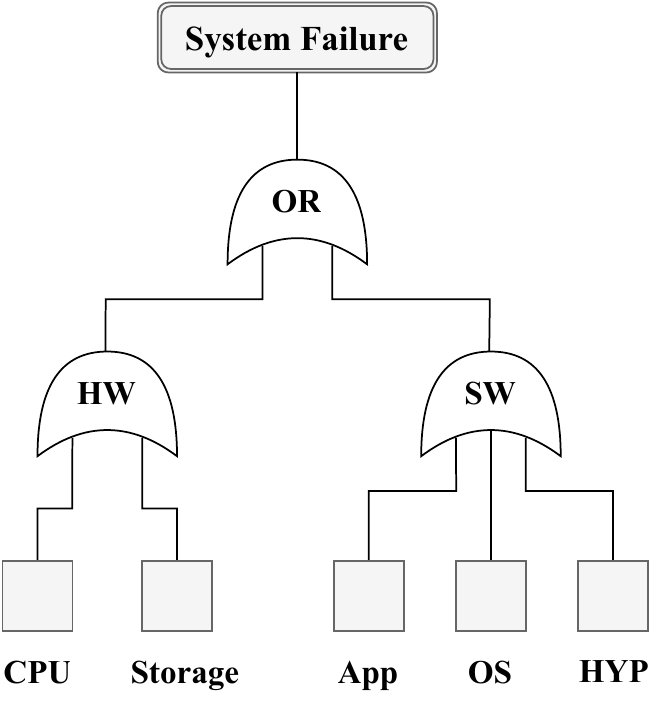}
	\caption{Fault Tree (FT) representation of a virtualized network node~\cite{multilev4}.}
	\label{fig:ft2_super}
\end{figure}

 A fault tree (FT) serves as a visual tool to represent the combination of events that can cause the occurrence of a system failure. On top of the FT there is the Top Event (TE), which acts as a root and represents the top-level failure that a system may reach. The construction of the FT starts from the TE and follows a deductive reasoning, progressing from general to specific, in a ``top-down'' manner. When analyzing the various elementary events contributing to the TE, this process is called fault tree analysis (FTA). 
 In the FT representation, the components of a system are represented as nodes and can be connected through logical operators. For instance, \textit{OR} gates are used to represent components connected in series, whereas \textit{AND} gates are used to represent components connected in parallel. Other logical operators (NOT, XOR, etc.) are supported by the FT formalism.
Accordingly, the FT representation is particularly suited to capture dependencies among components of a virtualized infrastructure. 
An explanatory example is offered in Fig.~\ref{fig:ft2_super}, where an FT is used to model a virtualized network node (see~\cite{multilev4}) made of two parts: the Hardware (HW) part including physical equipment such as CPU, storage, etc., and the Software (SW) part including the software applications (App), the Operating System (OS), and the Hypervisor (HYP). The TE is generically denoted as \textit{System Failure}.
It is straightforward to see that the virtualized network node fails if the hardware part fails \textit{OR} the software part fails. Similarly, the hardware part fails if the CPU fails  \textit{OR} the storage fails. Likewise, the software part fails if the application fails \textit{OR} the operating system fails \textit{OR} the hypervisor fails. 
In practice, the FT representation is useful when it is easy to break a system to model (a virtualized network node in the example above) into its main sub-components.
Thus, it is easy to express the TE in Fig.~\ref{fig:ft2_super} as:
\beq
TE=\underbrace{CPU \lor Storage}_{\textnormal{HW}} \lor \underbrace{App \lor OS \lor HYP}_{\textnormal{SW}},
\eeq
where the ``$\lor$'' symbol denotes the logical $OR$. Obviously, the availability of the virtualized node ($A_{VN}$) modeled in Fig.~\ref{fig:ft2_super} can be easily derived by applying the $NOT$ operator to the TE, namely:
\beq
A_{VN}=\overline{TE}.
\eeq

Notably, a reliability model for a cloud environment where FTs are extensively used is proposed in~\cite{ft1}. In this work, the authors associate the FTs to different states, where various cases (e.g., physical server available, virtual machines available/not available, etc.) are taken into account. 
FTs are exploited in~\cite{ft2} to model virtualized environments using a distributed multi-agent approach. Each agent autonomously evaluates the health of a virtual machine by analyzing triggered events (errors or other occurrences) using fault trees to capture the reliable (or unreliable) state of a system. 
A cloud-based decision support system is proposed in~\cite{ft3}, where a fault tree analysis is applied to model some distributed industrial applications according to the standard IEC $61499$. 
A fault tree model of the $5$G core network fault is advocated in~\cite{ft4}, where the tree is built by considering key nodes of the $5$G core infrastructure including the SMF and the AMF.
Authors in~\cite{ft5} advance a failure-aware virtual machine (VM) migration technique based on FTA to create a fault-tolerant cloud data center. The method involves constructing a fault tree to represent potential server failure events. In particular, by analyzing the fault tree, server failures are predicted, allowing virtual machines to be proactively migrated to alternative servers before any failure occurs.

Again using FTA concepts, authors in~\cite{ft6} suggest a distributed approach where each virtualized node can assess and predict its own health status. In their framework, each node can proactively decide whether to accept future jobs, delegate jobs to its replicated instances, or initiate a live migration process. The model is then evaluated using real Xen log traces.
Also authors in~\cite{ft7}  use FTA, in this case to monitor the health of each host within a fault-tolerant system. VMs are proactively migrated from less healthy hosts. The proposed methodology is evaluated against various failure scenarios, demonstrating superior throughput compared to state-of-the-art methods. The work in~\cite{ft8} employs FTs to identify repairs and failures of virtual machines, thereby reducing energy consumption and costs. Remarkably, the authors consider a dynamic FTA which incorporates temporal and dynamic aspects of the system behavior.

Note that RBDs series and parallel connections of components can be easily turned into the corresponding FT representation through the usage of \textit{NOT}, \textit{AND}, and  \textit{OR} logical operators. Considering again the RBD representation of VDCs introduced in Fig.~\ref{fig:rbd2_super}, the FT of a generic cluster (made of parallel cloud servers) can be written as:
\beq
\overline{Clu}=\overline{CS_1} \land \dots  \land  \overline{CS_m},
\label{eq:ft_clu_par}
\eeq
where $\overline{X}$ denotes the element failure (namely, element \textit{NOT} working), and the ``$\land$'' symbol denotes the logical \textit{AND}. Thus, it is easy to conclude that a cluster failure occurs when all cloud servers fail. Similarly, the FT representation of the whole VDC (meant as a series of clusters) can be written as:
\beq
\overline{VDC}=\overline{Clu_1}  \lor  \dots  \lor  \overline{Clu_n}.
\label{eq:ft_clu_ser}
\eeq
Also in this case, it is easy to deduce that a VDC failure (namely, VDC \textit{NOT} working) occurs when one or more clusters fail.

We finally mention a particular type of FT, namely the fault tree with repeated events (FTRE), where one or more events occur multiple times at different locations within the tree structure.  In a typical fault tree, each event is expected to appear only once, representing a unique point of failure or cause in the system being analyzed. However, in an FTRE, some events are duplicated, appearing in various branches of the tree. This repetition can happen due to complex dependencies or interactions within the system, where the same failure mode affects multiple parts of the system.

We conclude this section by pinpointing some pros and cons characterizing the FT representation.

\begin{itemize}
	\item [\textbf{Pros}]
	\item FTs help in systematically breaking down complex systems into more manageable sub-components.
	\item FTs offer a visual representation of the logical relationships between different components and their failure modes, by making it easier to understand failure pathways and their interdependencies.
	\item By analyzing the FT, critical components or failure points (i.e., those that significantly impact system reliability) can be easily identified, thus effective maintenance strategies can be designed.
	\item [\textbf{Cons}]
	\item For large and complex systems, FTs can become very intricate and difficult to manage.
	\item FTs do not easily handle partial failures which are common in real-world systems.
	\item Introducing repeated events increases the complexity of the fault tree, making it harder to analyze.
\end{itemize}

 \subsection{Continuous-Time Markov Chains (CTMCs)}
\label{sec:ctmc}

\begin{figure}[t]
	\centering
	\captionsetup{justification=centering}
	\includegraphics[scale=1.4]{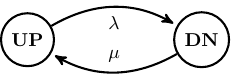}
	\caption{A simple two-state CTMC. The system is working when in state UP. In case of failure, the system passes to the DN (down) state with an exponential transition ruled by $\lambda$. After a repair occurs, the system comes back to the UP state with an exponential transition ruled by $\mu$.}
	\label{fig:ctmc}
\end{figure}

The Continuous-Time Markov Chain (CTMC) represents one of the most versatile formalisms to deal with reliability/availability concepts~\cite{trivedi-bobbio}. More in general, CTMCs are a mathematical tool used to model systems that undergo transitions from one state to another in a continuous time setting. Let us consider a simple system that can be in one of two distinct states depicted as circles as schematically shown in Fig.~\ref{fig:ctmc}: the state ``UP'' denotes a correctly working system, whereas the state ``DN'' (down) denotes a failed system. The system can move from one state to another. For instance, due to a failure, the system might change from UP to DN state via a \textit{transition} (the directed arc from UP to DN state).  Each transition has a rate, which determines how quickly the system is likely to move from one state to another. This rate is a key part of what makes the model ``continuous-time''. Unlike discrete models where changes happen at fixed intervals, in a CTMC, changes can happen at any moment. 
The time the system spends in a state before transitioning to another state is random, and is modeled using exponential distributions, which means that the probability of transitioning soon is high initially but decreases over time. 

We recall that the exponential distribution has the \textit{memoryless} property, meaning that the future behavior of the system only depends on its current state and not on how it arrived there (Markov property). This simplifies analysis and modeling. 
By considering the simple (two-state) CTMC model in Fig.~\ref{fig:ctmc}, we can see that once in state UP, after a failure occurs, the system passes into state DN with an exponentially-distributed time with transition rate $\lambda$. Likewise, once in state DN, after a repair occurs, the system comes back into state UP with an exponentially-distributed time with transition rate $\mu$. When dealing with availability and reliability aspects, in the majority of cases (and for the sake of simplicity) such rates are assumed to be constant over time (see, for instance, the useful lifetime zone into the bathtub curve previously discussed in Sect.~\ref{sec:rel_ava_concepts}), thus the CTMC is also called \textit{homogeneous} CTMC. Unless otherwise specified in the paper, the term CTMC refers to a homogeneous CTMC. 
The steady-state availability of the simple CTMC in Fig.~\ref{fig:ctmc} can be easily derived by considering the so-called balance equations (also known as the Kolmogorov forward equations~\cite{trivedibook}). The simple system to derive the steady-state availability of the CTMC in Fig.~\ref{fig:ctmc} follows:
\begin{align}
	\label{eq:baleq}
	\left\{
	\begin{array}{l} 
		{\begin{array}{ll} 
				\hspace{-0.2cm} \lambda \cdot \pi_{UP} = \mu \cdot \pi_{DN},
		\end{array}}
		\\
		\\
		\pi_{UP} + \pi_{DN} = 1,
	\end{array} 
	\right.
\end{align}
where $\pi_{UP}$ and $\pi_{DN}$ represent the steady-state probabilities to be in states up and down, respectively.
The first equation in~\eqref{eq:baleq} represents the balance equation stating that the rate at which the system enters a given state must be equal to the rate at which it leaves the state, whereas the second equation is a normalization condition meaning that the sum of the steady-state probabilities of the whole system must amount to $1$.  
Now, if we interpret the system steady-state availability ($A_{sys}$) as the probability to be in state UP, the solution of~\eqref{eq:baleq} leads to:
\beq
\label{eq:avasys}
A_{sys}= \pi_{UP} = \frac{\mu}{\lambda + \mu} = \frac{\textnormal{MTTF}}{\textnormal{MTTF}+\textnormal{MTTR}}, 
\eeq 
which corresponds to the basic definition~\eqref{eq:ava} since $\lambda=\textnormal{1/MTTF}$ and $\mu=\textnormal{1/MTTR}$.
 
\begin{figure}[t]
	\centering
	\captionsetup{justification=centering}
	\includegraphics[scale=0.31]{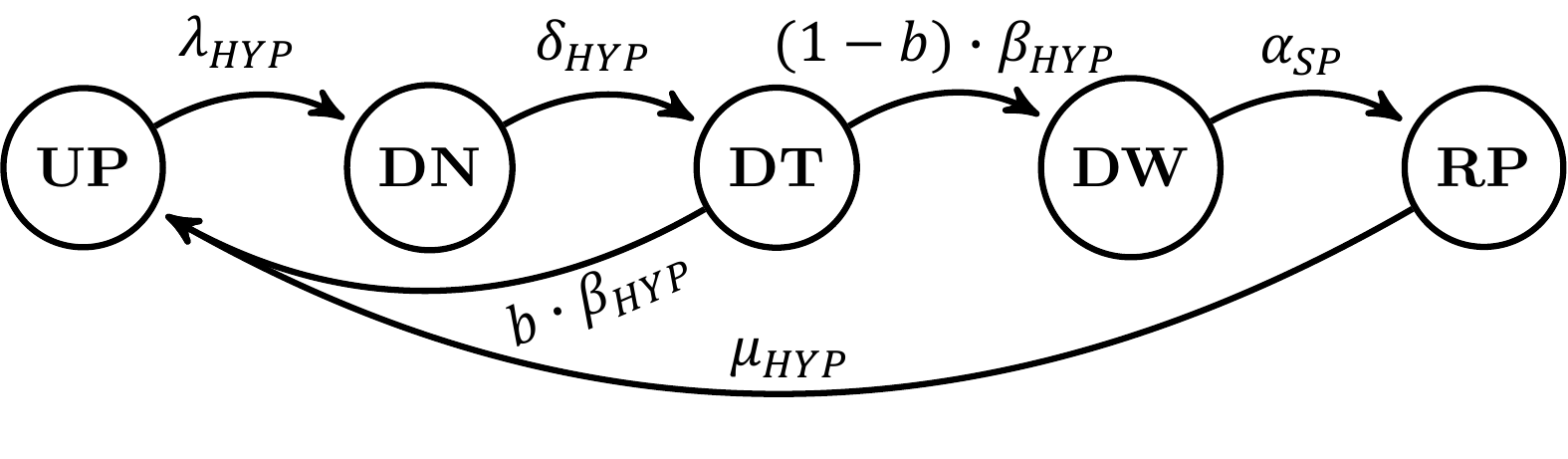}
	\caption{Continuous Time Markov Chain (CTMC) representation of a hypervisor~\cite{multilev1}.}
	\label{fig:ctmc2_super}
\end{figure}

Let us explore how to better use CTMCs to model the hypervisor, which constitutes one of the crucial elements of virtualized networks. Such modeling is described in~\cite{multilev1}, and the corresponding graphical representation is shown in Fig.~\ref{fig:ctmc2_super}. 
Starting from a working condition (UP state), the model can enter a fault condition (DN state) with rate $\lambda_{HYP}$. After a failure detection with rate $\delta_{HYP}$ the model reaches state DT, and the hypervisor is ready to be rebooted (to recover the failure) with rate $\beta_{HYP}$. With a probability $b$, the reboot is successful with the model coming back to the UP state. Conversely, with a probability ($1-b$), the reboot fails and the model enters the DW state prompting the summoning of a repair person at rate $\alpha_{SP}$. The model reaches the RP state when the repair begins and returns to the UP state upon completion of the repair, with an MTTR amounting to $\/\mu_{HYP}$. 

Other examples of works employing CTMCs to model reliability/availability aspects of virtualized networks are provided.
Authors in~\cite{ctmc1} exploit CTMCs to model the reliability behavior of $5$G base stations (BSs). Specifically, the state of a BS can be: healthy (all parameters of BSs configured with optimal value), sub-optimal (one or more parameters of BSs misconfigured), outage (the BS is not working).  
A highly-available load balancing scheme in an SDN scenario is proposed in~\cite{ctmc2}. Such scheme is built through the CTMC formalism, and is governed by a control policy which exploits a set of horizontal scaling thresholds to fairly distribute the network traffic between SDN controllers. 

A virtualized two-node cluster recovery model in which both software and hardware failures may occur is presented in~\cite{ctmc3}. The Markov model of the cluster embeds different states including active, standby, unstable, rejuvenation, VM failure, complete failure. 

The impact of rejuvenation onto a CTMC-based availability modeling of virtualized infrastructures is analyzed also in~\cite{ctmc3bis,ctmc3ter}.
A CTMC model to predict the rate of requests forwarded to public cloud providers by a federation of hybrid clouds is  proposed in~\cite{ctmc4}. Each member of the federation can share workload or resources to achieve cost savings while meeting QoS requirements. When all members have identical QoS requirements, authors provided theoretical results demonstrating that sharing all resources is the optimal strategy.
Authors in~\cite{ctmc5} start from the consideration that the VM migration enhances network robustness but also leads to service downtime and increased end-to-end delay. Accordingly, to examine the impact of VM failure, migration, and recovery, they define three states for the VMs in an edge server embodied into a CTMC structure. 
A survivability model of a virtualized system after a service breakdown occurrence by relying onto CTMCs is presented in~\cite{ctmc6}. The virtualized system structure includes three main components: a main host (which contains the hypervisor running an active VM), a backup host (a spare host replacing the main host when a VM migrates), a management host (responsible for controlling the whole cloud environment).
Moreover, a CTMC-based modeling is considered in~\cite{ctmc7}, where the availability of multiple cluster systems is investigated, in~\cite{ctmc8}, where the authors propose an analytical model of middle box functions in the realm of virtualized networks, and in~\cite{ctmc9} where a Markov chain model is employed to characterize the unavailability-aware model of virtualized and physical backup resources. 

Some pros and cons characterizing CTMCs:

 \begin{itemize}
 	\item [\textbf{Pros}]
 	\item CTMCs can provide both transient and steady-state analysis, allowing for evaluation of system performance over time and in the long run.
 	\item CTMCs allow modeling of state-dependent behaviors and transitions, providing a more accurate representation of systems that have varying failure and repair rates depending on their current state.
	\item Relying on solid theoretical foundation, CTMCs provide consistent methods for analysis and prediction.
 	\item [\textbf{Cons}]
 	\item Modeling very large systems with many states and transitions can become exceedingly complex and computationally intensive.
 	\item CTMCs typically assume that state transitions follow an exponential distribution, and thus sojourn times are also exponentially distributed, which may not always be appropriate for all systems or components.
	\item Solving the balance equations for large CTMCs often requires numerical methods, which can be computationally demanding and may involve approximation techniques.
 \end{itemize}
 
  \subsection{Markov Reward Models (MRMs)}
 \label{sec:mrm}
 
 A Markov Reward Model (MRM) enhances the capabilities of a CTMC by incorporating an attribute called a \textit{reward}, which can be assigned to the states, the transitions, or both~\cite{mrm_seminal,mrmbook}. 
 The reward typically corresponds to a performance metric or a cost related to a state, or some characteristic of the state or transition. Moreover, incorporating a reward into CTMC models offers a cohesive framework to define and calculate metrics that describe the system behavior relevant to the modeler.
 Although MRMs have been superseded by Stochastic Reward Nets (see Sect.~\ref{sec:srn}) which offer greater expressiveness and compactness, some works still use them in the field of reliability/availability of cloud networking, as outlined in the following. 
 Authors in~\cite{mrm1} introduce a performability analysis of cloud computing architectures, where MRMs are employed to get QoS measurements. MRMs are also exploited to analytically characterize failures in VMs~\cite{mrm1ter,mrm1quater}, and to characterize online cloud gaming~\cite{mrm2,mrm2bis}.
A performance/dependability MRM-based model is developed in~\cite{mrm3} to design a reliable cloud radio access network able to provide service continuity in case of baseband units failure.
 
Pros and cons of MRMs do not substantially change with respect to CTMCs.
 
 \subsection{Multi-State Systems (MSSs)}
\label{sec:mss}

Multi-State Systems (MSSs) are models used in reliability/availability engineering to represent systems that can exist in multiple states, each reflecting different levels of performance, capacity, or operational effectiveness. Unlike binary-state systems, which are limited to ``working'' and ``failed'' states, MSS can encompass a broader spectrum of conditions, such as fully operational, partially operational, and completely failed states~\cite{mss1}. This allows for a more realistic representation of complex systems, particularly those that degrade gradually or have multiple failure modes. Actually, a binary system can be considered the simplest case of an MSS having two distinguished states (perfect functioning and complete failure)~\cite{mss2}.
Remarkably, MSSs can be exploited in the domain of virtualized networks, where it is common to experience degradation with some VMs failing while others remain operational. 
Similarly to the CTMCs, the reliability and performance analyses through MSSs are conducted by examining the probabilities of being in each state and the rates of transitions between states. Moreover, being such structures more complex than classic CTMCs, various approaches have been proposed for their solution.

For example, authors in~\cite{mss3} and~\cite{mss4} adopt an approach based on multi-valued decision diagrams (MDDs) to evaluate the availability of cloud resources modeled via MSSs. MDDs are efficient graph-based data structures for symbolic representation and manipulation of multi-valued logical functions. 
Another approach to deal with MSSs is offered by the Universal Generating Function (UGF) and its variants, which provide a compact representation of the performance distribution of multi-state systems. Instead of enumerating all possible states and their probabilities, the UGF encapsulates this information in a polynomial-like function, significantly reducing the computational burden~\cite{levibook,ushakov}.
A particular variant of UGF dubbed Multi-dimensional UGF is profitably employed in~\cite{mss5,mss6,mss7,mss8,mss9,mss10} to model the degradation of softwarized nodes in presence of concurrent service providers in virtualized environments. A further refinement is presented in~\cite{mss11} to address an availability-aware virtual network function placement. 
\begin{figure}[t]
	\centering
	\captionsetup{justification=centering}
	\includegraphics[scale=0.45]{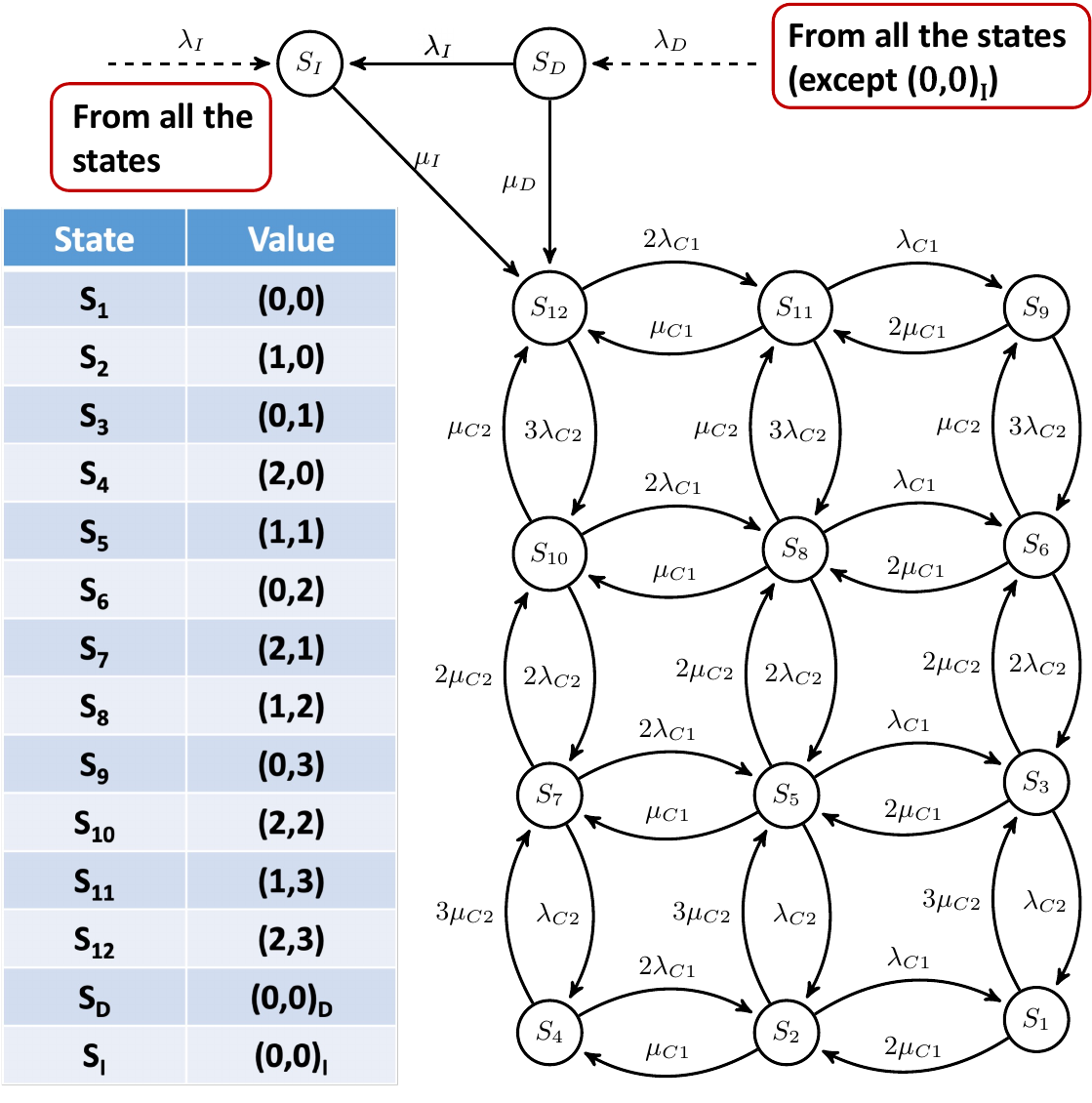}
	\caption{Multi-State System (MSS) applied to a three-layered containerized node shared between two providers~\cite{mss7}. Each state represents the number of working instances for each provider.} 
	\label{fig:mss}
\end{figure}  
As an example, Fig.~\ref{fig:mss} shows an MSS model (including $14$ states) of a containerized node adopted in~\cite{mss7}. Such a containerized node is represented as a classic three-layered structure including: an infrastructure layer (embodying hardware equipment), a Docker layer (embodying the container engine), and the Application layer (embodying the offered services through a number of containerized instances). Moreover, the containerized node is shared between two service providers, with one having $2$ containerized instances and the other having $3$ containerized instances, respectively.
Each state $S_{i}$ ($i=1,\dots,12$) models the working/failed behavior of \textit{containerized instances} associated to two service providers. For instance, the state $S_{11}=(1,3)$ denotes that provider $1$ has only $1$ instance working, whereas provider $2$ has $3$ instances working.  
It is evident that, starting from the fully operational state $S_{12}$ (where all containerized instances for both providers are working), two transitions are possible: $i)$ to state $S_{11}$ at rate $2\lambda_{C1}$, indicating that one of the two instances from provider $1$ may fail; $ii)$ to state $S_{10}$ at rate $3\lambda_{C2}$, indicating that one of the three instances from provider $2$ may fail. Conversely, the repair rate are $\mu_{C1}$ (from $S_{11}$ to $S_{12}$) and $\mu_{C2}$ (from $S_{10}$ to $S_{12}$) as one containerized instance can be repaired (rebooted) at a time. 
Moreover, states $S_I$ and $S_D$ represent the failure of Infrastructure and Docker layer, respectively. When the system is in either of these states, no working containerized instances are present, so both states have a value of $(0,0)$. 

Relevant pros and cons of the MSS formalism include:
 
 \begin{itemize}
 	\item [\textbf{Pros}]
 	\item MSSs allow for the modeling of systems that can exist in multiple states of performance, reflecting more realistic scenarios of gradual degradation and partial failures.
 	\item MSSs allow to evaluate a range of performance metrics beyond simple reliability, including availability, maintainability, and quality of service.
 	\item Tools like UGF provide efficient computational methods for analyzing MSSs, making it feasible to handle complex systems.
    \newline
 	\item [\textbf{Cons}]
 	\item Similarly to CTMCs, also MSSs typically assume that state transitions follow an exponential distribution.
 	\item While tools like UGF help, the computational effort needed to analyze large and complex MSSs can still be significant, especially for systems with many components and states.
 	\item There are only a few ready-to-use tools supporting the MSS-based analysis, potentially limiting the accessibility to this formalism.
 \end{itemize}

 \subsection{Stochastic Petri Networks (SPNs)}
\label{sec:spn}

Stochastic Petri Nets (SPNs) are an extension of the classical Petri net framework, incorporating stochastic (random) elements to model and analyze systems with probabilistic behavior. They are particularly useful in performance evaluation, reliability analysis, and in the study of systems where timing and randomness play crucial roles.
SPNs can be also considered as an evolution of CTMCs since, from an SPN, we can derive the corresponding Markov-based representation. Before analyzing (through a practical example) the relation between CTMCs and SPNs, it is useful to introduce some concepts and terminology.

\begin{figure}[t]
	\centering
	\captionsetup{justification=centering}
	\includegraphics[scale=0.48]{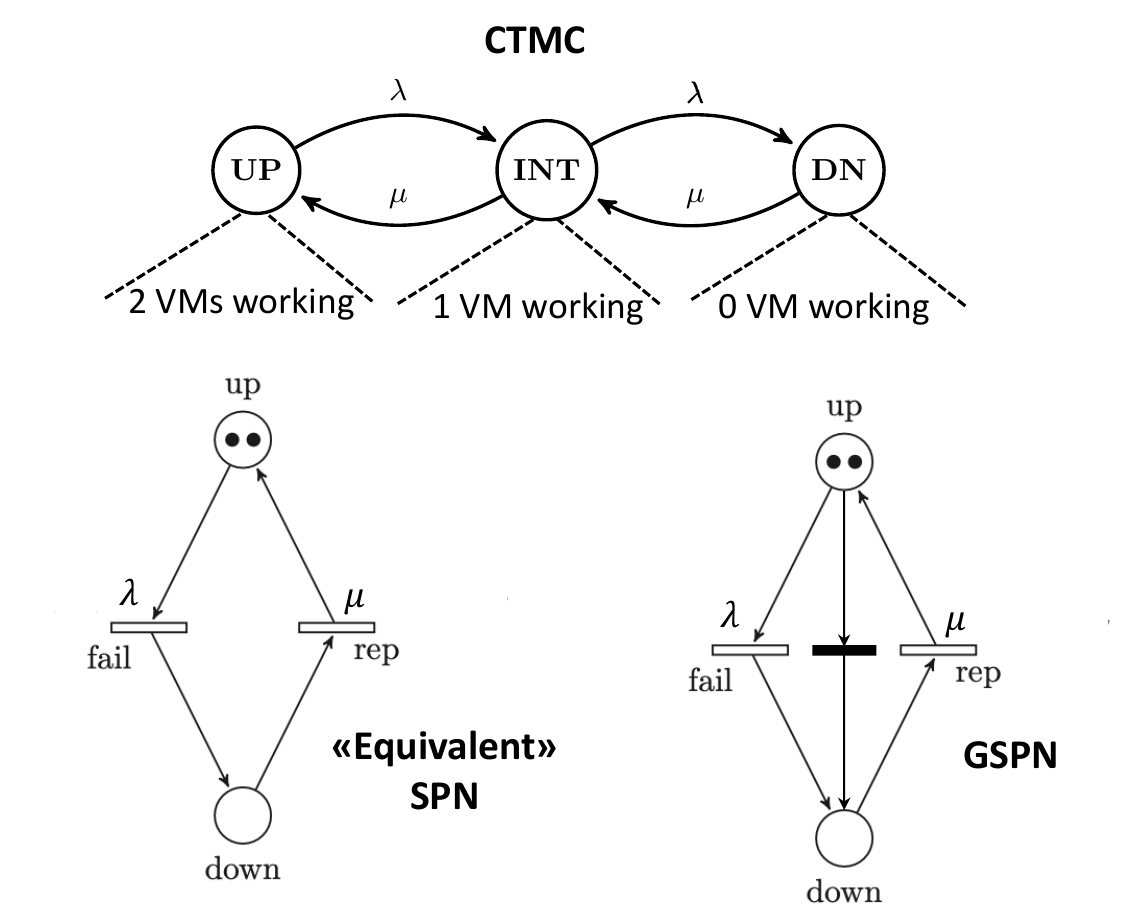}
	\caption{A CTMC (top figure) and its ``equivalent'' SPN (bottom-left figure) and Generalized SPN (bottom-right figure).} 
	\label{fig:spn}
\end{figure}

First of all, an SPN can be represented as a bipartite graph containing the following elements: 
\begin{itemize}
    \item \textit{Places}: represent conditions or states of the system (e.g., working/failed) and are depicted as circles.
 \item \textit{Timed Transitions}: represent the actions (e.g. passing from one place to another) typically governed by exponential probability distributions. Each timed transition has a rate parameter typically denoted by $\lambda$ (when passing from a ``working place'' to a ``failed place''), or by $\mu$  (when passing from a ``failed place'' to a ``working place'') which denote the exponentially distributed rate. The rate influences how frequently the transition fires. They are depicted as thick and unfilled rectangles.
	\item \textit{Tokens}: reside in places and represent the presence of a condition or the availability of a resource. They are depicted as dots (or numbers) inside a place. A particular distribution of tokens inside the SPN is known as \textit{marking}.
	\item \textit{Arcs}: connect places to transitions (input arcs) and transitions to places (output arcs), indicating the flow between conditions and events.
\end{itemize}
Let us consider Fig.~\ref{fig:spn} where, in the upper panel, we depict a CTMC representation of a simple virtualized system consisting of 2 VMs. When the system is functioning perfectly, both VMs are operational, and the model is in the UP state. When one VM fails, the model reaches an intermediate (INT) state of partial functioning at rate $\lambda$. Similarly, if the remaining VM also fails, the model reaches the completely down state (DN) at rate $\lambda$. The system returns to the INT state when one VM is repaired, and to the UP state when both VMs are repaired.

This simple CTMC model can be transformed into an ``equivalent'' SPN, as shown in the bottom-left panel of Fig.~\ref{fig:spn}. The two tokens in the \textit{up} place denote a fully operational state with both VMs working. When the transition named \textit{fail} is \textit{fired}, one token is transferred at rate $\lambda$ to the \textit{down} place, resulting in partial functionality with only one VM operational. 
When also the second token in the up place is transferred to the down place, the system has completely failed, with both VMs non-operational. 
Conversely, when the transition named \textit{rep} is \textit{fired}, one token is transferred at rate $\mu$ from the down place back to the up place. Another firing of \textit{rep} transfers the second token from the down place to the up place. The SPN offers a more compact representation of the CTMC, as the token mechanism automatically incorporates the intermediate state represented in the CTMC. Due to its flexibility, this formalism is widely used in the literature addressing reliability and availability in virtualized networks, as outlined below.
Authors in~\cite{spn1} propose a new approach to assess the availability of SFC placement by taking into account different redundancy strategies using SPNs. In particular, they devise an algorithm for automated generation and solving of small SPN models for SFC availability.  
SPNs are used to model a virtualized network in~\cite{spn2}, where the authors consider heterogeneous network load conditions. The aim is to promote the use of virtualization technology to effectively respond to various load conditions. 
SPNs are adopted also to evaluate the availability of a virtual firewall~\cite{spn3} (being a critical node of a virtualized network), to characterize a Mobile Edge Computing (MEC) scenario and analyze its performance~\cite{spn4}, to predict GPUs performance in a mobile cloud environment~\cite{spn5}, to capture the stochastic characteristics of cloud services and translating them into the elements of an availability model~\cite{spn6}, to evaluate steady-state availability during disaster recovery in cloud-based environments~\cite{spn7}, to evaluate the availability of OpenStack Swift and CephFS (two popular cloud storage services)~\cite{spn8}, to capture performability metrics in Infrastructure as a Service (IaaS) systems~\cite{spn9}, to evaluate dependability metrics in cloud data centers with focus onto the life cycle of running VMs~\cite{spn10}, to assess an availability planning for a MEC system~\cite{spn11}, and to carry on SFC reliability assessment along with a reliability optimization algorithm~\cite{spn12}.

A variant of SPNs is named Coloured SPNs (CSPNs), and introduces the concept of ``color'' to distinguish between different types of tokens. CSPNs are particularly useful for systems with multiple types of resources or entities that need to be modeled distinctly.
Some works experimented CSPNs into availability analysis of virtualized infrastructures as shown in~\cite{cspn1}, where the authors evaluate the availability of inter and intra-data center deployments of cloud services, in~\cite{cspn2} where an availability analysis involving resource utilization in cloud computing is carried on, in~\cite{cspn3}, where the authors evaluate the SFC availability under different elasticity strategies, and in~\cite{cspn4}, where a resilience evaluation of 5$G$ virtualized networks is performed. 

We conclude this section by outlining the key advantages and disadvantages of the SPNs.

 \begin{itemize}
	\item [\textbf{Pros}]
	\item SPNs can naturally capture system architecture with components and dependencies, describing the interactions between components and the stochastic nature of failures and repairs.
	\item With a more compact representation of the failure/repair behavior of components w.r.t. CTMCs, SPNs provide a versatile tool to model complex real-world systems.
	\item Several software tools (see Sect.~\ref{sec:tools}) are available for creating, analyzing, and simulating SPN models, making it easier for researchers and practitioners to apply SPNs to real-world problems.
 
	\item [\textbf{Cons}]
	\item The graphical nature of SPNs can make formal verification and validation harder compared to the algebraic representations of CTMCs.
	\item The analysis of SPNs, especially for large models, can be computationally intensive. This can make it challenging to perform detailed simulations and analyses within a reasonable time frame.
	\item Transitions in SPNs follow exponential distributions, which might not be suitable for all types of stochastic processes. This can limit the accuracy of models where other distributions (e.g., Weibull) are more appropriate.
\end{itemize}

 \subsection{Generalized Stochastic Petri Networks (GSPNs)}
 \label{sec:gspn}
 
A useful generalization of SPNs, namely Generalized SPNs (GSPNs), is introduced in~\cite{gspn_seminal}. Differently from standard SPNs, GSPNs allow for another type of transitions (in addition to time transitions) named \textit{immediate transitions} which fire in zero time once they are enabled. Such transitions, depicted as thin filled rectangles, do not effectively contribute to the time behavior of the system, thus they can be eliminated  from the corresponding Markov chain. A simplified version of a GSPN is shown in the bottom-right panel of Fig.~\ref{fig:spn}. This GSPN is derived from the SPN in the same figure, with the difference that now an immediate transition connects the up and down places. Such a transition is useful for representing ``catastrophic'' situations where, for example, the whole node crashes, leading to the immediate failure of both VMs. This results in the immediate transfer of the two tokens from the up place to the down place.
The introduction of immediate transitions (present also in other formalisms, as detailed in the following sections) allows to model a series of behaviors frequently encountered in the domain of virtualized networks, where nodes are often designed as ``nested'' structures (see Fig.~\ref{fig:vnf_models}). It means that, as the hardware layer fails, all the layers on top automatically fail. This behavior can be effectively captured through the usage of immediate transitions, as shown through a practical example in subsection~\ref{sec:srn}. 
A set of works employing GSPNs is outlined below. 

Authors in~\cite{gspn1} evaluate the availability of VNFs in active-standby configurations through GSPNs. The aim is to contrast such a model with an availability evaluation algorithm based on a network evolution model.  
The work in~\cite{gspn2} leverages GSPNs for performability evaluation of private cloud computing environments that adopt NoSQL DBMS as the storage system. Models are presented to jointly estimate throughput and availability, being critical QoS indicators. GSPNs are also employed to perform risk assessment of software defined networks with the emphasis on denial of service attacks~\cite{gspn3}. A coloured version of GSPN is also used in~\cite{cgspn1} to model the operational process of a cloud data center, considering the failure and resource usage of each involved node.

Pros and cons associated to GSPNs are outlined below.
\begin{itemize}
	\item [\textbf{Pros}]
	\item GSPNs can effectively model complex systems with multiple components and interactions, capturing both stochastic timing and logical dependencies.
	\item GSPNs introduce the immediate transitions which enable the representation of urgent or high-priority events that occur without delay, providing more accurate modeling of critical failure or repair events.
	\item There are several established tools (see Sect.~\ref{sec:tools}) that support the construction, simulation, and analysis of GSPNs, facilitating their practical application.
	\item [\textbf{Cons}]
	\item Similarly to SPNs, also GSPNs offer less control on formal verification and validation than CTMCs, as many states can be hardly accessible due to the formalism compactness.
	\item Effective use of GSPN tools requires familiarity with the formalism and the specific tool's features, which can involve a steep learning curve for new users.
	\item Similarly to SPNs, timed transitions in GSPNs typically follow exponential distributions, which might not be suitable for all types of stochastic processes. 
\end{itemize}

 \subsection{Stochastic Activity Networks (SANs)}
\label{sec:san}
Stochastic Activity Networks (SANs) are another powerful modeling formalism used to describe and analyze systems that exhibit stochastic behavior~\cite{san_seminal}. SANs are based on the concept of \textit{activity}, a stochastic event that modifies the system state. Activities are
analogous to transitions in GSPNs, since they can be timed
or instantaneous. 

Remarkably, timed activities can follow
a wide range of probability distributions (e.g., exponential,
uniform, normal, etc.), providing more flexibility in modeling
different types of stochastic behavior. In contrast, similarly to
the immediate transitions in GSPNs, the instantaneous activities allow for immediate state changes without time delay.
In SANs we also find the \textit{gates}, which provide a control logic to define when activities are enabled and how they affect the system state. An input gate controls whether an activity is enabled, whereas an output gate defines how the system state is updated after an activity completes.	
Currently, the main limitation of such a formalism is the scarcity of tools available (as far as we know, the only tool available is M{\"o}bius (see Sect.~\ref{sec:tools})). Nonetheless, some authors have already exploited SANs to model reliability/availability in virtualized environments. 
  \begin{figure*}[t]
 	\centering
 	\captionsetup{justification=centering}
 	\includegraphics[scale=0.32]{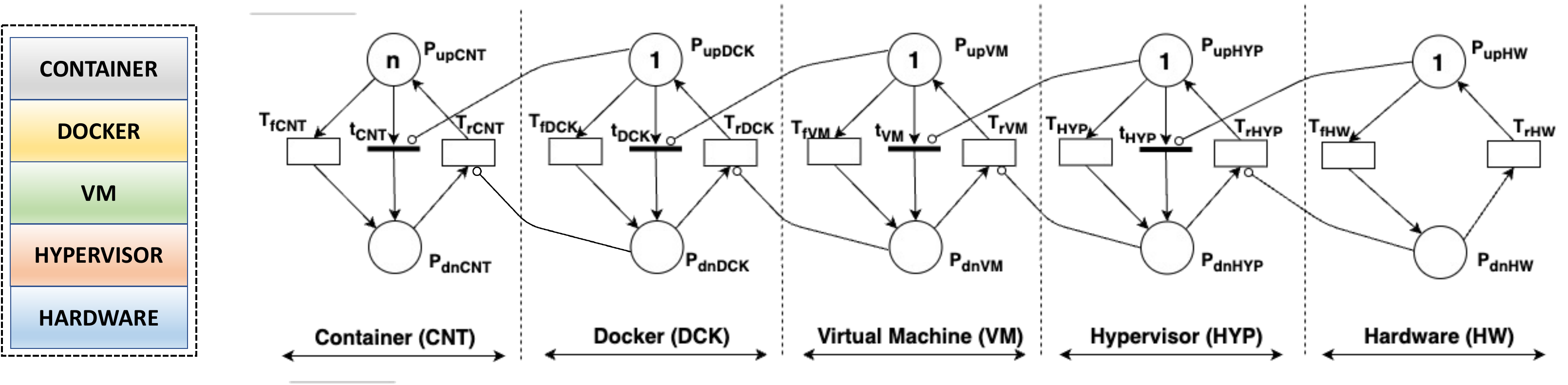}
 	\caption{Five-layer structure of a containerized node (on the left) and the corresponding SRN model (on the right)~\cite{srn_fig}.}
 	\label{fig:srn2_super}
 \end{figure*} 
In~\cite{san0}, authors offer an interesting perspective about the failure dynamics of the SDN controller, where a SAN model is used to characterize aspects such as software aging, different software failure modes, and external failures of the operating systems and of the underlying hardware. SANs are employed also in~\cite{san0bis} to assess and quantify the availability of NFV-supported services. 
In~\cite{san1}, authors derive a high-availability model of the NFV-MANO by characterizing it in terms of failures and repairs. In particular, the SAN availability model of the MANO involves a multi-layered architecture including the hardware, the operating system, the Docker daemon, and the MANO software, where software rejuvenation processes are taken into account. In a further refinement~\cite{san1bis}, the same authors include also an SDN-based infrastructure (considered as part of the VIM), and employ SANs to perform an availability analysis aimed at deriving useful redundancy configurations.
SANs is used also in~\cite{san2} to characterize and assess the performance and power consumption of virtualized servers in a cloud computing environment. More specifically, through SANs, the authors represent physical servers in three different power consumption and provisioning delay modes. The main idea is to leave the SAN model to decide to vertically scale the VMs, thus helping the overall system to save power by preserving acceptable performance. 

An availability analysis of the virtualized Evolved Packet Core (vEPC), representing one of the most interesting NFV use cases, is performed in~\cite{san3}. In particular, some crucial nodes of the vEPC architecture (e.g., MME, S-GW, P-GW, MANO) are separately modeled using the SAN formalism.
An evaluation of IaaS cloud data centers is proposed in~\cite{san3new}, where SAN modeling is employed to investigate the impact of different characteristics on service availability.
SANs are also used in~\cite{san3new2} to characterize the availability of some components of a $5$G-MEC system, including Radio Units, Distributed Units, MEC hosts, etc. 
An availability assessment of dispatching policies in IaaS cloud data centers is advanced in~\cite{san3bis} and
~\cite{san4}, where authors propose a SAN-based availability model of NFV-enabled network services under different modes (Standard Availability, Cold/Hot Protection).


Some pros and cons associated to the SANs follow.

\begin{itemize}
	\item [\textbf{Pros}]
	\item SANs provide greater flexibility than GSPNs, as activities allow for modeling non-Markovian systems that require general probability distributions.
	\item SANs use input and output gates to control the enabling and effects of activities, allowing detailed and complex control over the system state changes and interactions.
	\item SANs naturally support hierarchical and modular modeling, facilitating the construction and analysis of large, complex systems by breaking them down into smaller, manageable sub-models.
	\item [\textbf{Cons}]
	\item Like SPNs, SANs use a graphical representation that, while intuitive for modeling, can complicate formal validation when compared to the more structured algebraic approach of CTMCs.
	\item The detailed and flexible nature of SANs can lead to very complex models that can be difficult to design, manage, and understand.
	\item Very few tools are currently available to manage SANs.
\end{itemize}
 
  \subsection{Stochastic Reward Networks (SRNs)}
 \label{sec:srn}

Similarly to SANs, the Stochastic Reward Networks (SRNs) are a further refinement of GSPNs~\cite{srn_seminal1,srn_seminal2}. The \textit{reward} concept comes from the MRMs, where a reward rate is attached to each state of the Markov chain. In particular, SRNs introduce the reward functions which map states or transitions to real numbers, quantifying aspects like cost, performance, reliability, or availability. Rewards can be accumulated over time, providing a measure of the system behavior over its operational lifetime.
SRNs increase the modeling power of the GSPNs by adding new elements including: guard functions, inhibitory arcs, marking dependent arc multiplicities, general transition priorities, and reward rates at the net level. 
A \textit{Guard Function} is a Boolean function associated with a transition. The transition is considered enabled only if the guard function is true. 
Similarly to the guard function, an \textit{Inhibitory Arc} is an arc which prevents a transition to be fired unless a given condition is met.  
\textit{Marking dependent arc multiplicities} allow either the number of tokens required for the transition to be enabled, or the number of tokens removed from the input place, or the number of tokens placed in an output place to be a function of the current marking of the model. 
Timed and immediate transitions are also present in SRNs to model actions occurring in a finite time and instantaneous events, respectively. 

To better understand how to employ the SRN formalism to model some aspects of virtualized network infrastructures, let us consider the SRN shown in Fig.~\ref{fig:srn2_super}. This SRN characterizes a containerized network node made of $5$ nested layers~\cite{srn_fig}, starting from the lowest layer: hardware, hypervisor, virtual machine, Docker, and container. 
In this structure, we can recognize $5$ $P_{up}$ [$P_{dn}$] places denoting the five layers working [failed], respectively. We also note that all the $P_{up}$ places contain $1$ token, except for the $P_{upCNT}$ which contains $n$ tokens since the authors assume the presence of $n$ working containers within the node. Timed transitions are represented as thick unfilled rectangles and denoted by $T_f$ [$T_r$] to encode failure [repair] events. Immediate transitions are represented as thin filled rectangles and denoted by $t$. 
To analyze the evolution of this SRN it is useful to start from an initial working condition where all containers work, namely, $n$ tokens are in the place $P_{upCNT}$. In the event of a single container failure, the transition $T_{fCNT}$ is triggered, moving a token from $P_{upCNT}$ to $P_{dnCNT}$. In contrast, when the container is repaired, the transition $T_{rCNT}$ is triggered, returning the token to $P_{upCNT}$. This process repeats in the case of multiple container instance failures, transferring the appropriate number of tokens to $P_{dnCNT}$.
In the event of a Docker failure, the transition $T_{fDCK}$ is triggered and the (unique) token in $P_{upDCK}$ is transferred to $P_{dnDCK}$. Obviously, in this case all the containers on top will fail. In the SRN model, this situation is translated into ``deactivating'' the inhibitory arc which connects $P_{upDCK}$ and $t_{CNT}$, thus $t_{CNT}$ causes the immediate transferring of all $n$ tokens from $P_{upCNT}$ to $P_{dnCNT}$.
Moreover, we also note an inhibitory arc from $P_{dnDCK}$ to $T_{rCNT}$. This prevents tokens in $P_{dnCNT}$ from returning to $P_{upCNT}$ in case Docker is down (i.e., when the token is in $P_{dnDCK}$). 
Similar reasoning holds true in case of virtual machine, hypervisor, or hardware fault where a cascade effect can be observed. Notably, the only layer with no immediate transition is the hardware layer since no further dependencies from other lower layers must be considered.  
The steady-state availability is obtained by:
\beq
A=\sum_{m \in \mathcal{M}} r_m \pi_m,
\label{eq:ava_srn}
\eeq
where $\mathcal{M}$ is the set of all possible markings (namely, the set of all possible tokens distributions) that can be split in a subset of ``up'' states (having reward $r_m=1$) and ``down'' states  (having reward $r_m=0$), and where $\pi_m$ denotes the probability of the model to be in state $m$. 

We note that the number of works exploiting such formalism (eventually combined with others as discussed in Sect.~\ref{sec:multilev} further ahead) is significantly larger than the number of works exploiting SANs. This is probably due to the greater availability of tools for SRNs with respect to those available for SANs.
 
SRNs are the main formalism employed  in~\cite{srn10quater,srn10quinquies,srn10sexies}, where stochastic models of SDN-based infrastructures are introduced to analyze the pertinent availability.
Authors in~\cite{srn1} employ the SRNs to perform an availability analysis of an IaaS cloud architecture. Adopting this approach, they compare two cases, a monolithic SRN case where an IaaS cloud infrastructure is represented by using one and big SRN, and small sub-cases made of reduced SRN sub-models representing three \textit{pools} of VMs: hot (running), warm (turned on, but not ready) and cold (turned off). The advantage of SRN sub-modeling is demonstrated also in~\cite{srn1b} to analyze the IaaS cloud availability. A stochastic modeling of an IaaS infrastructure is proposed also in~\cite{srn1bis} and in~\cite{srn1ter}, where some SRN sub-models are combined to capture the monolithic IaaS structure. 
SRNs are also adopted in~\cite{srn2} to characterize some computer grid resources aimed at evaluating both performance and availability. Operations such as VM placement and migration are characterized through SRNs in~\cite{srn3}, where different scheduling methods are described as reward functions to perform an adaptive evaluation. 
An availability modeling of a hypervisor with three types of rejuvenation techniques (cold-VM, warm-VM, Migrate-VM) is presented in~\cite{srn4}. 

SRNs have also been employed to model active and backup servers in virtualized environments in~\cite{srn4bis}, and to represent cloud data centers with disaster tolerance in~\cite{srn4ter}.
An ``experience availability'' analysis is proposed to evaluate online cloud service in terms of both availability and response time in~\cite{srn5}, where the SRN formalism is extensively used. 
Furthermore, SRNs are used to model: systems with software rejuvenation enabled by the VM migration schedule~\cite{srn6}, availability and security aspects involved into VM migration processes~\cite{srn7}, software defined storage systems~\cite{srn8}, virtualized systems including hypervisor's rejuvenation policies~\cite{srn8bis}, reliability of drones in a fog/edge environment~\cite{srn9}, QoS-related performance of cloud data centers~\cite{srn9bis}, performance and availability aspects related to the IMS infrastructure~\cite{srn10,srn10bis}.

Interestingly, SRNs have also been exploited to automate the construction of failure/repair models in virtualized infrastructures. In~\cite{srnaut1}, authors present a component-based availability modeling framework (named Candy) to build an availability model from system specifications described  by the so-called Systems Modeling Language (SysML) which is then connected to the corresponding SRN model. A semi-automated framework (HASFC) to deal with the availability management of NFV-MANO is described in~\cite{srnaut2}. The HASFC framework, through the SRN availability modeling, is able to build a highly available service function chain that the MANO can deploy. Tools aimed at facilitating the deployment of reliable service function chains have also been presented in~\cite{srnaut3} and in~\cite{srnaut4}, where SRN is exploited to design the availability models whereas the Model-driven Engineering (MDE) paradigm aims to produce interpretable models by means of specific semantic rules. 

Some remarkable pros and cons associated to SRNs may be highlighted:
\begin{itemize}
	\item [\textbf{Pros}]
	\item Differently from GSPNs and SANs, SRNs provide built-in mechanisms to evaluate regime and transient rewards, useful for system performance and cost analysis.
	\item SRNs (similar to SANs) allow to model both Markovian (exponential) and non-Markovian (general) behaviors through reward structures.
	\item Many tools support SRNs for stochastic modeling and reward-based performance analysis.
	\item [\textbf{Cons}]
	\item Similarly to GSPNs and SANs, SRNs offer less control over individual states compared to CTMCs. This can result in some states being less accessible.
	\item Constructing and analyzing SRNs requires specialized knowledge in stochastic processes, reward structures, and the use of related software tools, which may necessitate training and experience.
	\item Some advanced SRN features (e.g., impulse rewards) depend on specific tool support, which may not be universally available.
\end{itemize}

\subsection{Non-Markovian Processes (NMPs)}
\label{sec:nmp}

In contrast to Markov processes, which satisfy the Markov property of absence of memory, non-Markovian processes exhibit memory effects, meaning that the probability of transitioning to a future state depends on the entire history of the system evolution up to the present moment.
Accordingly, time dependencies in non-Markovian processes can be complex and may not follow simple exponential decay as in Markovian processes. Instead, event durations and inter-event times may follow arbitrary probability distributions.
In order to exploit the flexibility of non-Markovian models, also compact formalisms (and related tools) such as SPNs, SANs, SRNs, can be conveniently modified to embed non-Markovian features~\cite{nmp1,nmp2}.
In this section we briefly describe $3$ types of NMPs (along with pertinent related works): Semi-Markov processes, Markov Regenerative Processes, and Phase-Type Expansions. 

 \subsubsection{Semi-Markov Processes (SMPs)}
\label{sec:smp}

A semi-Markov process is a stochastic process that generalizes the Markov process by allowing the holding (sojourn) time in each state to follow any probability distribution, not necessarily exponential\cite{mrmbook}. In reliability and availability engineering, semi-Markov processes are used to model systems where the assumption of exponentially distributed times (as in traditional Markov processes) does not hold. This is common in practical applications where the time to failure, repair times, or maintenance durations follow more complex distributions.
In the field of reliability/availability of virtualized infrastructures, recent works have employed SMPs to: model availability and reliability of service chains~\cite{smp1,smp2,smp3,smp4,smp5}, characterize the availability of virtualized nodes in presence of rejuvenation~\cite{smp6}, describe the resilience of unmanned aerial vehicles in a multi-access edge computing environment~\cite{smp7}, assess the availability of cloud-based environments~\cite{smp8,smp9,smp10,smp11,smp12}. 

\begin{figure}[t]
	\centering
	\captionsetup{justification=centering}
	\includegraphics[scale=0.34]{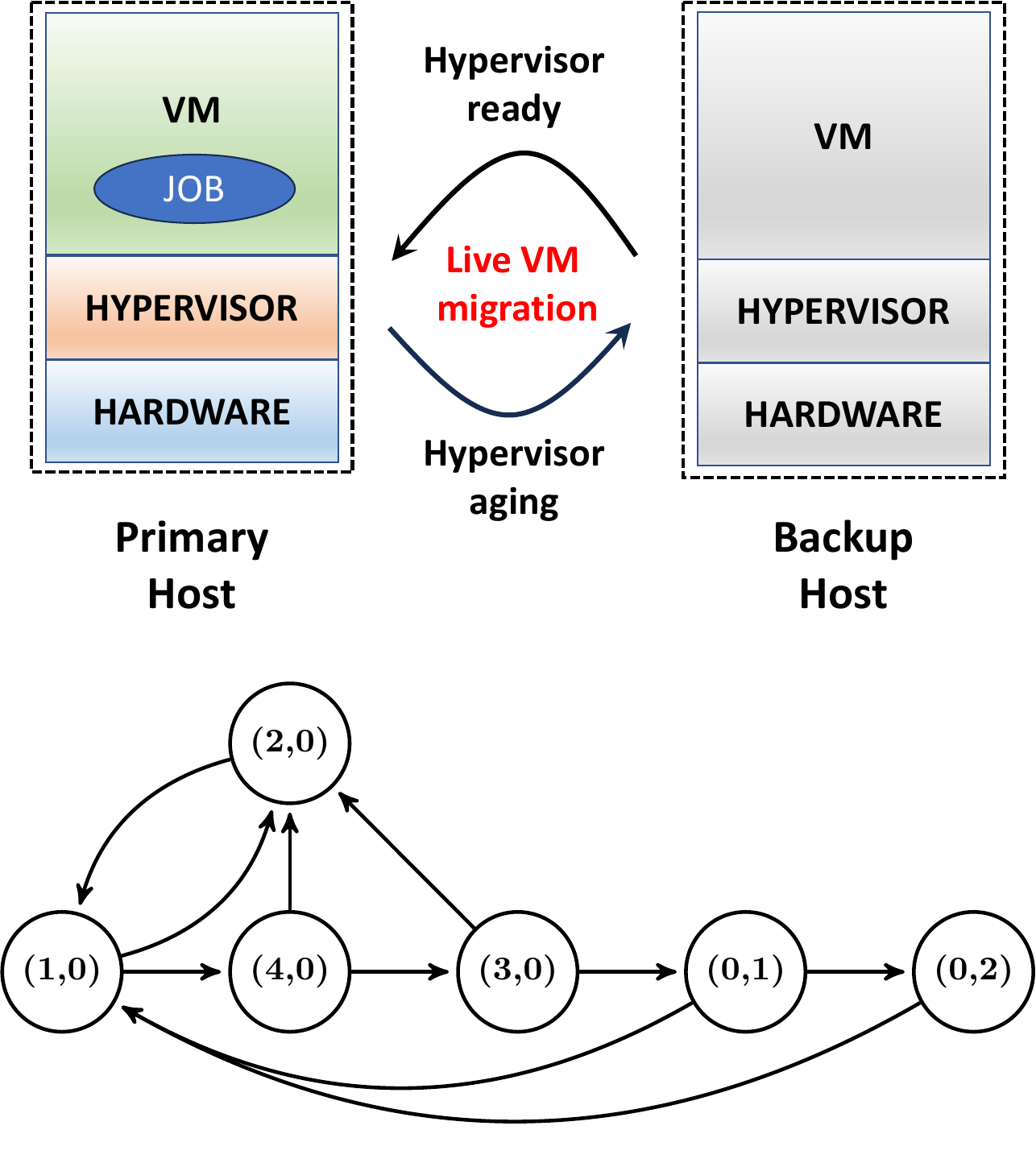}
	\caption{Virtualized system with live VM migration~\cite{smp6}: technological (top) and SMP model (bottom).}
	\label{fig:smp}
\end{figure}

Figure~\ref{fig:smp} shows the technological representation (top part) and the corresponding SMP model (bottom part) of a virtualized system that incorporates live VM migration techniques for rejuvenation, aimed at more accurately estimating job completion time~\cite{smp6}. The system consists of two hosts, a \textit{primary host} with robust computational capabilities hosting an active VM, and a \textit{backup host} with limited computational capacity to support live VM migration.
If hypervisor aging is detected during job execution on the primary host, the backup host is selected to take over the job. The SMP model encodes the system state as a $2$-tuple index $(i,j)$, where $i$ and $j$ represent the states of the primary and backup hosts, respectively. The following states are defined: $(0)$ indicates that the job is not running; $(1)$ signifies that the job is running on a host whose robustness is ensured through hypervisor reboot or fix; $(2)$ represents a host failure caused by a hypervisor crash; $(3)$ indicates that the job can be migrated via live VM migration; and $(4)$ denotes degraded host performance due to hypervisor aging.
In the SMP model depicted in Fig.~\ref{fig:smp}, all state transitions follow non-exponential (general) distributions. Specifically, most transitions are modeled as \textit{hypo-exponential}, reflecting the increasing failure rate typically associated with software aging over time.

 \subsubsection{Markov Regenerative Processes (MRGPs)}
\label{sec:mrp}
 
A Markov regenerative process can be considered as a generalization of many stochastic processes, including SMPs~\cite{cinlarbook,kulkarnibook}. In reliability and availability engineering, MRGPs are particularly useful for modeling systems where certain events cause the system to ``reset'' its probabilistic behavior, while between these events, the system behavior can be more complex. This is common in systems with periodic maintenance, inspections, or other interventions that effectively restart the probabilistic model.
More formally, in an MRGP, the Markov property holds only when the process enters a subset of specific states called regeneration states~\cite{trivedi-bobbio}. 
Attempts in using MRGPs in the realm of virtualized networks include~\cite{mrp1}, where a Markov regenerative process is exploited to build an analytical model of an application service in a virtualized system,~\cite{mrp2}, where an MRGP model to capture the states granularity of an application server is considered, and ~\cite{mrp3}, where the authors propose a performability model for RAID storage systems using MRGP to compare different RAID architectures.

 \subsubsection{Phase-Type Expansion (PTE)}
 
The Phase-Type Expansion is a mathematical technique used in reliability and availability engineering to model complex systems with non-Markovian behavior by decomposing them into a series of simpler, Markovian sub-models known as phase-type distributions. This technique is particularly valuable for analyzing systems with diverse lifetimes or repair durations.  
Some works exploiting this concept in the field of virtualized networks include~\cite{pte1}, where the authors apply a phase-type expansion to assess the availability of VMs subject to two rejuvenation policies (cold-VM and warm-VM),~\cite{pte2}, where the hypervisor time to failure is characterized through phase-type distributions,~\cite{pte3}, where PTE is used in the availability modeling of fault tolerant cloud systems. 
 
\label{sec:pte}
 
Some pros and cons associated to non-Markovian approaches follow:
\begin{itemize}
	\item [\textbf{Pros}]
	\item NMPs can model failure/repair events with any probability distribution (e.g., Weibull, log-normal, gamma), which is more realistic for many real-world systems, where the exponential distribution assumption of Markovian models does not hold.
	\item NMPs are versatile and can be applied to systems with complex temporal behaviors that cannot be captured by memoryless properties of Markovian models.
	\item NMP can use phase-type distributions to approximate any arbitrary distribution, enabling detailed and flexible modeling of time-to-event data.
	\item [\textbf{Cons}]
	\item Constructing non-Markovian models is more complex than building Markovian models. It requires detailed knowledge of the underlying processes and appropriate distribution fitting.
	\item Simulations and analytical solutions can take much longer to execute compared to Markovian models, potentially limiting their practical applicability in time-sensitive scenarios.
	\item There are fewer specialized software tools available for non-Markovian models compared to Markovian models, which can limit the ease of implementation and analysis.
\end{itemize} 
 
  \subsection{Multi-level models}
\label{sec:multilev} 
 
\begin{figure}[t]
	\centering
	\captionsetup{justification=centering}
	\includegraphics[scale=0.4]{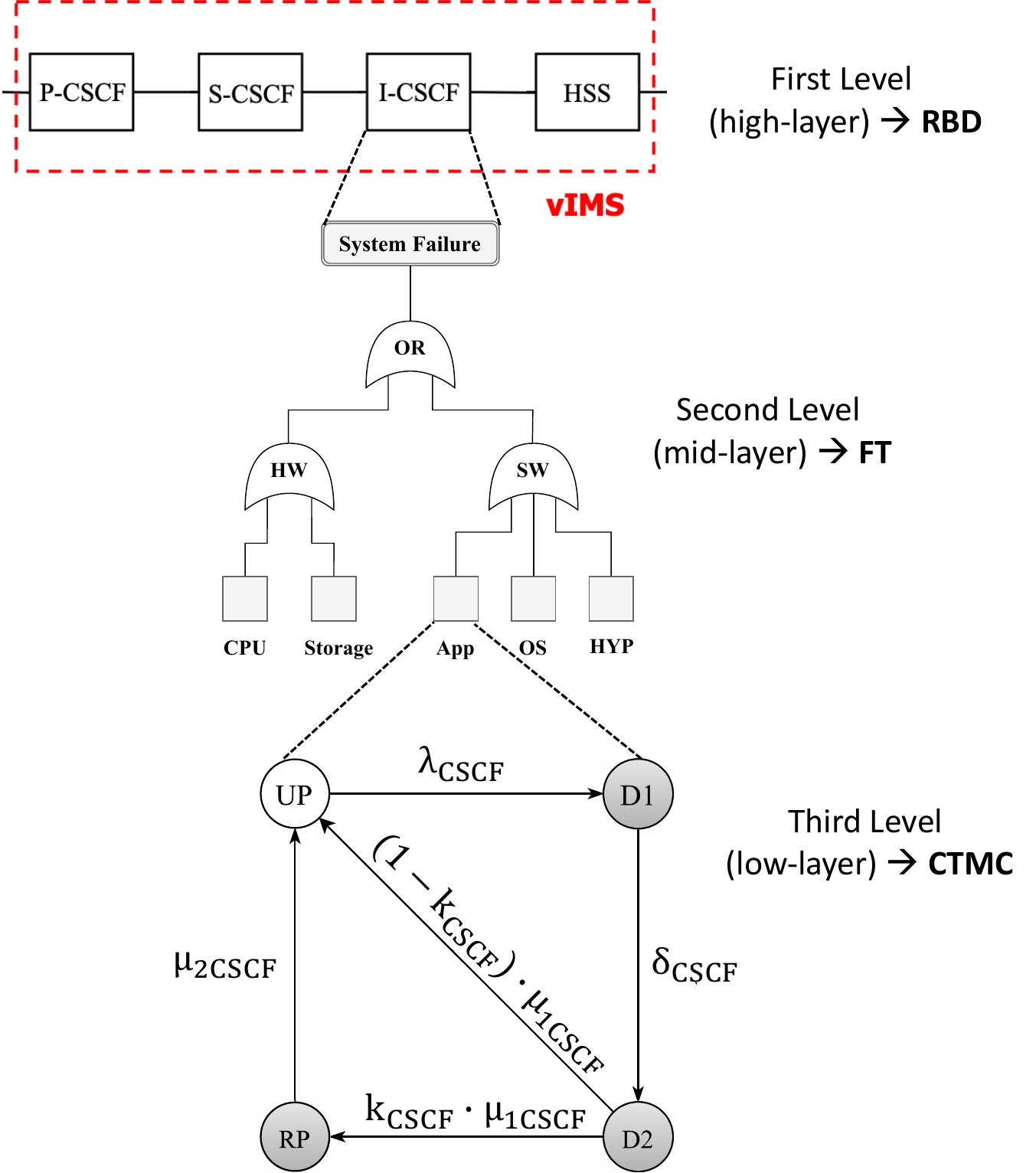}
	\caption{Multi-level formalism to evaluate the availability of a virtualized IP Multimedia Subsystem (vIMS) infrastructure~\cite{multilev4}, where multiple formalisms are combined together: RBD to model high-level interconnections among nodes (first level), FT to model sub-components dependencies (second level), CTMC to model failure/repair behavior of sub-components.}
	\label{fig:multilev}
\end{figure}

All the formalisms seen in the previous subsections have their own peculiarities and allow for modeling specific reliability/availability aspects of a system. However, given the complexity of real-world systems, it is difficult for a single formalism to appropriately capture the whole system behavior. Thus, it is possible to combine different formalisms to capture various aspects of a system, resulting in the so-called multi-level models~\cite{trivedi-bobbio,multilev_seminal}. 
In~\cite{multilev4}, three different formalisms are adopted to characterize the availability of a virtualized IP Multimedia Subsystem (vIMS) infrastructure as shown in Fig.~\ref{fig:multilev}. The vIMS system forms an SFC with $4$ nodes (P-CSCF, S-CSCF, I-CSCF, HSS) connected in series as represented in the topmost part of Fig.~\ref{fig:multilev}. This ``macro-structure'' can be modeled using RBDs, i.e., the first level of availability modeling. The second level of modeling employs a fault tree structure to capture the interdependencies among hardware and software components within each node. For a fault tree description of the sub-model in Fig.~\ref{fig:multilev}, refer to Sect.~\ref{sec:ft}. The third (and last) level of modeling uses a CTMC to characterize each single sub-system of hardware and software modules. In particular, a $4$-state CTMC models the software application (App) sub-system on top of the I-CSCF node. Starting from a perfectly functioning condition, the App enters state $D1$ with rate $\lambda_{CSCF}$ as a consequence of a failure not yet detected. Once the failure is detected, the sub-system enters state $D2$ with a detection rate $\delta_{CSCF}$. At this point, a recovery procedure starts with repair rate $\mu_{1CSCF}$, and two events can occur: $i)$ the recovery procedure is successful and the sub-system comes back to the $UP$ state with probability $(1-k_{CSCF})$; $ii)$ the recovery procedure is unsuccessful and the sub-system enters the $RP$ state with probability $k_{CSCF}$ before reaching the $UP$ state with rate $\mu_{2CSCF}$. In this work, the multi-level formalism is managed via SHARPE (see Sect.~\ref{sec:tools}) which provides an efficient way to solve a model which incorporates more than one formalism. 
Some existing works (see below) adopt this flexible approach to deal with reliability/availability issues of complex systems by combining the advantages offered by different techniques. 

Authors in~\cite{multilev1} adopt a two-level approach to model a virtualized system where fault trees are employed in the upper level to capture the nodes dependencies, and CTMC is adopted in the lower level to characterize the availability of sub-systems (e.g., CPU, memory, etc.). The FT/CTMC coupling is also used in~\cite{multilev2} to evaluate the availability of virtual machines within a Cloud infrastructure, e.g., in~\cite{multilev3} to build an availability model of a virtual triple modular redundant system with applications in cloud computing, and in~\cite{multilev5} to characterize an edge network from a resiliency point of view. 
An RBD/SPN combination is used in~\cite{multilev7}~\cite{multilev8} and in~\cite{multilev9} for a dependability evaluation of data center infrastructures, and in~\cite{multilev10,multilev10bis} to assess performance and availability of storage services in a private cloud.   
A detailed availability analysis of virtualized network environments supported by the combination of RBD and SPN formalisms is presented in~\cite{multilev10ter}. In particular, both nodes and links in a virtualized network environment are modeled through SPNs, for two possible scenarios: a ``no-redundancy mapping scenario'', where virtual nodes are mapped onto physical nodes and virtual links are mapped onto physical paths, and a ``hot-standby mapping scenario'', where each virtual node is mapped onto two physical nodes (primary+spare) and each virtual link is mapped onto four physical paths, corresponding to the Cartesian product of the physical nodes (primary-primary, primary-spare, spare-primary, spare-spare).  

The RBD/SRN couple is adopted in:~\cite{multilev6} for availability assessment of virtualized data centers, in~\cite{multilev6bis,multilev6ter} to evaluate steady-state availability of an IMS infrastructure deployed in a containerized environment, in~\cite{multilev6quater} for availability evaluation of an SFC managed by the Virtualized
Infrastructure Manager (VIM).
RBD/SRN is also adopted in ~\cite{multilev6quinquies} to characterize the availability of $5$G virtualized architectures where AMF, SMF, and UPF nodes need to be conveniently replicated to ensure both a given performance (in terms of end-to-end delay constraint) and a given availability target (identifed as the five-nines).
FTs and SRNs are employed in~\cite{multilev18} for a container availability analysis.
RBDs and CTMC are used in a combined manner to characterize, from an availability point of view, multimedia services in a cloud environment as shown in~\cite{multilev11,multilev12,multilev13,multilev14}.
A three-layered model is used for reliability and availability evaluation of cloud data center networks in~\cite{multilev16}, where: reliability graphs (an ``intermediate'' formalism between RBD and FT) are used to model the system network topology at the top layer, FTs are used to model the architecture of the subsystems at the middle layer, and SRNs are used to capture the behaviors and dependency of the components in sub-systems at the bottom layer. A similar hierarchical structure is considered in~\cite{multilev17} to model availability in SDNs, where failure modes and recovery behaviors include link failures at network level, and software and hardware failures at network device level. A combination of RBD and MRM is adopted in~\cite{multilev19} to model a mobile cloud environment also considering four distinct redundancy mechanisms, and in~\cite{multilev20} to analyze the warm- standby replication mechanism in a virtualization environment.

Some pros and cons associated to multi-level models include:
\begin{itemize}
	\item [\textbf{Pros}]
	\item Multi-level modeling enables the reuse of existing sub-models, allowing different components or subsystems to be analyzed independently and then integrated into a larger system model, improving scalability.
	\item Different formalisms have different strengths and weaknesses. Combining them allows to leverage the strengths of each formalism to effectively model different aspects of a system's behavior .
	\item Multi-level formalisms enable to represent system behavior at different levels of detail, from high-level system architecture to detailed component interactions, allowing for a more granular analysis.
	\item [\textbf{Cons}]
	\item Combining multiple formalisms can result in complex models that are challenging to develop, understand, and maintain. Managing the complexity of such models requires significant expertise and computational resources.
	\item Integrating different formalisms and tools can be non-trivial, as they may use different modeling languages, assumptions, and data formats. Ensuring seamless integration and data consistency across different formalisms can be a significant challenge.
	\item Combining different formalisms increases the risk of inconsistencies or inaccuracies between different parts of the model. Ensuring consistency and coherence across different levels of the model requires special attention to detail and thorough validation processes.
\end{itemize} 

\vspace{1em}
We conclude this section by providing a concise \textit{mapping} between the analyzed references and the adopted formalisms summarized in: Table ~\ref{tab:non_state_space} (works adopting non-state space formalisms), Table~\ref{tab:state_space_classic} (works adopting Markovian state-space formalisms), Table~\ref{tab:state_space_petri} (works adopting Markovian (Petri-based) state-space formalisms), Table~\ref{tab:state_space_nonmarkov} (works adopting non-Markovian state-space formalisms), Table~\ref{tab:multilev} (works adopting multi-level formalisms).

\onecolumn

\begin{table}[h]
    \centering
    \footnotesize
    \renewcommand{\arraystretch}{1.2}
    \caption{Works using non-state space formalisms.}
    \label{tab:non_state_space}
    \begin{tabular}{|c|c|p{13cm}|}
        \hline
        \rowcolor[rgb]{ .91,  .91,  .91} \textbf{Formalism} & \textbf{Paper} & \textbf{Brief Description} \\
        \hline
        
        \multirow{10}{*}{\textbf{RBD}}  
        & Ahmed et al.~\cite{rbd1} & Reliability assessment of a virtual data center (VDC) in a cloud computing architecture  \\
        \cline{2-3} & Santos et al.~\cite{rbd2} & Availability/Energy-aware SFC placement, where the chain is modeled via RBD \\
        \cline{2-3} & Jansen et al.~\cite{rbd2bis} & Reliability analysis of virtualized and physical nodes via RBD series/parallel rules \\
        \cline{2-3} & Wang et al.~\cite{rbd2ter} & Availability model of VNFs composing a Service Function Chain \\
        \cline{2-3} & Chantre et al.~\cite{rbd2quater} & Series/parallel redundant models in 5G NFV-based networks \\
        \cline{2-3} & Ding et al.~\cite{rbd2quinquies} & Reliability services in NFV with series/parallel VNF modeling \\
        \cline{2-3} & Li et al.~\cite{rbd2sexies} & Availability-aware provision of SFC in a MEC environment with redundant VNF modeling \\
        \cline{2-3} & Fan et al.~\cite{rbd2septies} & VNF series/parallel modeling for availability analysis of service chains \\
        \cline{2-3} & Wang et al.~\cite{rbd2octies} & Availability characterization of physical and virtual resources in SFCs via series/parallel models \\
        \cline{2-3} & Souza et al.~\cite{rbd3} & Availability evaluation of NFV datacenters whose components are modeled via RBD \\
        \hline
        
        \multirow{8}{*}{\textbf{FT}}  
        & Li et al.~\cite{ft1} & Reliability model for cloud computing with FTs modeling various fault states of VMs \\
        \cline{2-3} & Butoi et al.~\cite{ft2} & FTA applied in virtualized environments with a distributed and multi-agent approach \\
        \cline{2-3} & Dai et al.~\cite{ft3} & FT modeling of IEC 61499 function blocks executed on both physical controllers and cloud services \\
        \cline{2-3} & Zhu et al.~\cite{ft4} & FT modeling of 5G core network faults \\
        \cline{2-3} & Paulraj et al.~\cite{ft5} & Design of a fault-tolerant cloud data center, where FTA is used to study the VMs failure process \\
        \cline{2-3} & Butoi et al.~\cite{ft6} & VMs failures characterization in an IaaS-based environment via FTA \\
        \cline{2-3} & Paulraj et al.~\cite{ft7} & Analysis of fault-tolerant cloud systems via FT modeling \\
        \cline{2-3} & Mani et al.~\cite{ft8} & FT usage to identify repairs and failures of VMs and to reduce energy consumption in cloud computing \\
        \hline 
    \end{tabular}
\end{table}

\footnotesize
\renewcommand{\arraystretch}{1.2}
\begin{table}[h]
    \centering
    \caption{Works using stochastic Markovian state-space formalisms.}
    \label{tab:state_space_classic}
    \begin{tabular}{|c|c|p{13cm}|}
        \hline
        \rowcolor[rgb]{ .91,  .91,  .91} \textbf{Formalism} & \textbf{Paper} & \textbf{Brief Description} \\
        \hline
        
        \multirow{11}{*}{\centering \textbf{CTMC}} 
        & Farooq et al.~\cite{ctmc1} & Availability/Reliability modeling of 5G base stations with different states (e.g., working, failed) \\
        \cline{2-3} & Escheikh et al.~\cite{ctmc2} & CTMC modeling of load balancers to distribute network traffic among SDN controllers \\
        \cline{2-3} & Thein et al.~\cite{ctmc3} & Markov model of virtualized two-node cluster with various states (active, standby, rejuvenated, etc.) \\
        \cline{2-3} & Thein et al.~\cite{ctmc3bis} & Markov model of Active/Standby VMs with four states (healthy, unstable, rejuvenation, failure) \\
        \cline{2-3} & Thein et al.~\cite{ctmc3ter} & VM-based software rejuvenation for high availability of application server systems \\
        \cline{2-3} & Song et al.~\cite{ctmc4} & CTMC model to predict the rate of requests forwarded to cloud providers by a federation of hybrid clouds \\
        \cline{2-3} & Liu et al.~\cite{ctmc5} & Three-state (failure, migration, recovery) CTMC model of a VM in an edge server \\
        \cline{2-3} & Changa et al.~\cite{ctmc6} & Survivability model of a virtualized system after service breakdown through CTMC \\
        \cline{2-3} & Hong et al.~\cite{ctmc7} & Availability assessment of multiple cluster nodes with common mode failures via Markov models \\
        \cline{2-3} & Wakuda et al.~\cite{ctmc8} & Markov model for unavailability analysis of middlebox functions with multiple backup servers \\
        \cline{2-3} & Nozomi et al.~\cite{ctmc9} & VM backup allocation model to minimize unavailability using Markov models to analyze transition states \\
        \hline
        
        \multirow{6}{*}{\centering \textbf{MRM}}  
        & Kirsal et al.~\cite{mrm1} & Performance (QoS) and availability (failure/repair) of cloud systems via MRM modeling \\
        \cline{2-3} & de S. Matos et al.~\cite{mrm1ter} & Steady-state availability analysis of a virtualized server via MRM \\
        \cline{2-3} & Sun et al.~\cite{mrm1quater} & MRM used to analyze the correlation between reliability and performance in a cloud service \\
        \cline{2-3} & Hossfeld et al.~\cite{mrm2} & Online cloud gaming characterization via MRM models \\
        \cline{2-3} & Hossfeld et al.~\cite{mrm2bis} & User-centric MRM models to characterize Google Stadia cloud gaming \\
        \cline{2-3} & Sana et al.~\cite{mrm3} & Performance/dependability MRM-based model to design a reliable cloud radio access network \\
        \hline 
        
        \multirow{9}{*}{\centering \textbf{MSS}}  
        & De Simone et al.~\cite{mss7} & Performability assessment of multi-provider IMS architecture via MSS modeling of containers \\
        \cline{2-3} & Mo et al.~\cite{mss3} & Availability assessment of cloud computing systems via multi-valued decision diagrams to represent MSSs \\
        \cline{2-3} & Mo et al.~\cite{mss4} & Multi-valued decision diagram approach for performability of cloud nodes with different state probabilities \\
        \cline{2-3} & Di Mauro et al.~\cite{mss5} & MUGF approach to deal with SFC availability, benefiting from MSS representation of virtual nodes \\
        \cline{2-3} & De Simone et al.~\cite{mss6} & Performability evaluation of a container-based IMS infrastructure via MSS modeling \\
        \cline{2-3} & De Simone et al.~\cite{mss8} & MSS formalization of containerized network functions belonging to multi-tenant SFCs \\
        \cline{2-3} & Di Mauro et al.~\cite{mss9} & MSS modeling of an SDN controller supervising different virtual provider instances \\
        \cline{2-3} & Di Mauro et al.~\cite{mss10} & Reliability assessment of an SDN controller through MSS modeling \\
        \cline{2-3} & Arakawa et al.~\cite{mss11} & MSS formalization of VNFs belonging to SFCs aimed at availability evaluation \\
        \hline 
    \end{tabular}
\end{table}
\normalsize

\footnotesize
\begin{table}[h]
    \centering
    \caption{Works using stochastic Markovian (\textit{Petri-based}) state-space formalisms.}
    \label{tab:state_space_petri}
    \begin{tabular}{|c|c|p{13cm}|}
        \hline
       \rowcolor[rgb]{ .91,  .91,  .91} \textbf{Formalism} & \textbf{Paper} & \textbf{Brief Description} \\
        \hline
        \multirow{15}{*}{\textbf{SPN}} 
        & Santos et al.~\cite{spn1} & SPN to model redundancy strategies for SFC availability \\
        \cline{2-3} & Schoenen et al.~\cite{spn2} & SPN modeling of network virtualization resource pooling \\
        \cline{2-3} & Zabala et al.~\cite{spn3} & SPN modeling of virtualized firewall aimed at performance analysis \\
        \cline{2-3} & Carvalho et al.~\cite{spn4} & SPN to represent a MEC scenario and analyze its performance \\
        \cline{2-3} & Silva et al.~\cite{spn5} & Mobile cloud computing performance prediction via SPN \\
        \cline{2-3} & Jammal et al.~\cite{spn6} & Availability model of cloud services via SPN formalism \\
        \cline{2-3} & Mendonça et al.~\cite{spn7} & SPN and fault injection to evaluate availability of cloud-based disaster recovery solutions \\
        \cline{2-3} & Ghosh et al.~\cite{spn8} & OpenStack Swift and CephFS reliability modeling through SPN \\
        \cline{2-3} & Silva et al.~\cite{spn9} & Performability evaluation of disaster-tolerant IaaS infrastructures via SPN modeling \\
        \cline{2-3} & Brilhante et al.~\cite{spn10} & Dependability assessment of the cloud-based platform Eucalyptus \\
        \cline{2-3} & Brito et al.~\cite{spn11} & Availability assessment of a MEC system via SPN characterization \\
        \cline{2-3} & Rui et al.~\cite{spn12} & Reliability assessment of SFCs with VNF migration strategy \\
        \cline{2-3} & Jammal et al.~\cite{cspn1} & High-availability evaluation of critical cloud services with scoring systems + Colored SPN \\
        \cline{2-3} & Rizwan Ali et al.~\cite{cspn2} & Availability analysis of resource utilization in cloud computing via Colored SPN modeling \\
        \cline{2-3} & Dong et al.~\cite{cspn3} & Availability evaluation of SFCs under different elasticity strategies \\
        \cline{2-3}  & Li et al.~\cite{cspn4} & Resilience evaluation of virtualized $5$G and beyond networks \\
        
        \hline
        \multirow{4}{*}{\textbf{GSPN}} 
        & Zhu et al.~\cite{gspn1} & Availability evaluation of VNFs in active-standby configurations through GSPN \\
        \cline{2-3} & Rodrigues et al.~\cite{gspn2} & Performability evaluation of private cloud computing via GSPN modeling of NoSQL DBMS \\
        \cline{2-3} & Almutairi et al.~\cite{gspn3} & GSPN model for Denial of Service attacks in SDN \\
        \cline{2-3} & Liu et al.~\cite{cgspn1} & Colored GSPN for cloud data center reliability evaluation \\
        \hline
        \multirow{10}{*}{\textbf{SAN}} 
        & Vizarreta et al.~\cite{san0} & Failure dynamics of SDN controller via SAN modeling \\
        \cline{2-3} & Tola et al.~\cite{san0bis} & SAN availability models for network services and VNFs \\
        \cline{2-3} & Tola et al.~\cite{san1} & SAN availability model for NFV-MANO component with software rejuvenation \\
        \cline{2-3} & Tola et al.~\cite{san1bis} & Availability and sensitivity analyses of NFV services via SAN modeling \\
        \cline{2-3} & Entezari-Maleki et al.~\cite{san2} & SAN modeling aimed at evaluating VMs performance via scale up/down operations \\
        \cline{2-3} & Gonzalez et al.~\cite{san3} & Availability assessment of virtual Evolved Packet Core using SAN \\
        \cline{2-3}  &   Roohitavaf al.~\cite{san3new} & Availability modeling and evaluation of cloud virtual data centers \\
   	\cline{2-3}  & Pathirana al.~\cite{san3new2} & SAN              availability modeling of crucial components in $5$G-MEC         systems \\
	\cline{2-3} & Ataie et al.~\cite{san3bis} & Availability       evaluation of dispatching policies in cloud environments via    SAN modeling \\
    \cline{2-3}  & Tola et al.~\cite{san4} & Availability modes characterization in NFV-enabled network services \\
        
        \hline
        \multirow{25}{*}{\textbf{SRN}} 
        & Di Mauro et al.~\cite{srn_fig} & SRN modeling of containerized IMS aimed at availability evaluation \\
        \cline{2-3} & Longo et al.~\cite{srn1} & SRN availability modeling for IaaS cloud infrastructures \\
        \cline{2-3} & Ghosh et al.~\cite{srn1b} & SRN sub-modeling for IaaS cloud availability analysis \\
        \cline{2-3} & Ataie et al.~\cite{srn1bis} & Two SRN models to evaluate performance, availability, and power consumption of IaaS cloud \\
        \cline{2-3} & Liu et al.~\cite{srn1ter} & Monolithic and interacting SRN sub-models for IaaS availability analysis \\

  \cline{2-3}  & Entezari-Maleki et al.~\cite{srn2} & Performability evaluation via SRN modeling of grid resource computing \\
			\cline{2-3} & Liu et al.~\cite{srn3} &  Availability study of VMs with placement and migration modeled using SRN \\			
			\cline{2-3}  & Machida et al.~\cite{srn4} & SRN Availability model of virtualized servers aimed at analyze VMM rejuvenation \\
			\cline{2-3} & Shojaee et al.~\cite{srn4bis} & Availability analysis of cloud virtualization via SRN \\
        \cline{2-3} & Nguyen et al.~\cite{srn4ter} & SRN availability modeling of cloud data centers with disaster recovery \\
        \cline{2-3} & Cao et al.~\cite{srn5} & Experience-availability model of online cloud services via SRN \\
        \cline{2-3} & Torquato et al.~\cite{srn6} & Interacting SRN models for availability evaluation of VM migration \\
        \cline{2-3} & Torquato et al.~\cite{srn7} & SRN availability and security modeling of VM migration \\
        \cline{2-3} & Di Mauro et al.~\cite{srn8} & SRN availability modeling of OpenStack Swift (Software-defined storage) \\
        \cline{2-3} & Nguyen et al.~\cite{srn8bis} & Failure/Recovery analysis via SRN models of virtualized servers \\
        \cline{2-3} & Zhang et al.~\cite{srn9} & Performability analysis of drone computing via SRN characterization \\

\cline{2-3} & Di Mauro et al.~\cite{srn10} & Performability evaluation of SFCs via automated REST-based framework  \\
			\cline{2-3}  & Di Mauro et al.~\cite{srn10bis} & SRN modeling of different containerized deployments of IMS \\
			\cline{2-3}  & Paing~\cite{srn10quater} &  SRN modeling of a cluster of SDN controllers   \\			
			\cline{2-3}   & Han et al.~\cite{srn10quinquies} & SRN availability/reliability model for power consumption in SDN-based infrastructures  \\
			\cline{2-3}  & Nguyen et al.~\cite{srn10sexies} & SRN modeling for SDN architectures \\
			\cline{2-3}  & Machida et al.~\cite{srnaut1} &  Activity diagrams/SRNs mapping for cloud service management   \\
            \cline{2-3}  & Di Mauro et al.~\cite{srnaut2} &  Automated framework for SFC high-availability management via SRN   \\
			\cline{2-3}  & Di Mauro  et al.~\cite{srnaut3} &  Automated generation of availability models for SFCs assisted by SRN   \\
			\cline{2-3}    & Di Mauro  et al.~\cite{srnaut4} &  EMF tool for automated generation of SFCs via SRN   \\

        \hline
    \end{tabular}
\end{table}

\footnotesize
\renewcommand{\arraystretch}{1.2}
\begin{table}[h]
    \centering
    \caption{Works using stochastic non-Markovian state-space formalisms.}
    \label{tab:state_space_nonmarkov}
    \begin{tabular}{|c|c|p{13cm}|}
        \hline
        \rowcolor[rgb]{ .91,  .91,  .91} \textbf{Formalism} & \textbf{Paper} & \textbf{Brief Description} \\
        \hline
        
        \multirow{12}{*}{\centering \textbf{SMP}} 
& Li et al.~\cite{smp1} & Semi-Markov model to characterize VNF and SFC suffering from software aging \\
			\cline{2-3}   & Liu et al.~\cite{smp2} & Semi-Markov models to capture behaviors of VMs and VMMs with software aging \\
			\cline{2-3}  & Bai et al.~\cite{smp3} & Transient and steady-state availability of MEC-SFC services via SMP \\
			\cline{2-3}    & Bai et al.~\cite{smp4} & SMP to evaluate the impact of rejuvenation strategies on the SFC service reliability/availability \\
			\cline{2-3}    & Bai et al.~\cite{smp5} & SMP to analyze the effectiveness of proactive rejuvenation techniques in micro service chains \\
			\cline{2-3}  & Bai et al.~\cite{smp6} & SMP to model application services availability in virtualized systems \\
			\cline{2-3}  & Bai et al.~\cite{smp7} & SMP to characterize resilience of Unmanned Aerial Vehicle-based MEC services \\
			\cline{2-3}   & Ivanchenko et al.~\cite{smp8} & Availability of IaaS services via SMP modeling \\
			\cline{2-3}   & Ivanchenko et al.~\cite{smp9} & Markov/Semi-Markov models aimed at availability assessment of cloud server systems \\
			\cline{2-3}   & Mengistu et al.~\cite{smp10} & SMP models for availability/reliability prediction for volunteer cloud systems \\
			\cline{2-3}    & Ivanchenko et al.~\cite{smp11} & SMP approach to characterize sudden failures of physical machines in IaaS \\
   			\cline{2-3}  & Kharchenko et al.~\cite{smp12} & SMP approach for availability modeling of cloud and IoT systems \\

        \hline
        
        \multirow{3}{*}{\centering \textbf{MRGP}}  
& Bai et al.~\cite{mrp1} & Availability modeling of application services, VM, and VMMs, via MRGP  \\
			\cline{2-3}    & Ning et al.~\cite{mrp2} & MRGP-based performance analysis of computing systems suffering from software aging  \\
			\cline{2-3}     & Machida et al.~\cite{mrp3} & Performability of RAID storage systems via MRGP modeling \\
\hline
        \multirow{3}{*}{\centering \textbf{PTE}}  
        & Okamura et al.~\cite{pte1} & PTE to analyze the pointwise availability of virtual-machines with software rejuvenation policies   \\
			\cline{2-3}   & Bruneo et al.~\cite{pte2} & PTE-based evaluation for the optimal rejuvenation policy maximizing VMM availability  \\
			\cline{2-3}    & Mani et al.~\cite{pte3} & Availability analysis of fault tolerant cloud systems via PTE \\
        \hline 
    \end{tabular}
\end{table}
\normalsize

\footnotesize
\renewcommand{\arraystretch}{1.2}
\begin{table}[h]
    \centering
    \caption{Works using multi-level formalisms.}
    \label{tab:multilev}
    \begin{tabular}{|c|c|p{13cm}|}
        \hline
        \rowcolor[rgb]{ .91,  .91,  .91} \textbf{Formalism} & \textbf{Paper} & \textbf{Brief Description} \\
  \hline
        \multirow{4}{*}{\textbf{FT/CTMC}} 
        & Kim et al.~\cite{multilev1} & Hierarchical availability modeling (FT high level, CTMC low level) \\
        \cline{2-3} & Jhawar et al.~\cite{multilev2} & Fault tolerance mechanisms in IaaS cloud environments \\
        \cline{2-3} & Paharsingh et al.~\cite{multilev3} & Availability modeling of Triple Modular Redundant System (TMR) via FT/CTMC \\
        \cline{2-3} & Nguyen et al.~\cite{multilev5} & Resiliency evaluation of medical edge networks via FT/CTMC modeling \\
        \hline
        \multirow{6}{*}{\textbf{RBD/SPN}} 
        & Silva et al.~\cite{multilev7} & Tool to evaluate dependability of data centers combining RBD and SPN \\
        \cline{2-3} & Callou et al.~\cite{multilev8} & Combination of RBD and SPN to assess sustainability and dependability of data centers \\
        \cline{2-3} & Sousa et al.~\cite{multilev9} & Hierarchical RBD/SPN approach to evaluate dependability of the Eucalyptus platform \\
        \cline{2-3} & Bezerra et al.~\cite{multilev10} & Performability of storage services in private clouds via RBD/SPN \\
        \cline{2-3} & Bezerra et al.~\cite{multilev10bis} & Hierarchical strategy to evaluate availability and performance of cloud storage services \\
        \cline{2-3} & Lira et al.~\cite{multilev10ter} & RBD/SPN availability modeling of virtualized architectures \\
        \hline
        \multirow{5}{*}{\textbf{RBD/SRN}} 
        & Torquato et al.~\cite{multilev6} & Virtualized datacenter availability model via RBD/SRN \\
        \cline{2-3} & Di Mauro et al.~\cite{multilev6bis} & Availability analysis of containerized IMS in a cloud environment via hierarchical modeling \\
        \cline{2-3} & Di Mauro et al.~\cite{multilev6ter} & Performability evaluation of network service chains with delay constraints \\
        \cline{2-3} & Di Mauro et al.~\cite{multilev6quater} & SFC availability modeling and evaluation combining RBD and SRN \\
        \cline{2-3} & De Simone et al.~\cite{multilev6quinquies} & Performability assessment of 5G network nodes of Open5GS architecture via combined formalisms \\
        \hline
        \multirow{1}{*}{\textbf{FT/SRN}} 
        & Sebastio et al.~\cite{multilev18} & Availability evaluation of container management systems via FT/SRN \\
        \hline
        \multirow{2}{*}{\textbf{RBD/MRM}} 
        & Araujo et al.~\cite{multilev19} & Mobile cloud computing modeling with distinct redundancy mechanisms \\
        \cline{2-3} & Dantas et al.~\cite{multilev20} & Analysis of warm-standby replication procedure into virtualized networks \\
        \hline
        \multirow{4}{*}{\textbf{RBD/CTMC}} 
        & Melo et al.~\cite{multilev11} & Availability models for cloud environments designed for video streaming services \\
        \cline{2-3} & Melo et al.~\cite{multilev12} & Sensitivity analysis of Video on Demand streaming service in cloud \\
        \cline{2-3} & Melo et al.~\cite{multilev13} & Availability and sensitivity evaluation of Video on Demand hosted by Eucalyptus platform \\
        \cline{2-3} & Bezerra et al.~\cite{multilev14} & RBD/CTMC modeling of video services in cloud \\
        \hline
        \multirow{3}{*}{\textbf{3-level}} 
        & Di Mauro et al.~\cite{multilev4} & Availability model of VNFs composing a Service Function Chain via RBD/FT/CTMC \\
        \cline{2-3} & Nguyen et al.~\cite{multilev16} & Reliability and availability evaluation of cloud data centers via Reliability graphs/FT/SRN modeling \\
        \cline{2-3} & Nguyen et al.~\cite{multilev17} & Reliability and availability evaluation of SDN via Reliability graphs/FT/SRN modeling \\
        \hline
    \end{tabular}
\end{table}
\normalsize

\twocolumn
\section{Modeling Tools}
\label{sec:tools}

In this section we propose an overview of the most useful software tools to deal with reliability and availability modeling. We have selected a subset of tools that: $i)$ are both commercial and open-source/free, $ii)$ are officially and currently supported (or, at least, not discontinued), $iii)$ support a variety of formalisms to model reliability and availability aspects related to virtualized networks. Beyond a concise description, we provide a summary of the main features for each tool in Table~\refeq{tab:tools} further ahead. Moreover, for each tool, we provide reference to its main web page. 

\subsection{SHARPE}
SHARPE\footnote{https://sharpe.pratt.duke.edu/} (Symbolic Hierarchical Automated Reliability and Performance Evaluator)~\cite{tool_sharpe} has been developed at the Duke University (US) and is a versatile hierarchical modeling tool designed for analyzing stochastic models related to reliability, availability, performance, and performability. It both supports analytical solutions and simulations. 
This tool supports a variety of modeling techniques including: Markov chains (irreducible, acyclic, phase-type), semi Markov chains, reliability block diagrams, fault trees, reliability graphs, single-chain product-form queueing networks, multiple-chain product-form queueing networks, generalized stochastic Petri nets. Many of these formalisms can be also combined in a hierarchical way (e.g., high-level RBDs and low-level Markov Chains). Steady-state, transient and interval measures can be computed. SHARPE allows both GUI and textual interaction, is compatible with most OS platforms (Windows, Linux, Solaris), and, in general, with any JVM-based platform. It is free for academic use upon agreement.

\subsection{TimeNET}
TimeNET\footnote{http://timenet.tu-ilmenau.de} (Timed Net Evaluation Tool)~\cite{tool_timenet} has been developed at the Technische Universit{\"a}t of Berlin (Germany), and later maintained at the Universities of Dortmund and Leipzig. The tool has been optimized to deal with many aspects related to SPNs.
In particular, it provides support for: Deterministic and stochastic Petri nets (DSPNs) which allow exponentially distributed and deterministic firing delays, general SPNs (referred to as extended DSPNs) which allow for non-exponentially distributed firing delays, and DSPNs with concurrently enabled deterministic transitions (referred to as concurrent DSPNs).  
Depending on the SPN class, TimeNET supports analytical and numerical solutions. Moreover it allows for steady-state and transient analyses. TimeNET is equipped with a convenient user interface, and it is compatible with most OS platforms (Windows, Linux, Solaris) and with any JVM-based platform. It is free for non-commercial use.

\subsection{SPNP}
SPNP\footnote{https://trivedi.pratt.duke.edu/software\_packages/spnp} (Stochastic Petri Net Package)~\cite{tool_spnp} has been developed at the Duke University (US) and has been customized to deal with transient and steady-state analysis of GSPNs. It is mainly textual (very powerful for C-skilled users). 
SRNs are specified through CSPL (C based SRN Language) being an extension of the C  language with additional constructs to describe SRN models. Then, SRN specifications are converted into an MRM that is solved to compute steady-state, transient, and sensitivity measures.
SPNP is compatible with most OS platforms (Windows, Linux, Solaris), and is free for academic use upon agreement.

\subsection{Mercury}
Mercury\footnote{https://www.modcs.org/}~\cite{tool_mercury} is a tool developed and maintained by the MoDCS research group (Brazil) which supports a variety of modeling techniques aimed at performance and dependability analyses, including: FTs, RBDs, SPNs, CTMCs, DTMCs, Energy Flow Models (EFMs). Mercury allows to evaluate models characterized both by exponentially and non-exponentially distributed firing delays. It has a dedicated SPN editor and evaluator allowing for simulation and numerical analysis techniques and transient/steady-state evaluation. Differently from other tools, Mercury also embeds an Energy Flow Model editor and evaluator useful to evaluate attributes such as costs, sustainability, power consumption and sustainability of cloud-related infrastructures. Mercury can be used through CLI or GUI and is compatible with most OS platforms (Windows, Linux). To download the tool, a preliminary registration to the MoDCS web portal is needed.

\subsection{WebSPN}
WebSPN\footnote{https://webspn.unime.it/}~\cite{tool_webspn} is a modeling tool developed at the University of Messina (Italy). It has been mainly conceived for the analysis of non-Markovian SPNs, with the possibility of analyzing a wide class of models with multiple preemptive repeat different, preemptive resume, and preemptive repeat identical concurrently enabled generally distributed transitions. It is based on the client-server paradigm: the client side is dedicated to manage the design of models, whereas the server side is in charge to solve the models via some analyses engines written in C/C++. It is possible to separately install the client and server parts onto Linux-based platforms. The tool is free for non-commercial uses. 

\subsection{GreatSPN}
GreatSPN\footnote{https://www.di.unito.it/$\sim$greatspn/index.html/}~\cite{tool_greatspn} is a GUI-based tool developed at the University of Torino (Italy). The tool is aimed at modeling, validating, and evaluating performance of distributed systems through Generalized SPNs and their colored extension (Stochastic well-formed Nets). GreatSPN has a modular structure and allows to run different analysis modules on different machines in a distributed computing environment. The various modules are written in C, and use special storage techniques to save memory for intermediate result files and for program data structures.
GreatSPN is available for free for universities and non-profit organizations upon a license agreement, and is compatible with INTEL and PPC architectures.

\subsection{M{\"o}bius}
M{\"o}bius\footnote{https://www.mobius.illinois.edu/}~\cite{tool_mobius} is a tool developed at the University of Illinois Urbana-Champaign (US). The tool is aimed at modeling the behavior of complex systems with particular focus on reliability, availability, and performance of computer and network systems. 
Supported models include stochastic extensions of Petri nets, Markov chains and their extensions, and stochastic process algebras. Depending on the model, the tool supports both analytical solutions and simulations.
M{\"o}bius is available upon staff approval, and is compatible with Windows, Linux, and MacOS platforms.

\subsection{CPN IDE}
CPN IDE\footnote{https://cpnide.org/}~\cite{tool_cpn} formerly CPN Tools is a tool  originally developed at the Aarhus University (Denmark) and then transferred to Eindhoven University of Technology (The Netherlands). It has been conceived to deal with coloured and non-coloured Petri nets. It is compatible with platforms supporting JVM and offers a new JavaScript-based CPN editor which communicates through a REST interface with a Java-based controller. All the components are freely available.

\subsection{QPME}
QPME\footnote{https://se.informatik.uni-wuerzburg.de/software-engineering-group/tools/qpme/download/}~\cite{tool_qpme} is a tool made of two components: QPE (QPN Editor) and SimQPN (Simulator for QPNs). QPE provides a user-friendly graphical tool, whereas SimQPN provides a discrete-event simulation engine. This tool combines the modeling power of queueing networks and generalized SPNs. 
QPME is compatible with most OS platforms (Windows, Linux, Solaris), and is open-source.

\subsection{APNN-Toolbox}
APNN-Toolbox\footnote{https://ls4-www.cs.tu-dortmund.de/APNN-TOOLBOX/}~\cite{tool_apnn} is born as an open toolset to define a common exchange interface called Abstract Petri Net Notation (APNN) developed at the Dortmund University (Germany). It is equipped with a GUI to define and manage Petri nets. Quantitative analysis focuses on continuous time Markov chains (CTMCs) and their transient  and steady-state behavior. Kronecker representations are used to face the state space explosion problem.

\subsection{Snoopy}
Snoopy\footnote{https://www-dssz.informatik.tu-cottbus.de/DSSZ/Software/Snoopy}~\cite{tool_snoopy} is a tool written in C++ and developed at the University of Technology in Cottbus (Germany). This tool provides a Petri net framework where models can be hierarchically organized, allowing to deal with large networks. 
Snoopy is compatible with most OS platforms (Windows, Linux, Mac), and is distributed free of charge for academic use.

\begin{table*}[h]
	\centering
	\caption{Modeling tools and their main features. The first $14$ rows (in white) enumerate open-source/freeware tools, whereas the last $4$ rows (in light gray) enumerate commercial tools.}
	\small
	\renewcommand{\arraystretch}{1.3}
	\begin{tabular}{ p{2.7cm} | p{1.8cm}|p{4.3cm}|p{5cm}}
		\hline	
		
		\textbf{Authors/Company} &  \textbf{Tool} &  \textbf{Features} &  \textbf{Supported Methods} \\  \hline
		
		Sahner \& Trivedi \cite{tool_sharpe} & SHARPE & GUI, specification language, Win/Linux/Solaris and any JVM-based platform, free for academic use upon agreement & FTs, RBDs, Markov models, Semi-Markov models, MRMs, SPNs, SRNs, Multi-level models  \\ \hline
		
		German et al. \cite{tool_timenet} & TimeNET & GUI, Win/Linux/Solaris and any JVM-based platform, free for non-commercial use & SPNs, SRNs, non-Markovian SPNs  \\ \hline
		
		Ciardo et al. \cite{tool_spnp} & SPNP & Textual, specification language (CSPL), exploits the power of C language, Win/Linux/Solaris platforms, free for academic use upon agreement & Generalized SPNs, SRNs  \\ \hline
		
		Maciel et al. \cite{tool_mercury} & Mercury & GUI, Win/Linux, free for non-commercial uses & FTs, RBDs, SPNs, CTMCs, DTMCs, EFMs  \\ \hline
		
		Longo et al. \cite{tool_webspn} & WebSPN & GUI, Client/Server model, Linux, free for non-commercial uses & non-Markovian SPNs  \\ \hline
		
		Marsan et al. \cite{tool_greatspn} & GreatSPN & GUI, INTEL/PPC architectures, free for academic uses & Generalized SPNs, Stochastic Well-formed Nets  \\ \hline
		
		Courtney et al. \cite{tool_mobius} & M{\"o}bius & GUI, Win/Linux/MacOS, free for non-commercial uses & SPNs/SANs, Markov Chains, Stochastic Process Algebras   \\ \hline
		
		Verbeek \& Fahland \cite{tool_cpn} & CPN IDE & GUI, JVM-based platforms via REST interface, free & Coloured \& non-Coloured Petri Nets   \\ \hline
		
		Kouned et al. \cite{tool_qpme} & QPME & GUI, Win/Linux/Solaris, open source & Generalized SPNs  \\ \hline
		
		Bause et al. \cite{tool_apnn} & APNN-Toolbox & GUI, Win/Linux/Solaris and JVM-based platforms, open source & Abstract Petri Net Notation  \\ \hline
		
		Rohr et al. \cite{tool_snoopy} & Snoopy & GUI, Win/Linux/MacOS, free for academic uses &  Stochastic, Continuous and hybrid Petri nets \\ \hline
		
		Butler \cite{tool_sure} & SURE & Textual/GUI, abstract language, Win/Linux/SPARC platforms, NASA public domain license & 
		Semi-Markov models  \\ \hline
		
		Dingle et al. \cite{tool_pipe2} & PIPE2 & GUI, JVM-based platforms, MIT license &  Generalized SPNs  \\ \hline
		
		Paolieri et al. \cite{tool_oris} & ORIS Tool & GUI, Win/Linux/MacOS, open source &  SPNs, non-Markovian SPNs  \\ \hline

		\rowcolor[RGB]{229,228,226}
		Reliasoft & BlockSim\textregistered & GUI, Windows platforms, .NET Framework, Commercial license &  FMEA/FMECA, FTs, RBDs, Markov Chains, Maintainability analysis  \\ \hline
		
		\rowcolor[RGB]{229,228,226}
		Isograph  & Reliability Workbench\textregistered & GUI, Windows platforms, Commercial license & FMEA/FMECA, FTs, RBDs, Markov Chains, System Safety assessments   \\ \hline
		
		\rowcolor[RGB]{229,228,226}
		BQR & BQR Care\textregistered & GUI, Windows platforms, Commercial license &  FMEA/FMECA, FTs, RBDs, System Safety assessments   \\ \hline
		
		\rowcolor[RGB]{229,228,226}
		ITEM  & ITEM ToolKit\textregistered & GUI, Windows platforms, Commercial license &  FMEA/FMECA, FTs, RBDs, Maintainability analysis  \\ \hline

	\end{tabular}
	\label{tab:tools}
\end{table*}

\subsection{SURE}
SURE\footnote{https://shemesh.larc.nasa.gov/people/rwb/sure.html} (Semi-markov Unreliability Range Evaluator)~\cite{tool_sure} is born as a modeling tool developed at NASA (National Aeronautics and Space Administration). 
It is a reliability analysis tool useful to calculate upper and lower bounds on state probabilities for a large class of semi-Markov models. It has been customized to analyze fault-tolerant reconfigurable systems. A Java version of this tool (called WinSURE) works on Linux, Mac, and Windows operating systems, and it is distributed under the NASA public domain license.

\subsection{PIPE2}
PIPE2\footnote{https://pipe2.sourceforge.net/} (Platform Independent Petri net Editor 2)~\cite{tool_pipe2} is an open-source Java-based tool that supports the design and analysis of GSPN models. It was born as a postgraduate team programming project in the Department of Computing at Imperial College (London). The tool relies onto Performance Trees formalism, useful to provide graphical specification of performance queries. A Performance Tree query is represented as a tree structure that consists of nodes and connecting arcs. Being Java-based, this tool can be installed on a variety of software platforms. 

\subsection{ORIS Tool}
ORIS\footnote{https://www.oris-tool.org/}~\cite{tool_oris} has been developed at the University of Firenze (Italy). 
The tool allows to perform transient and steady-state analysis of Stochastic Time Petri Net and Markov Regenerative Processes, and transient analysis associated to semi-Markov processes, where the timed transitions can also be modeled in terms of non-exponential distributions. The tool supports different operating system (Windows/Linux/MacOS) and provides some Java API useful to perform extensive reliability/availability analyses considering the combination of several parameters and model variants.

\subsection{Reliasoft BlockSim\textregistered}
Reliasoft BlockSim\footnote{https://help.reliasoft.com/blocksim21/content/current\_application\_intro.htm} is a commercial product which provides a uniform framework to manage an ample variety of analyses for both repairable and non-repairable systems including reliability analysis, maintainability analysis, availability analysis, reliability optimization, throughput calculation, process flow diagrams, resource allocation, life cycle cost estimation and other system analyses. In particular it supports FMEA (Failure Modes and Effects Analysis) and FMECA (Failure Modes, Effects, and Criticality Analysis), which are systematic methodologies used to identify potential failure modes in a system, process, or product and assess their impact. It supports Windows-based operating systems.

\subsection{Isograph Reliability Workbench\textregistered}
Isograph Reliability Workbench\footnote{https://www.isograph.com/software/reliability-workbench/} is a comprehensive suite of tools designed for reliability, availability, maintainability, and safety (RAMS) analysis.
It supports various formalisms and analysis, including: Reliability block diagram analysis, FTA, common cause and importance analysis, Event Tree Analysis (ETA) with multiple risk categories, Markov analysis including multi-phase modelling, FMEA and FMECA. It supports Windows-based operating systems.

\subsection{BQR Care\textregistered}
BQR Care\footnote{https://www.bqr.com/products/care} software provides a comprehensive suite for system reliability management, focusing on several key areas: Fault Tree Analysis, Life Cycle Cost Analysis, Reliability Block Diagrams analysis, Maintenance Optimization, FMEA and FMECA. It supports Windows-based operating systems.

\subsection{ITEM ToolKit\textregistered}
ITEM ToolKit\footnote{https://www.itemtoolkit.com/} offers comprehensive tools for reliability analysis, supporting various methodologies and standards, using a proprietary graphical user interface. This tool supports the following functionalities and activities: Reliability Prediction, Reliability Block Diagrams Analysis, Fault Tree Analysis, Event Tree Analysis, Markov analysis, Maintainability and Spares Analysis.  It supports Windows-based operating systems.

\section{Conclusion and open challenges}
\label{sec:conclusions}

This survey provides an in-depth examination of various aspects—including standards, techniques, and software tools—related to the availability and reliability modeling and analysis of virtualized network infrastructures. The primary goal is to offer a convenient reference for researchers, practitioners, and network designers involved in the modeling and deployment processes of reliable and highly available virtualized networks.

First, we highlight critical aspects of the standardization process handled by the European Telecommunications Standards Institute (ETSI), which has released a series of guidelines on reliability and availability issues related to virtualized networks. Next, we propose a taxonomy of formalisms adopted to model various reliability and availability aspects, along with a mapping of references related to the usage of these formalisms in virtualized infrastructures. When possible, we have avoided mathematical details, focusing instead on the crucial information necessary to understand how to employ these formalisms to model reliable virtualized networks.

Moreover, we present and discuss a comprehensive inventory of the most well-known software tools useful for managing these formalisms. For each tool, we have extracted the main features (supported operating systems, implemented formalisms, etc.).

Finally, we pinpoint a set of relevant open issues and research challenges which deserve a deeper investigation.
\subsection{Scalability Issues}

A major challenge in large-scale virtualized networks is accurately modeling the interplay between availability, reliability, and performance. Traditional modeling techniques often treat these aspects separately, which can lead to oversimplified assumptions and suboptimal designs, particularly as networks scale. In complex environments—such as large-scale cloud data centers or 5G network cores—failures, resource contention, and dynamic workload shifts create intricate dependencies that existing models struggle to capture.

To address these limitations, future research should focus on developing scalability-aware modeling techniques that explicitly incorporate the interdependencies between availability, reliability, and performance in large-scale virtualized infrastructures. This includes exploring hybrid analytical and simulation-based approaches capable of scaling with network complexity while maintaining predictive accuracy. Ensuring that these models align with industry standards, such as those defined by ETSI, will be crucial for their practical adoption.

\subsection{Fault Tolerance and Resilience}
As virtualized networks grow in complexity, ensuring fault tolerance becomes increasingly challenging due to the dynamic and distributed nature of these environments. Traditional resilience strategies, often designed for static and hardware-based infrastructures, may struggle to adapt to the failure modes of virtualized network elements, where faults can propagate unpredictably across software-defined components.

Existing formal methods often assume static configurations, making them less effective in modeling transient failures and cascading effects in cloud-native and edge computing environments. Future research should focus on designing adaptive fault-tolerance frameworks that leverage anomaly detection, self-healing mechanisms, and cross-layer resilience strategies.

\subsection{Security and Trust}
Security issues in virtualized networks are amplified due to the complexity of such systems. One significant challenge is addressing the reliability issues that can emerge throughout the VNF software lifecycle, from implementation and coding to verification and testing. Additionally, the redundancy that enhances reliability and availability also increases the attack surface, introducing new security vulnerabilities. Balancing the need for redundant systems with the imperative to secure these systems against potential threats is a critical area of future research. 

\subsection{Energy Efficiency}
As the demand for more reliable and available virtualized networks grows, also the energy consumption required to maintain them grows. Future challenges in this area include modeling the trade-offs between energy efficiency and reliability/availability through specific techniques. Indeed, while redundancy and robust fault tolerance are essential for maintaining high availability, they can also lead to increased energy usage. Research should also focus on integrating these models into energy-efficient technologies and exploring dynamic management strategies that adjust energy consumption based on real-time network conditions, all while adhering to ETSI guidelines.

\subsection{Automation and Orchestration}
The complexity of managing virtualized networks necessitates advanced orchestration tools that can automate deployment, management, and fault isolation processes. However, as these orchestration processes become more sophisticated and less controlled by human operators, ensuring that they do not compromise reliability and availability during operations is a significant challenge. Future research should focus on enhancing orchestration tools with service and network verification features leveraging real-time reliability and availability models, ensuring that automated processes adhere to ETSI standards and maintain service continuity throughout deployment and operation.

\subsection{Software Tools Enhancement}
Standardization of software tools across the industry is essential for consistency in network virtualization. A significant challenge is to promote the adoption of common formalisms and software tools. This includes developing ready-to-use reliability templates and APIs that are tailored to specific components like VNFs, VMs, and hypervisors. By standardizing these tools and enhancing their usability, practitioners can more easily implement and assess reliability and availability, leading to more robust and reliable virtualized networks.

\subsection{Machine Learning and Artificial Intelligence}
Machine learning (ML) and Artificial Intelligence (AI) offer promising opportunities for enhancing the reliability and availability of virtualized networks. Future challenges involve integrating ML and AI with formal reliability models, such as Markov chains or fault trees, to enable risk assessment, predictive maintenance, anomaly detection, and automated fault recovery. Additionally, ensuring the interoperability of ML/AI models with existing ETSI-standardized systems is crucial. Successfully addressing these challenges will lead to more intelligent, proactive management of network resources, improving overall performance and reducing downtime.

\subsection{Emerging Cloud and Network Virtualization Technologies}
Virtualization technologies are constantly evolving, and this evolution brings new challenges to the traditional techniques for reliability and availability modeling.  A current research trend for example consists in leveraging non-standard computing platform (accelerators, GPU, TPU, etc.) for NFV. Similarly, deployment flexibility is increasing (from virtual machines, to containers, from microservices to serverless and function-as-a-service compute models), hence bringing more pervasiveness and complexity to the interdependencies across the elements of the virtualized systems. The impact of these new technical direction on standaradization, formalisms and software tools is open for investigation.

\section{ACKNOWLEDGMENTS}
The work of Mario Di Mauro, Walter Cerroni, Fabio Postiglione, and Massimo Tornatore was partially supported by the European Union - Next Generation EU under the Italian National Recovery and Resilience Plan (NRRP), Mission 4, Component 2, Investment 1.3, CUPs E83C22004640001, D43C22003080001, and J33C22002880001, partnership on ``Telecommunications of the Future'' (PE00000001 - program “RESTART”).

\bibliographystyle{unsrt}
\bibliography{biblio}

\end{document}